\documentclass[referee]{raa}

\usepackage{amssymb}
\usepackage{lipsum}
\usepackage{gensymb}
\usepackage{listings}
\usepackage{verbatim}
\usepackage{microtype,siunitx,booktabs}
\usepackage{makecell}
\usepackage{array}
\usepackage{newtxtext,newtxmath}
\usepackage[T1]{fontenc}
\usepackage{amsmath}
\usepackage{natbib}
\usepackage{rotating}
\usepackage{multirow}
\usepackage{graphicx,times}
\usepackage{hyperref}
\usepackage{float}
\usepackage{xcolor}
\usepackage{orcidlink}
\bibpunct{(}{)}{;}{a}{}{,}

\begin{document}

\title{Chemo-dynamical Analysis of Eight UBC Open Clusters}

\volnopage{ {\bf 20XX} Vol.\ {\bf X} No. {\bf XX}, 000--000}
\setcounter{page}{1}

\author{
Remziye Canbay\inst{1^*,\orcidlink{0000-0003-2575-9892}}
\and Deniz Cennet \c{C}ınar\inst{2,\orcidlink{0000-0001-7940-3731}}
\and \"{O}zcan \c{C}al\i\c{s}kan\inst{2,\orcidlink{0009-0003-5839-8007}}
\and Bur\c{c}in Tan\i k \"{O}zt\"{u}rk \inst{3,\orcidlink{0000-0002-6372-2372}}
\and Seval Ta\c{s}demir\inst{2,\orcidlink{0000-0003-1339-9148}} 
\and Ege Eraydın\inst{4,\orcidlink{0009-0007-8615-9600}}
\and Sel\c{c}uk Bilir\inst{3,\orcidlink{0000-0003-3510-1509}}
}

\institute{
Akdeniz University, Faculty of Science, Department of Space Sciences and Technologies, 07058, Antalya, T\"{u}rkiye {\it rmzycnby@gmail.com} \\ 
\and
Istanbul University, Institute of Graduate Studies in Science, Programme of Astronomy and Space Sciences, 34116, Istanbul, Türkiye
\and
Istanbul University, Faculty of Science, Department of Astronomy and Space Sciences, 34119, Beyazıt, Istanbul, Türkiye \\
\and
Yeditepe University, Faculty of Arts and Sciences, Programme of Physics, 34640, Ata\c{s}ehir, Istanbul, Türkiye\\
}

\vs \no
 
\abstract{We present a comprehensive chemo-dynamical analysis of eight open clusters selected from the UBC catalog using Gaia DR3 data. These clusters are located at heliocentric distances of $\sim$2-5 kpc, probing relatively distant regions of the Galactic disk beyond the solar neighborhood. Cluster membership is determined using the \texttt{UPMASK} algorithm, while structural parameters are derived from radial density profiles through King model fitting combined with MCMC sampling. Their structural parameters reveal a diversity of internal configurations, from diffuse to moderately concentrated systems. Fundamental astrophysical parameters (extinction, distance, metallicity, and age) are obtained via Bayesian isochrone fitting based on \texttt{PARSEC} models. The clusters span a wide age range ($\sim$20 Myr to $\sim$5 Gyr), and show a broad metallicity distribution ($-0.34\leq {\rm [Fe/H]~(dex)}\leq +0.25$). Orbital analysis based on backward integrations shows that all clusters follow nearly circular orbits ($e \approx 0.03-0.09$) with low vertical distance from the Galactic plane ($Z_{\rm max}<0.4$ kpc), confirming their membership in the Galactic thin disk and indicating dynamically cold kinematics. The comparison between the inferred traceback early orbital radii and present-day guiding radii indicates modest radial displacements, with $\Delta R < 0.5$ kpc for the UBC sample. These offsets are consistent with mild radial redistribution expected for young and dynamically cold open clusters, rather than strong radial migration. Thus, the results suggest that radial migration should be considered when interpreting the present-day spatial and chemical distribution of these clusters, although the inferred migration amplitudes remain moderate. Our results further demonstrate that open clusters located at relatively large distances from the Sun can be characterized with high precision using Gaia DR3 data, and that mild radial redistribution should be considered when interpreting the present-day distribution of stellar populations in the Galactic disc.}

\authorrunning{Canbay et al.} 

\titlerunning{Chemo-dynamical Analysis of Eight UBC Open Clusters}  

\maketitle


\section{Introduction}
\label{sec:introduction}
Open clusters (OCs) are key systems for studying star formation, stellar evolution, and the structure of the Galaxy. Because their member stars share similar chemical compositions and have well-constrained ages, OCs provide reliable benchmarks for testing stellar evolution models \citep{Friel95, Lada2003}. Most Galactic OCs contain from a few dozen to a few thousand stars. Since these stars formed from the same molecular cloud, they generally have comparable chemical properties. This internal uniformity makes OCs well-suited to examining the relationships among stellar age, mass, and luminosity. In addition, OCs trace the chemical and dynamical evolution of the Milky Way. They offer constraints on processes such as radial migration, disk evolution, and the chemical enrichment of the interstellar medium \citep[e.g.,][]{Friel95, Jacobson2016, Donor2020, Cantat-Gaudin2020, Cantat-Gaudin_Anders2020, Netopil2022, Spina2022, Joshi24, Otto2026}. Studies of Galactic OCs therefore contribute not only to our understanding of stellar evolution, but also to improving distance estimates, refining cluster age determinations, and modeling the large-scale structure of the Galaxy. For these reasons, OCs remain key targets in both observational and theoretical astrophysics.

Before the Gaia era, studies of OCs in the Milky Way mainly relied on ground-based observations. Although thousands of clusters had been identified through classical photometric and astrometric surveys, many of them lacked reliable estimates of distance, age, and membership \citep[e.g.,][]{Dias2002, Kharchenko2005, Kharchenko2012, Kharchenko2013, Dias2014, Dias2018}. Wide-field sky surveys improved the situation to some extent. In particular, infrared surveys such as the Two Micron All Sky Survey \citep[2MASS;][]{Skrutskie2006} and the VISTA Variables in the Via Lactea survey \citep[VVV;][]{Minniti2010} made it possible to detect clusters in regions heavily affected by dust extinction and to identify distant or faint OCs \citep[e.g.,][]{Bica2003, Bilir2006a, Bilir2010, Borissova2014, Barba2015}. Observations in the infrared reduce the impact of interstellar extinction, allowing clusters in the Galactic plane and star-forming regions to be observed more effectively. As a result, distant systems and clusters projected against dense stellar backgrounds could be examined more systematically. Despite these advances, identifying OCs in distant or crowded regions remained difficult. The limited precision of pre-Gaia astrometric data restricted accurate membership determination and constrained detailed analyses of massive stellar groups.

The high-precision astrometry and photometry provided by the Gaia space telescope \citep{Gaia2016} have transformed studies of OCs. With the Gaia Data Release 2 \citep[Gaia DR2;][]{Gaia2018} and Gaia Early Data Release 3 \citep[Gaia EDR3;][]{Gaia2021}, \citet{Cantat-Gaudin2020} identified thousands of new clusters based on accurate trigonometric parallaxes and proper-motion measurements. These studies applied statistical membership selection and density-based methods to separate cluster members from field stars. In addition, advanced search algorithms and machine learning approaches have led to the detection of previously unknown low-density and distant clusters \citep[e.g.,][]{Castro-Ginard2018, Castro-Ginard2019, Castro-Ginard2020, Sim2019, Qin2023, Chi2023a, Chi2023b, Chi2024, Chi2025, Dias2025, Dias2026}. Using the Gaia Data Release 3 catalog \citep[Gaia DR3;][]{Gaia2023}, \citet{Hunt2021, Hunt2023, Hunt2024} developed methods that combine statistical and kinematic criteria to improve member selection, especially for distant and sparse clusters. These efforts have supported both the discovery of new OCs and the refinement of membership lists and dynamical parameters for known clusters. The accuracy of Gaia data has therefore not only increased the number of identified OCs but also improved the reliability of their fundamental and dynamical properties \citep[e.g.][]{Cantat-Gaudin2018, Cantat-Gaudin2020, Liu2019, Cantat-Gaudin_Anders2020, Dias2021, Hunt2023, Hunt2024}. Compared to pre-Gaia catalogs, current OC samples are both larger and more robust, providing stronger constraints on the structure of the Galactic disk and its star formation history.

The University of Barcelona Clusters (UBC) catalog is a set of newly identified Galactic OCs discovered using the high-precision astrometric data from the Gaia mission\citep{Gaia2016}. The OCs were detected using data-mining techniques applied to a multidimensional astrometric space, combining stellar positions, trigonometric parallaxes, and proper motions to identify stellar overdensities. Using data from Gaia DR2 \citep{Gaia2018}, the first study applying the Density-Based Spatial Clustering of Applications with Noise (\textsc{DBSCAN}) clustering algorithm \citep{Ester1996} together with a neural-network classifier capable of recognizing cluster sequences in color-magnitude diagrams (CMDs) led to the discovery of 23 new OCs \citep{Castro-Ginard2018}. The search was subsequently expanded toward the Galactic anti-center, revealing 53 additional OCs \citep{Castro-Ginard2019}. A systematic exploration of the Galactic disk using the Gaia DR2 later resulted in the identification of 582 new OCs, of which 570 were cataloged with the UBC designation (UBC 1-570) \citep{Castro-Ginard2020}. With the improved astrometric precision of Gaia Early Data Release 3 \citep[Gaia EDR3,][]{Gaia2021}, the cluster-search methodology was further refined using the {\sc OCfinder} algorithm, leading to the discovery of 628 additional systems, typically labeled UBC 1001-1628 \citep{Castro-Ginard2022}. These discoveries substantially increased the census of Galactic OCs and provided an important dataset for studies of Galactic structure and stellar populations.

\begin{figure*}[!ht]
\centering
\includegraphics[width=\linewidth]{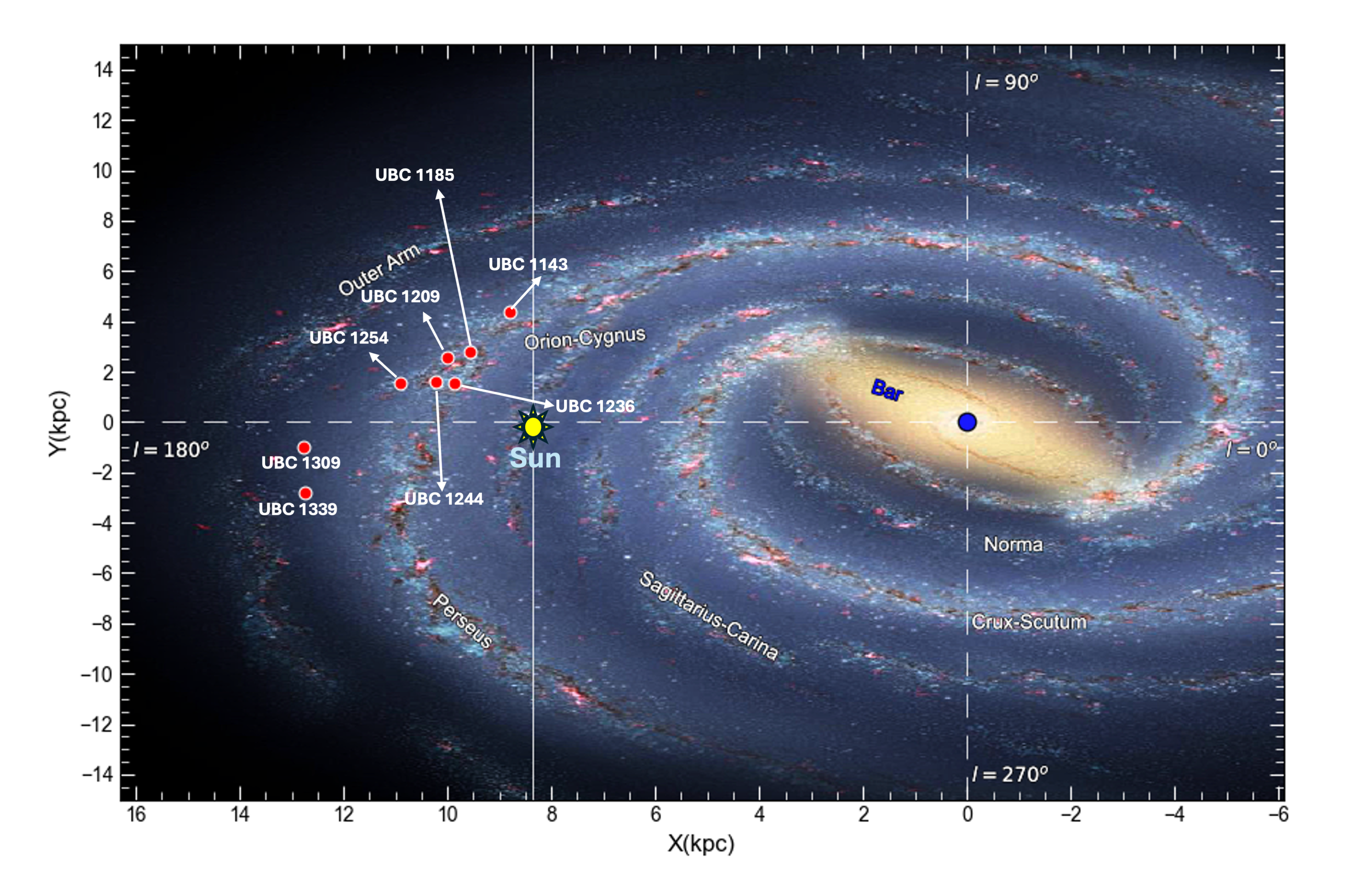}
\label{fig:location}
\caption{The positions of eight UBC OCs in the Milky Way Galaxy.}
\label{fig:location}
\end{figure*}

In this work, we present a detailed analysis of eight selected UBC OCs identified by \citep{Castro-Ginard2022} using Gaia EDR3 data \citep{Gaia2021}. We examine their structural and astrophysical parameters, kinematic characteristics, and Galactic orbital motions using advanced statistical and Bayesian methods. Considering their dynamical states, membership distributions, and kinematic consistency, we focus on UBCs 1143, 1185, 1209, 1236, 1244, 1254, 1309, and 1339. To better characterize their spatial configuration, Figure~\ref{fig:location} shows the distribution of these eight UBC OCs across the Galactic disk, highlighting their locations relative to the Sun and their placement within the Milky Way’s spiral structure. This study aims to characterize the evolutionary stages, structural stability, and orbital dynamics of the sampled OCs and to further evaluate the incidence of radial migration within the Galactic disk. 


\section{Data analysis}
\subsection{Photometric Data}

For the analysis of the selected OCs in this study, we adopted the central coordinates reported in the literature \citep{Castro-Ginard2020}. Photometric, astrometric, and spectroscopic data of sources within a $20^\prime$ radius around these centers were extracted from the Gaia DR3 catalog \citep{Gaia2023}. The photometric data include the $G$, $G_{\rm BP}$, and $G_{\rm RP}$ bands, while astrometric parameters consist of positions ($\alpha, \delta$), trigonometric parallaxes ($\varpi$), and proper-motion components ($\mu_{\alpha} \cos \delta, \mu_{\delta}$). Radial velocities ($V_{\rm R}$) were also taken into account; however, since Gaia DR3 radial velocities are only available for relatively bright stars ($G<14$ mag), they are not present for all sources in the cluster fields. Before the analysis, we applied basic quality filters to ensure reliable astrometric data. We kept only sources with a five-parameter astrometric solution (\texttt{astrometric\_params\_solved} $= 31$), so that both proper-motion components and trigonometric parallaxes were available. In addition, we imposed a constraint on the Renormalized Unit Weight Error ({\sc RUWE}), which measures the goodness-of-fit of the Gaia astrometric solution after correcting for its dependence on magnitude and color index \citep{Lindegren2021}. We also excluded sources with {\sc RUWE}$>1.4$, a commonly adopted threshold to remove sources with potentially unreliable or non-single-star astrometric solutions \citep{Gaia2021}. These cuts improve the overall data quality while keeping the sample as complete as possible. All parameters and their uncertainties were treated consistently throughout the study. The number of sources within a $20^\prime$ radius around each of the eight OCs is given in Table~\ref{tab:photometric_errors}.

\subsection{Determination of the Cluster Center}

\begin{figure*}[!ht]
\centering
\centering
\includegraphics[width=0.35\linewidth]{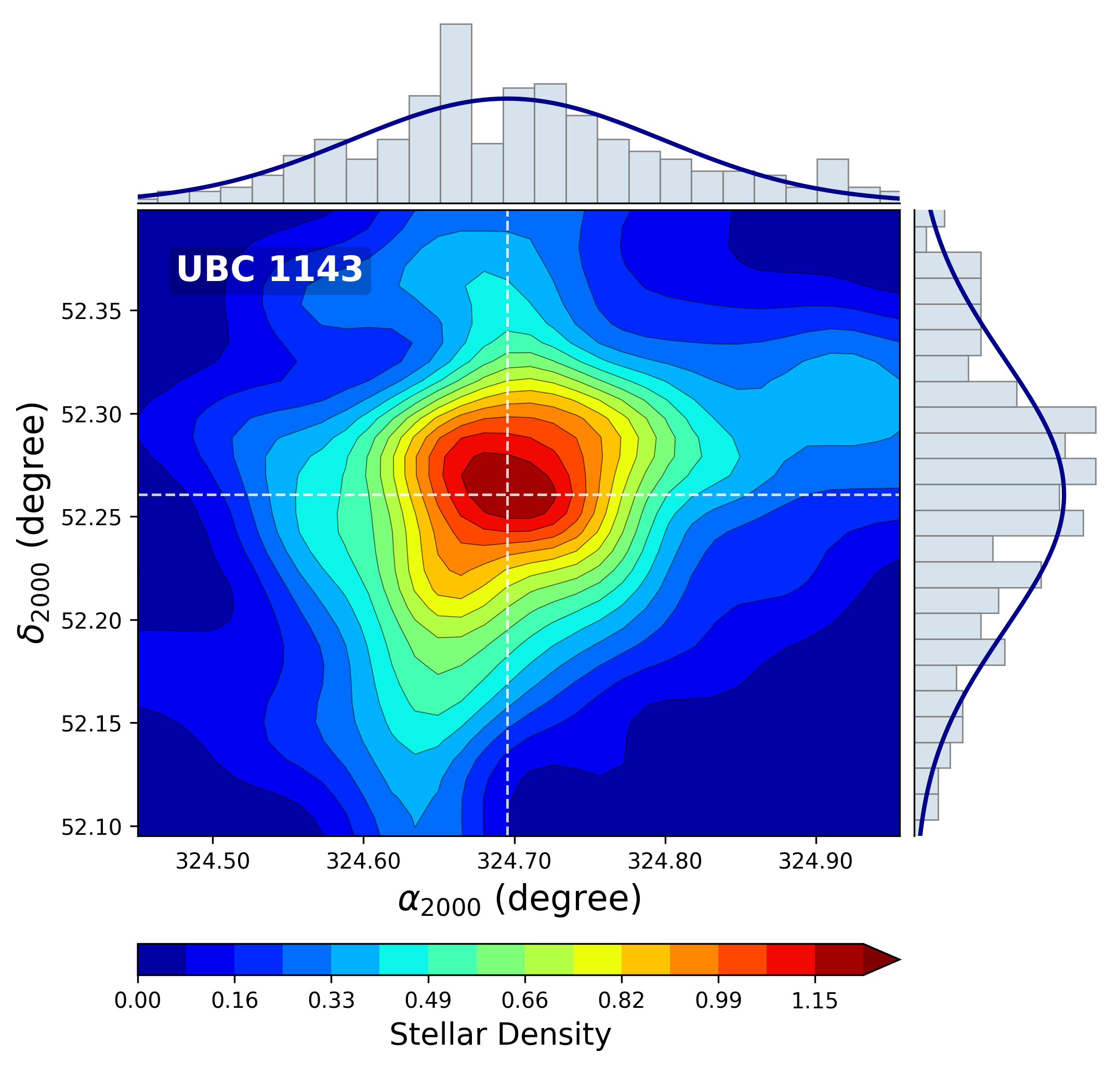}
\includegraphics[width=0.35\linewidth]{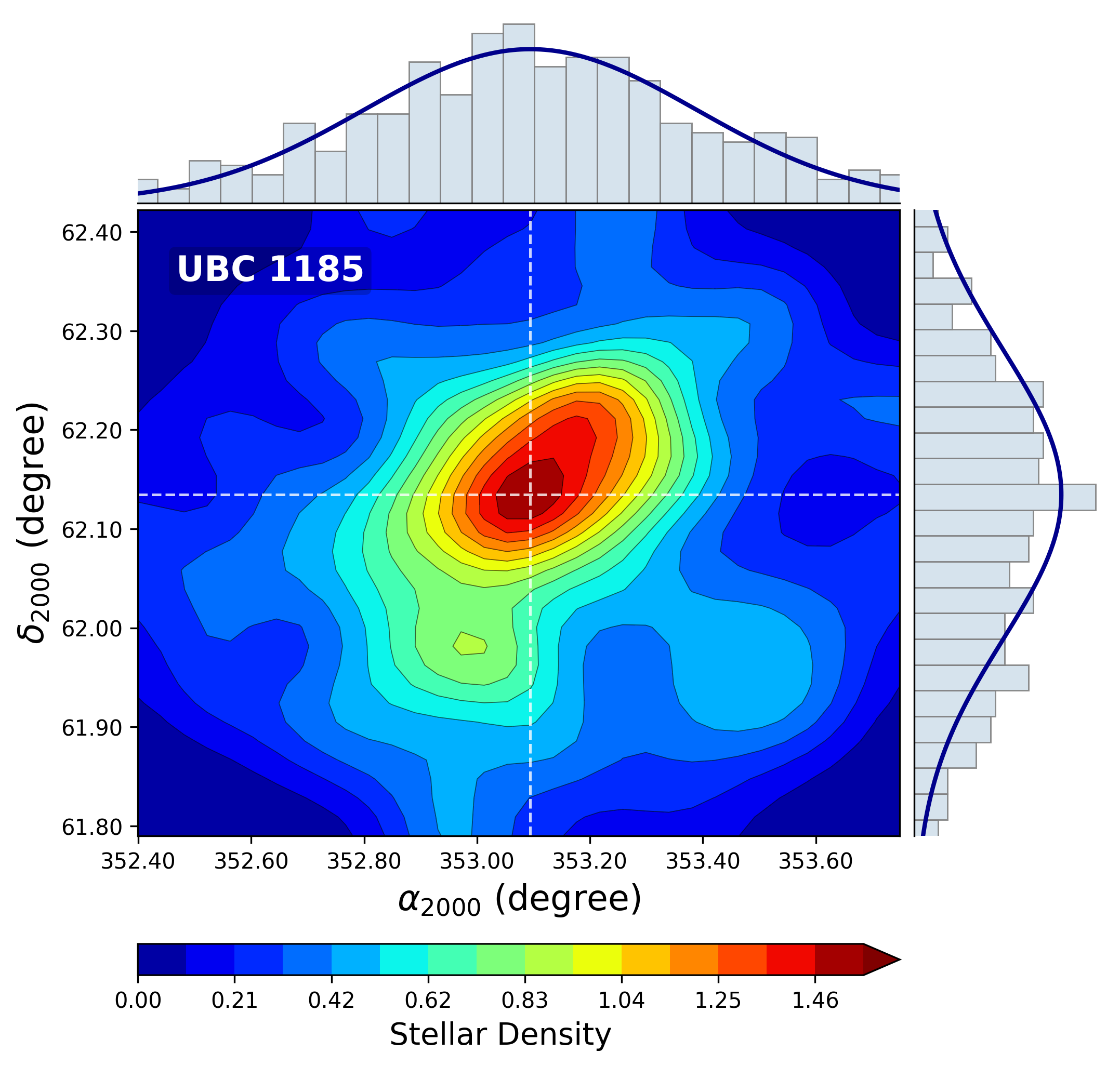}\\
\includegraphics[width=0.35\linewidth]{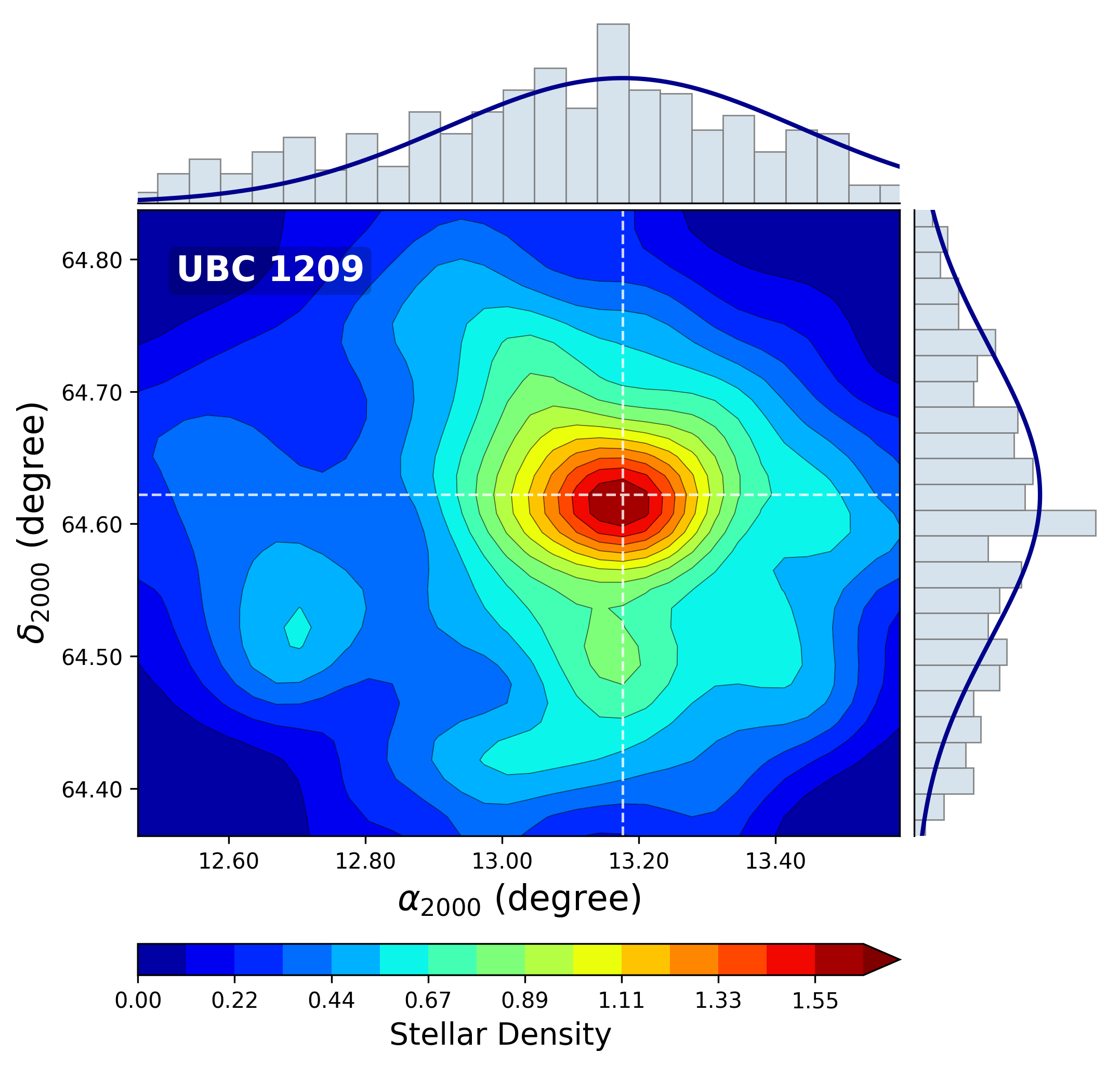}
\includegraphics[width=0.35\linewidth]{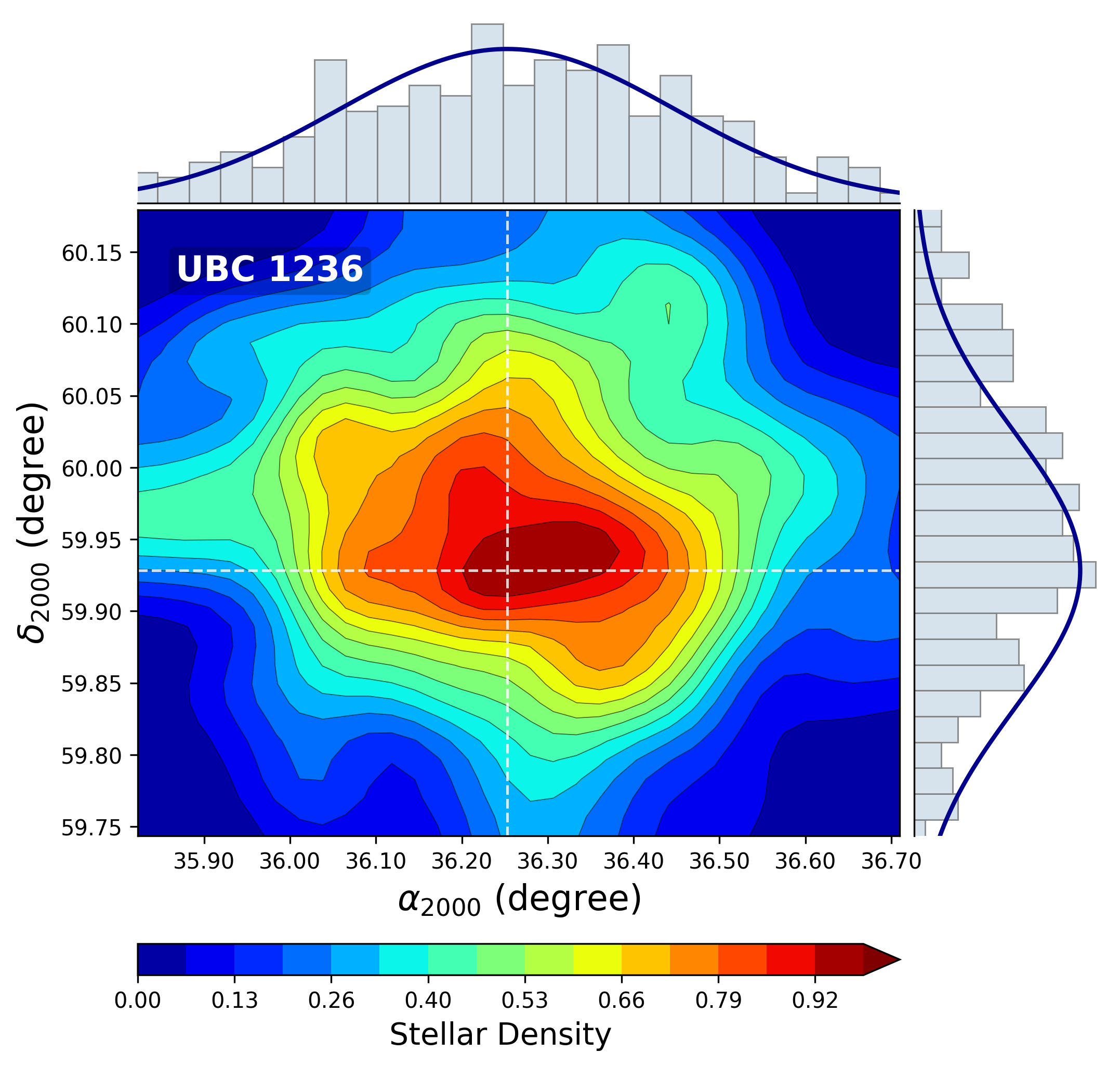}\\
\includegraphics[width=0.35\linewidth]{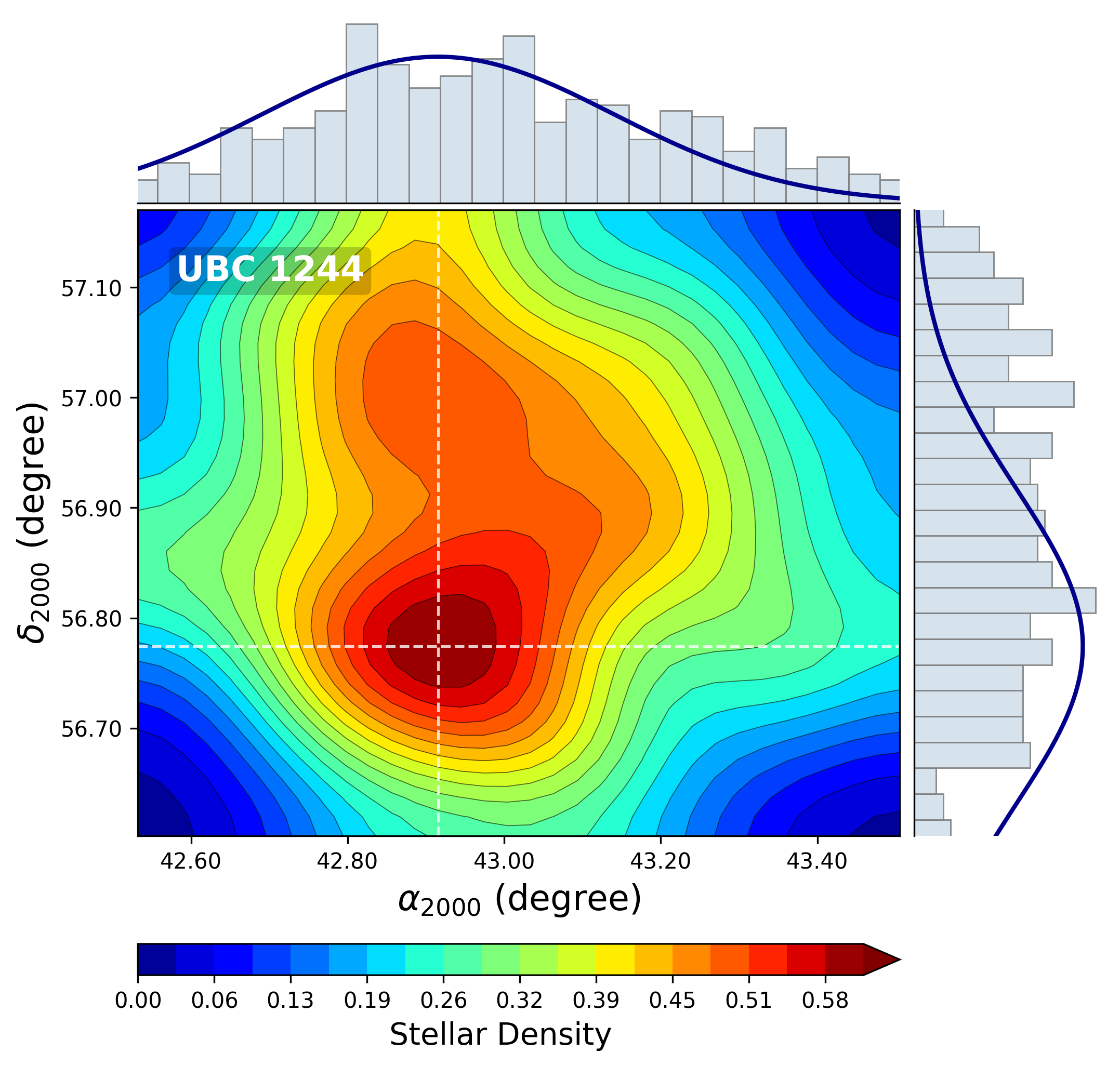}
\includegraphics[width=0.35\linewidth]{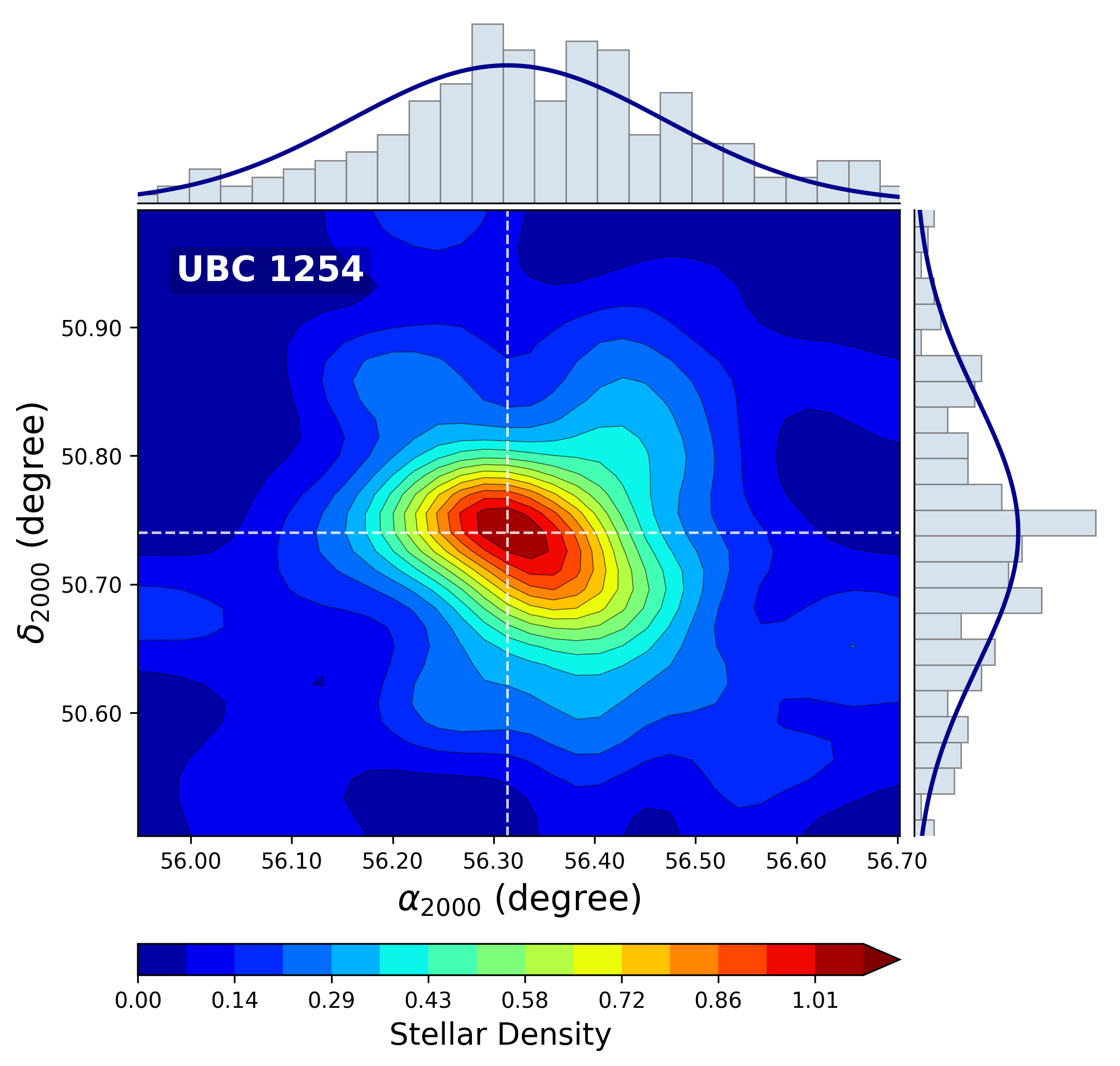}\\
\includegraphics[width=0.35\linewidth]{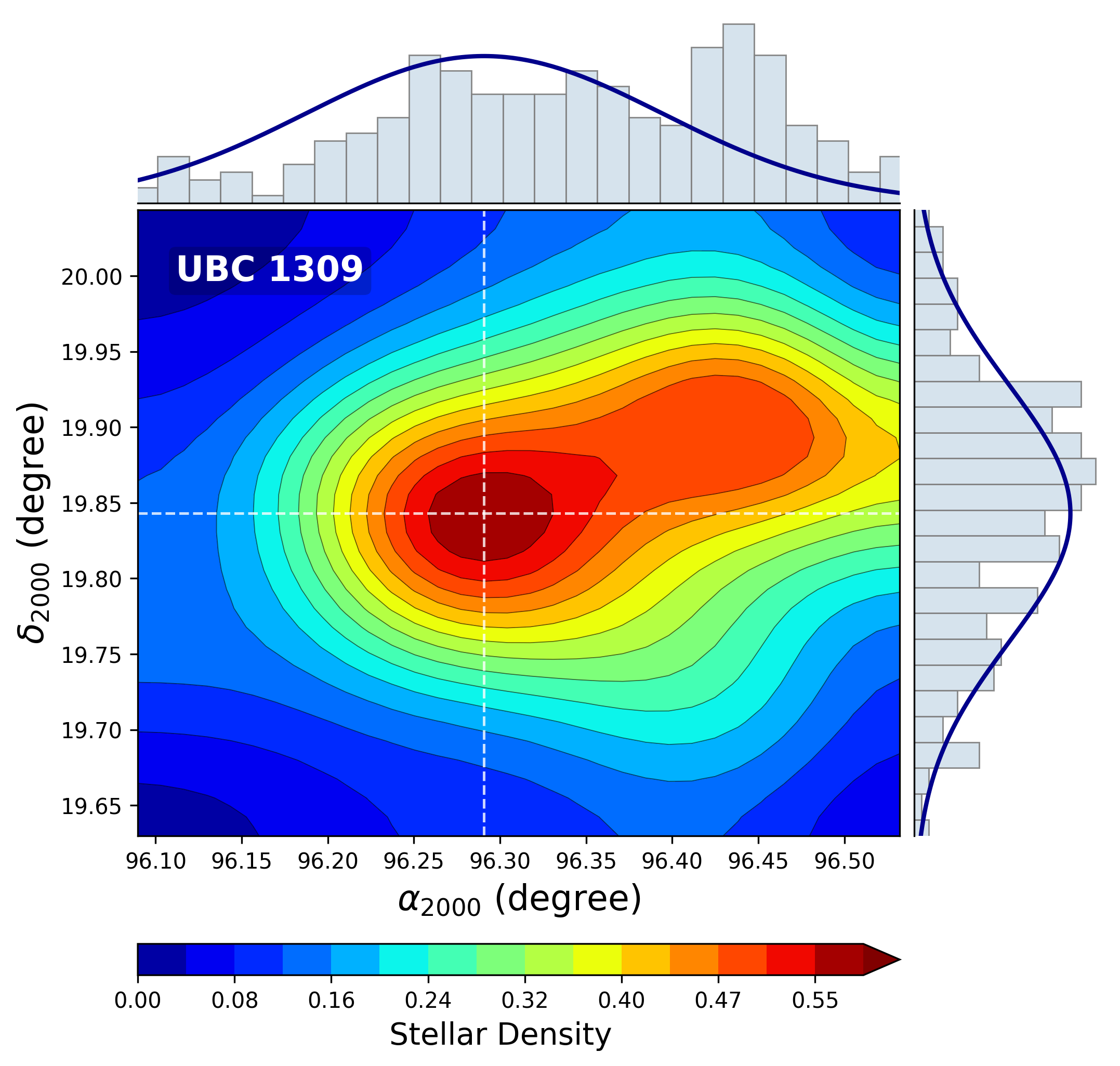}
\includegraphics[width=0.35\linewidth]{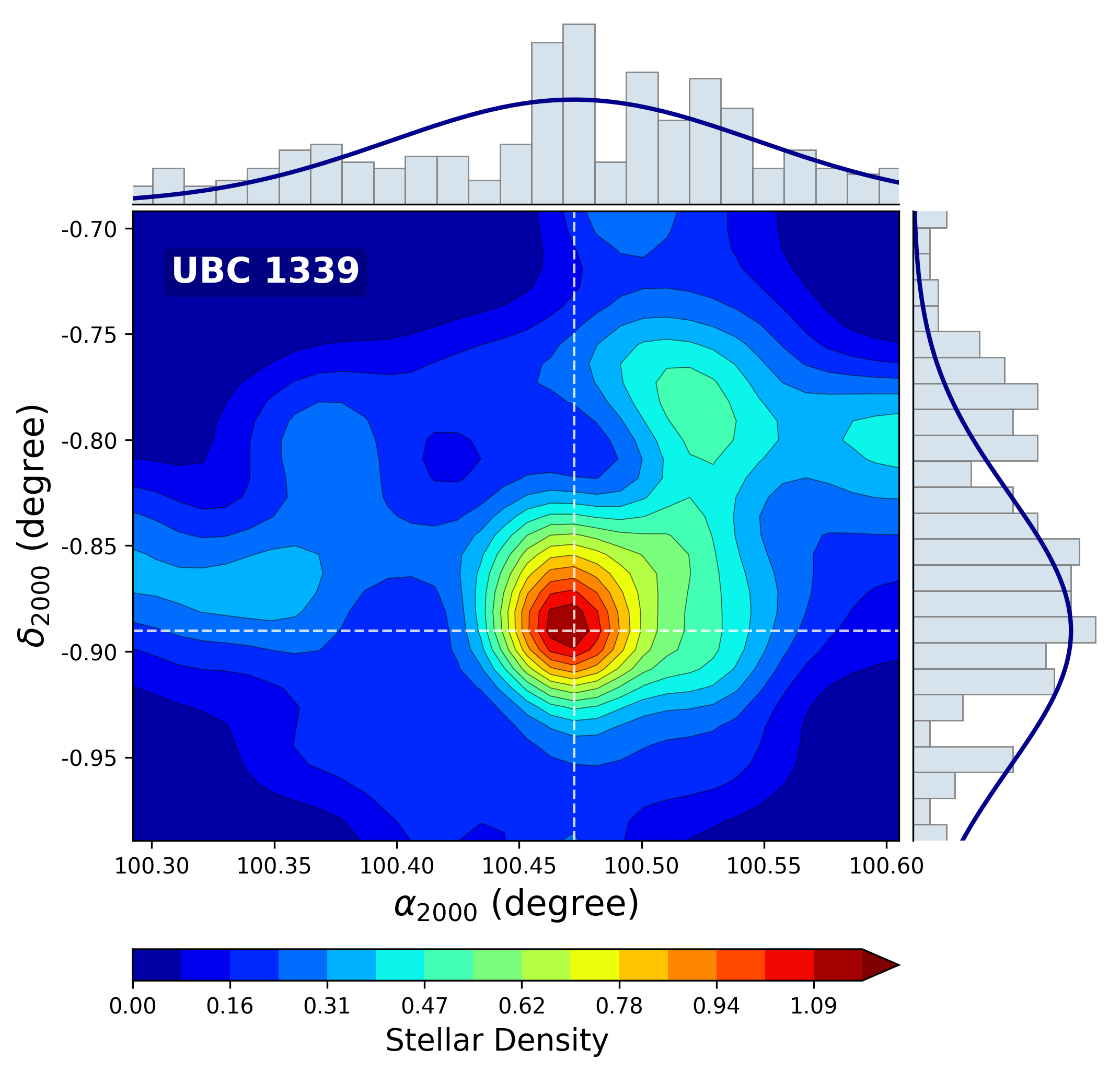}
\caption{Two-dimensional stellar-density distributions of the selected eight UBCs in equatorial coordinates. The upper and right-hand panels show the marginal stellar-count histograms in right ascension and declination, respectively, together with the corresponding Gaussian fits. The white dashed lines indicate the adopted cluster centres.}
\label{fig:cluster_centres}
\end{figure*}

The central coordinates of the eight UBCs were determined from the projected stellar-density distributions in equatorial coordinates. For each cluster, all stars within the adopted $20^\prime$ radius were retrieved from the Gaia database and examined in terms of their equatorial coordinates ($\alpha$, $\delta$) at the J2000 epoch. Since the cluster center is expected to coincide with the location of the highest stellar concentration, one-dimensional stellar-count histograms were constructed separately for $\alpha$ and $\delta$. To ensure a homogeneous determination for all targets, the sampled regions were divided into equal bins of 0.05 deg in both coordinates, and the number of stars in each bin was counted. Gaussian profiles were then fitted to the resulting $\alpha$ and $\delta$ histograms, and the peak positions of these fits were adopted as the refined cluster centers and listed in Table~\ref{King_para}. The corresponding two-dimensional stellar-density maps, together with the marginal distributions in $\alpha$ and $\delta$, are presented in Figure~\ref{fig:cluster_centres}.

\subsection{Photometric Completeness Limit}
\label{sec:Completeness}

The robust derivation of the structural and astrophysical parameters of the selected eight OCs depends on accurately determining their photometric $G$-band completeness limits. Establishing this threshold is a critical preliminary step, as incomplete sampling can introduce significant selection biases that propagate into subsequent analyses \citep[e.g.,][]{Salpeter1955, Moraux2003}. In particular, a well-defined completeness limit is essential to avoid artificial flattening of the luminosity and mass functions at faint magnitudes, which would otherwise distort the inferred initial mass function (IMF) slopes and compromise the reliability of stellar population parameters. This issue is especially important because low-mass main-sequence stars located near the detection boundary play a crucial role in constraining isochrone fitting and cluster age determinations.

\begin{figure*}
\centering
 \includegraphics[width=0.95\linewidth]{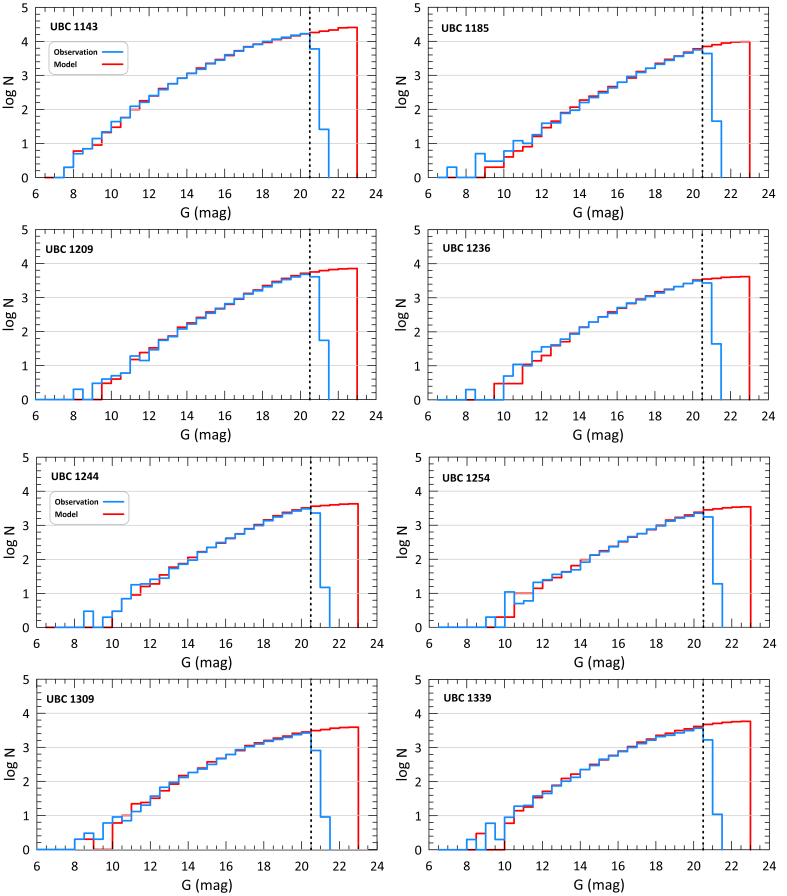}
\caption{Observed (blue) and model (red) star count histograms in the $G$-band for the eight OCs. The black dashed lines indicate the photometric completeness limits.}
\label{fig:completness_limit}
\end{figure*}

To assess the photometric completeness limit, stellar number counts were evaluated within successive bins of apparent magnitude in the Gaia $G$ band. As shown in Figure~\ref{fig:completness_limit}, the stellar counts exhibit a steady increase with magnitude, reaching a pronounced peak at approximately $G=20.5$ mag. Beyond this magnitude, a noticeable decline in the number of detected sources is observed, signaling the onset of photometric incompleteness. Accordingly, stars fainter than $G=20.5$ mag were excluded from the final sample, thereby ensuring the robustness of the derived parameters for selected OCs.

To determine the photometric completeness limits quantitatively, synthetic stellar catalogs were generated for the regions covering selected UBCs using the Besançon Galaxy Model\footnote{\url{https://model.obs-besancon.fr/}} \citep{Robin2003}. This Galactic population synthesis framework incorporates the three-dimensional interstellar extinction map developed by \citet{Marshall2006}, allowing for a realistic treatment of line-of-sight extinction effects. The simulations were performed over the apparent magnitude range $6<G~{\rm (mag)} \leq 23$, selected to closely reproduce the observed magnitude distribution of the detected stars while accounting for spatially varying extinction. For each OC field, a solid angle of 0.45 square degrees was adopted in the model calculations.

The resulting synthetic stellar distributions were subsequently contrasted with the observational data, as shown in Figure~\ref{fig:completness_limit}. At bright magnitudes, where observed stellar counts exceed those predicted by the model, the nearly dominant magnitude ranges associated with the OC populations are delineated, indicating a completeness of nearly 100\%. Conversely, the magnitude at which the model begins to exceed the observed counts systematically marks the onset of photometric incompleteness and hence defines the effective completeness threshold. This comparison demonstrates that the photometric completeness limits used in our analysis are robust and do not suffer from significant stellar loss across the magnitude range considered. Determining the completeness limit of Gaia data via the synthetic star count method has proven successful in multiple studies of OCs documented in the literature \citep[e.g.,][]{Yontan2015, Tanik2025, Tasdemir2026, Bilir2026}.

Photometric uncertainties were quantified using the formal error estimates provided in the Gaia DR3 database, which were adopted as representative within discrete magnitude intervals. To this end, stars located within the cluster regions were grouped into successive $G$-band bins, and their corresponding mean apparent magnitudes together with mean $G_{\rm BP}-G_{\rm RP}$ color indices were examined. At the adopted completeness threshold of $G=20.5$ mag, the mean internal uncertainty in the $G$ magnitude was determined to be 0.007 mag, while the associated mean error in the $G_{\rm BP}-G_{\rm RP}$ color index amounted to 0.121 mag. The variation of the mean photometric uncertainties over the entire sampled $G$-magnitude range for the selected OCs is listed in Table~\ref{tab:photometric_errors}. As shown in the table, both the mean $G$-band magnitude errors and the $G_{\rm BP}-G_{\rm RP}$ color index uncertainties increase toward fainter magnitudes, reflecting the expected decline in photometric precision. Nevertheless, within the adopted magnitude limits for the analysis of the eight OCs, the mean magnitude and color index uncertainties remain comparable and sufficiently low, ensuring the reliability of the derived astrophysical parameters.

\begin{table*}[h]
\centering
\small
\caption{Gaia magnitude and color index errors calculated across different $G$-band magnitude intervals for sources in the directions of eight UBC OCs.}
\label{tab:photometric_errors}
\resizebox{\linewidth}{!}{%
\begin{tabular}{c|ccc|ccc|ccc|ccc}
\toprule
 & \multicolumn{3}{c|}{\textbf{UBC 1143}} & \multicolumn{3}{c|}{\textbf{UBC 1185}} & \multicolumn{3}{c|}{\textbf{UBC 1209}} & \multicolumn{3}{c}{\textbf{UBC 1236}} \\\cmidrule(lr){2-4} \cmidrule(lr){5-7} \cmidrule(lr){8-10} \cmidrule(lr){11-13}
$G$ (mag) & $N$ & $\sigma_{{\rm G}}$ & $\sigma_{{G_{{\rm BP}}-G_{{\rm RP}}}}$ & $N$ & $\sigma_{{\rm G}}$ & $\sigma_{{G_{{\rm BP}}-G_{{\rm RP}}}}$ & $N$ & $\sigma_{{\rm G}}$ & $\sigma_{{G_{{\rm BP}}-G_{{\rm RP}}}}$ & $N$ & $\sigma_{{\rm G}}$ & $\sigma_{{G_{{\rm BP}}-G_{{\rm RP}}}}$ \\
\hline 
~~6--14& ~~~584 & 0.003 & 0.006 & ~~~254 & 0.003 & 0.006 & ~~~259 & 0.003 & 0.006 & ~~~231 & 0.003 & 0.005 \\
14--15 & ~~~735 & 0.003 & 0.006 & ~~~307 & 0.003 & 0.006 & ~~~339 & 0.003 & 0.005 & ~~~271 & 0.003 & 0.006 \\
15--16 & ~~1403 & 0.003 & 0.006 & ~~~609 & 0.003 & 0.006 & ~~~696 & 0.003 & 0.006 & ~~~543 & 0.003 & 0.007 \\
16--17 & ~~2486 & 0.003 & 0.009 & ~~1278 & 0.003 & 0.010 & ~~1313 & 0.003 & 0.009 & ~~~990 & 0.003 & 0.009 \\
17--18 & ~~4183 & 0.003 & 0.018 & ~~2440 & 0.003 & 0.019 & ~~2494 & 0.003 & 0.017 & ~~1762 & 0.003 & 0.018 \\
18--19 & ~~6480 & 0.003 & 0.038 & ~~4078 & 0.003 & 0.042 & ~~4085 & 0.003 & 0.038 & ~~2728 & 0.003 & 0.044 \\
19--20 & ~~9430 & 0.004 & 0.084 & ~~7001 & 0.004 & 0.088 & ~~6484 & 0.004 & 0.081 & ~~4158 & 0.004 & 0.089 \\
20--21 &  12587 & 0.010 & 0.219 &  10480 & 0.010 & 0.204 & ~~9125 & 0.009 & 0.193 & ~~6043 & 0.009 & 0.197 \\
21--23 & ~~1047 & 0.027 & 0.428 & ~~1161 & 0.026 & 0.430 & ~~1215 & 0.025 & 0.422 & ~~~852 & 0.025 & 0.418 \\
\hline
Total/Error & 38935 & 0.006 & 0.112 & 27608 & 0.007 & 0.126 & 26010 & 0.007 & 0.116 & 17578 & 0.006& 0.118 \\
\bottomrule
\end{tabular}%
}

\resizebox{\linewidth}{!}{%
\begin{tabular}{c|ccc|ccc|ccc|ccc}
\toprule
 & \multicolumn{3}{c|}{\textbf{UBC 1244}} & \multicolumn{3}{c|}{\textbf{UBC 1254}} & \multicolumn{3}{c|}{\textbf{UBC 1309}} & \multicolumn{3}{c}{\textbf{UBC 1339}} \\\cmidrule(lr){2-4} \cmidrule(lr){5-7} \cmidrule(lr){8-10} \cmidrule(lr){11-13}
$G$ (mag) & $N$ & $\sigma_{{\rm G}}$ & $\sigma_{{G_{{\rm BP}}-G_{{\rm RP}}}}$ & $N$ & $\sigma_{{\rm G}}$ & $\sigma_{{G_{{\rm BP}}-G_{{\rm RP}}}}$ & $N$ & $\sigma_{{\rm G}}$ & $\sigma_{{G_{{\rm BP}}-G_{{\rm RP}}}}$ & $N$ & $\sigma_{{\rm G}}$ & $\sigma_{{G_{{\rm BP}}-G_{{\rm RP}}}}$ \\
\hline 
~~6--14& ~~~192 & 0.003 & 0.006 & ~~~177 & 0.003 & 0.005 & ~~~317 & 0.003 & 0.006 & ~~~391 & 0.003 & 0.005 \\
14--15 & ~~~209 & 0.003 & 0.005 & ~~~157 & 0.003 & 0.005 & ~~~373 & 0.003 & 0.006 & ~~~438 & 0.003 & 0.006 \\
15--16 & ~~~454 & 0.003 & 0.006 & ~~~356 & 0.003 & 0.006 & ~~~636 & 0.003 & 0.008 & ~~~850 & 0.003 & 0.007 \\
16--17 & ~~~829 & 0.003 & 0.010 & ~~~675 & 0.003 & 0.010 & ~~1235 & 0.003 & 0.011 & ~~1526 & 0.003 & 0.011 \\
17--18 & ~~1531 & 0.003 & 0.020 & ~~1127 & 0.003 & 0.019 & ~~2057 & 0.003 & 0.022 & ~~2611 & 0.003 & 0.022 \\
18--19 & ~~2660 & 0.004 & 0.048 & ~~1962 & 0.003 & 0.047 & ~~2929 & 0.004 & 0.054 & ~~3995 & 0.004 & 0.050 \\
19--20 & ~~4308 & 0.005 & 0.097 & ~~3225 & 0.004 & 0.096 & ~~3851 & 0.006 & 0.117 & ~~5108 & 0.006 & 0.112 \\
20--21 & ~~5926 & 0.011 & 0.227 & ~~4220 & 0.010 & 0.216 & ~~4330 & 0.012 & 0.264 & ~~6420 & 0.013 & 0.256 \\
21--23 & ~~~482 & 0.027 & 0.449 & ~~~490 & 0.027 & 0.430 & ~~~~61 & 0.030 & 0.351 & ~~~195 & 0.029 & 0.448 \\
\hline
Total/Error & 16591 & 0.007 & 0.130 & 12389 & 0.007 & 0.126 & 15789 & 0.007 & 0.117 & 21534 & 0.007 & 0.120 \\
\bottomrule
\end{tabular}%
}
\end{table*}

\section{Determining the Structural Parameters}

The structural properties of stellar clusters can be effectively characterized by their radial density profiles (RDPs), which provide insight into their richness and degree of central concentration. In this study, RDPs were constructed using high-probability member stars selected from Gaia data. For each cluster, the surrounding region was divided into concentric annuli, and the stellar surface density in each annulus was computed as
\begin{equation}
\rho(r_i) = \frac{N_i}{A_i},
\end{equation}
where $N_i$ and $A_i$ denote the number of stars and the area of the $i$-th annulus, respectively. Uncertainties were estimated assuming Poisson statistics, i.e. $1/\sqrt{N_i}$. The resulting RDPs for representative clusters are shown in Figure~\ref{rdps}, while the full set is presented in Appendix~\ref{fig:rdps-append}.

To derive the structural parameters, we fitted the empirical surface density profile of \citet{King1962}, which describes the radial distribution of stars in bound systems. The adopted form is
\begin{equation}
\rho(r) = \rho_0 \left[
\frac{1}{\sqrt{1 + (r/r_{\rm c})^2}} -
\frac{1}{\sqrt{1 + (r_{\rm t}/r_{\rm c})^2}}
\right]^2 + \rho_{\mathrm{bg}},
\end{equation}
where $\rho_0$ is the central stars density, $r_{\rm c}$ is the core radius, $r_{\rm t}$ is the tidal radius, and $\rho_{\mathrm{bg}}$ represents the background star density.

The model parameters were estimated using a maximum-likelihood approach, adopting the likelihood function
\begin{equation}
\ln \mathcal{L} = -\frac{1}{2}\sum_i 
\left( \frac{\rho_i - \rho_{i,\mathrm{model}}}{\sigma_{\rho_i}} \right)^2 .
\end{equation}
Parameter exploration was performed using the \texttt{emcee} Markov Chain Monte Carlo (MCMC) sampler \citep{Foreman-Mackey2013}, assuming uniform priors. Convergence was assessed with the Gelman--Rubin diagnostic \citep{gelman1992}. The results obtained in this study are presented in Table~\ref{King_para}, while the analysis results for a representative OC, namely UBC 1143, are shown in Figure~\ref{rdps}. In addition, the RDPs of all studied OCs, along with their corresponding King model fits \citet{King1962}, are provided in the Appendix~\ref{fig:rdps-append}. For UBC~1236, UBC~1244, and UBC~1254, field star contamination becomes significant at fainter magnitudes; therefore, a limiting magnitude of $G = 18.5$ was adopted. RDPs for these clusters were constructed using stars brighter than this limit.

The concentration parameter was computed as $C = r_{\rm t}/r_{\rm c}$ following \citet{King1962}, providing a quantitative measure of the degree of central concentration and dynamical state of the clusters. The derived values span a wide range, indicating diverse structural properties among the analyzed UBC clusters, from relatively diffuse systems to more centrally concentrated ones. Overall, our results are broadly consistent with previous studies, although some differences are present. These discrepancies are likely due to variations in data selection, RDP construction, and fitting methodology, as well as the improved precision and homogeneity of the Gaia DR3 data.

\begin{figure*}[ht!]
\centering
\includegraphics[width=0.65\linewidth]{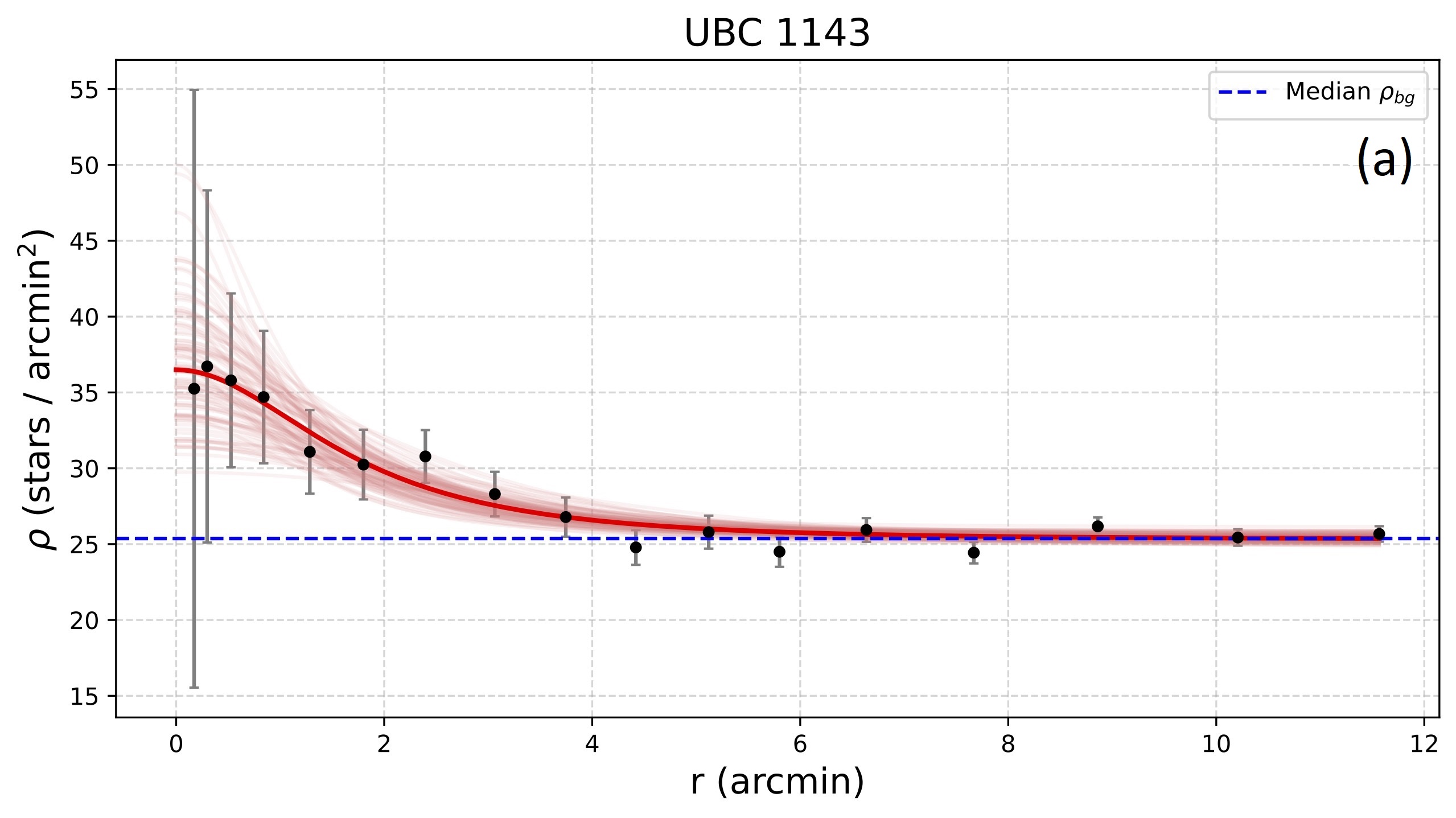}\\
\includegraphics[width=0.65\linewidth]{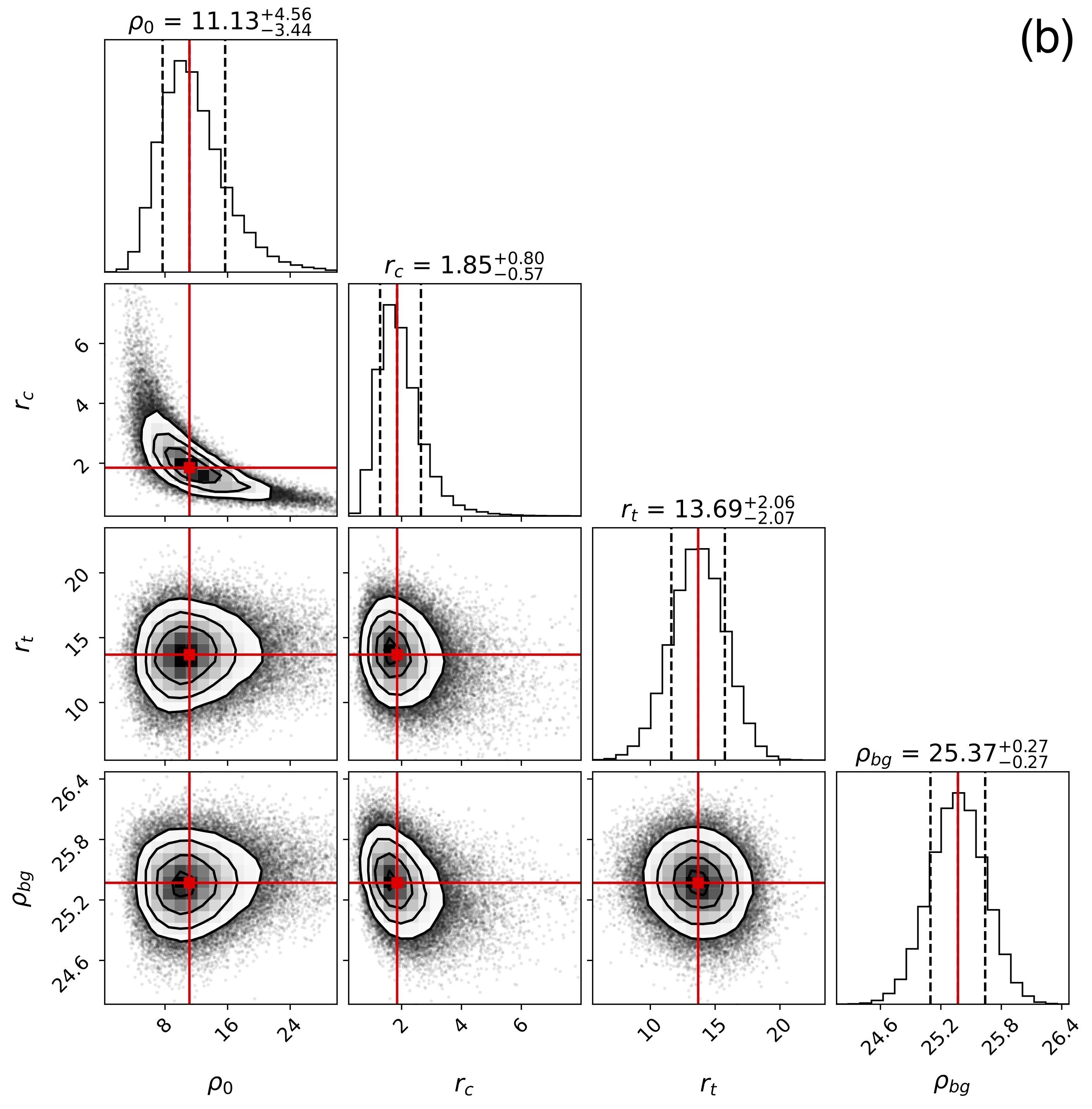}
\caption{RDP analysis of UBC~1143 as an example. (a) The observed stellar density profile as a function of radius is shown with black points and error bars. The red curve represents the best-fitting King model, while the shaded region indicates the posterior uncertainty. The horizontal dashed blue line marks the median background density level ($\rho_{\mathrm{bg}}$). (b) Posterior distributions and parameter correlations obtained from the MCMC analysis for the King model parameters: central density ($\rho_0$), core radius ($r_{\rm c}$), tidal radius ($r_{\rm t}$), and background density ($\rho_{\mathrm{bg}}$). The red lines indicate the median values, and the dashed red lines represent the 16th and 84th percentiles.}
\label{rdps}
\end{figure*}

\begin{table*}[ht]
\centering
\footnotesize
\caption{Structural parameters of the analyzed UBC OCs derived from fitting the \citet{King1962} model to their RDPs. Columns list the cluster as follows: Equatorial coordinates ($\alpha, \delta$), the  Galactic coordinates ($l, b$), radius ($r_{\rm t}$), core radius ($r_{\rm c}$), central density ($\rho_0$), background density ($\rho_{\rm bg}$), and concentration parameter ($C$).}
\label{King_para}
\setlength{\tabcolsep}{4pt}
\renewcommand{\arraystretch}{1.15}
\begin{tabular}{lccccccccc} 
\hline\hline 
Cluster & $\alpha_{\rm J2000}$ & $\delta_{\rm J2000}$ & $l$ & $b$ &  $ r_{\rm t}$ & $ r_{\rm c}$ & $\rho_0$ & $\rho_{\rm bg}$ & $C$ \\
\cmidrule(lr){4-5} \cmidrule(lr){6-7} \cmidrule(lr){8-9}
& (hh:mm:ss.ss) & (dd:mm:ss.ss) & \multicolumn{2}{c}{($\degree$)}  & \multicolumn{2}{c}{\rm (arcmin)} & \multicolumn{2}{c}{(stars arcmin$^{-2}$)} & \\
\hline 
UBC 1143 & 21:38:46.86 & +52:15:38.58 & 095.800 & -00.153 & $13.69^{+2.06}_{-2.07}$ & $1.85^{+0.80}_{-0.57}$ & $11.13^{+4.56}_{-3.44}$& $25.37^{+0.27}_{-0.27}$ & 7.40 \\
UBC 1185 & 23:32:22.65 & +62:09:13.81 & 113.850 & +00.684 & $16.77^{+0.99}_{-1.00}$ & $3.18^{+1.13}_{-0.78}$ & $8.51^{+2.36}_{-1.94}$ & $16.49^{+0.21}_{-0.22}$ & 5.27   \\
UBC 1209 & 00:52:42.39 & +64:37:19.87 & 123.071 & +01.750 & $16.77^{+1.00}_{-0.99}$ & $2.50^{+1.81}_{-1.05}$ & $4.78^{+3.05}_{-1.96}$ & $16.63^{+0.18}_{-0.18}$ & 6.71   \\
UBC 1236 & 02:25:07.13 & +59:56:29.64 & 134.412 & -00.822 & $15.22^{+3.40}_{-2.76}$ & $7.11^{+3.81}_{-2.87}$ & $2.48^{+1.26}_{-0.71}$ & $5.07^{+0.07}_{-0.08}$  & $2.14$ \\
UBC 1244 & 02:51:39.69 & +56:46:27.62 & 138.848 & -02.329 & $7.95^{+3.32}_{-1.68}$  & $3.19^{+1.81}_{-1.24}$ & $4.91^{+1.91}_{-1.30}$ & $5.08^{+0.09}_{-0.09}$  & $2.49$ \\
UBC 1254 & 03:45:20.72 & +50:44:25.19 & 148.788 & -03.187 & $15.35^{+5.29}_{-5.24}$ & $1.50^{+0.74}_{-0.47}$ & $6.47^{+2.99}_{-2.13}$ & $3.77^{+0.09}_{-0.10}$  & $10.23$ \\
UBC 1309 & 06:25:09.79 & +19:50:34.53 & 192.345 & +03.387 & $6.59^{+1.21}_{-1.34}$  & $4.91^{+1.58}_{-1.78}$ & $4.83^{+2.03}_{-1.22}$ & $11.89^{+0.15}_{-0.16}$ & 1.34   \\
UBC 1339 & 06:41:55.63 & -00:52:52.40 & 212.636 & -02.523 & $11.42^{+2.86}_{-2.50}$ & $5.74^{+1.46}_{-1.58}$ & $4.79^{+1.14}_{-0.92}$ & $11.49^{+0.16}_{-0.17}$ & 1.99   \\
\hline
\end{tabular}
\end{table*}

The clusters analyzed in this study were originally identified by \citet{Castro-Ginard2018, Castro-Ginard2019, Castro-Ginard2020} using Gaia data. These studies were primarily focused on cluster detection and membership assignment, and did not include structural parameter estimates based on RDP analysis or King model fitting. As a result, a direct comparison of structural parameters is not possible. In this context, the parameters derived in this study provide a homogeneous set of structural measurements for these UBC clusters.
We further compared our results with the physical radii reported by \citet{Hunt2024}. The $r_{\rm t}$ are broadly consistent, corresponding to typical values of $\sim$5–15 pc in \citet{Hunt2024}, in agreement with our results once the cluster distances are considered. The core radii, however, show larger discrepancies. For example, \citet{Hunt2024} report smaller core radii for UBC 1209 ($r_{\rm c} \sim 1.9$ pc) and UBC 1309 ($r_{\rm c} \sim 0.6$ pc), whereas our values are systematically larger for these systems. These differences are likely related to variations in membership selection, radial binning, and fitting methodology. In contrast, UBC 1236 shows consistently large core radii in both studies, suggesting an intrinsically extended inner structure. Overall, $r_{\rm t}$ are consistent within uncertainties, while core radii exhibit greater scatter, as commonly found for sparse and low-contrast OCs.

\section{Membership Determination}
\label{section:membership}
Cluster membership was identified using the 
Unsupervised Photometric Membership Assignment in Stellar Clusters algorithm \citep[\texttt{UPMASK};][]{Krone-Martins2014}, an unsupervised, model-independent approach that separates cluster stars from field contamination by analyzing spatial and kinematic coherence. This method combines $k$-means clustering with iterative significance testing in a multidimensional parameter space, allowing membership probabilities to be refined without imposing any predefined structural assumptions on the clusters. For the eight selected UBC OCs, stellar membership probabilities were derived by applying \texttt{UPMASK} within a five-dimensional parameter space defined by equatorial coordinates ($\alpha, \delta$), trigonometric parallax ($\varpi$), and proper-motion components ($\mu_{\alpha}\cos\delta$, $\mu_\delta$), along with their corresponding uncertainties. The procedure was executed iteratively, with 25 outer-loop iterations for each cluster to ensure stable convergence. The optimal number of clusters in the $k$-means step was determined by selecting configurations that maximized the statistical separation between cluster members and field stars. As a result, the final $k$-means values for the analyzed UBC OCs ranged from 17 to 20.

\begin{figure*}
\centering
\includegraphics[width=0.38\linewidth]{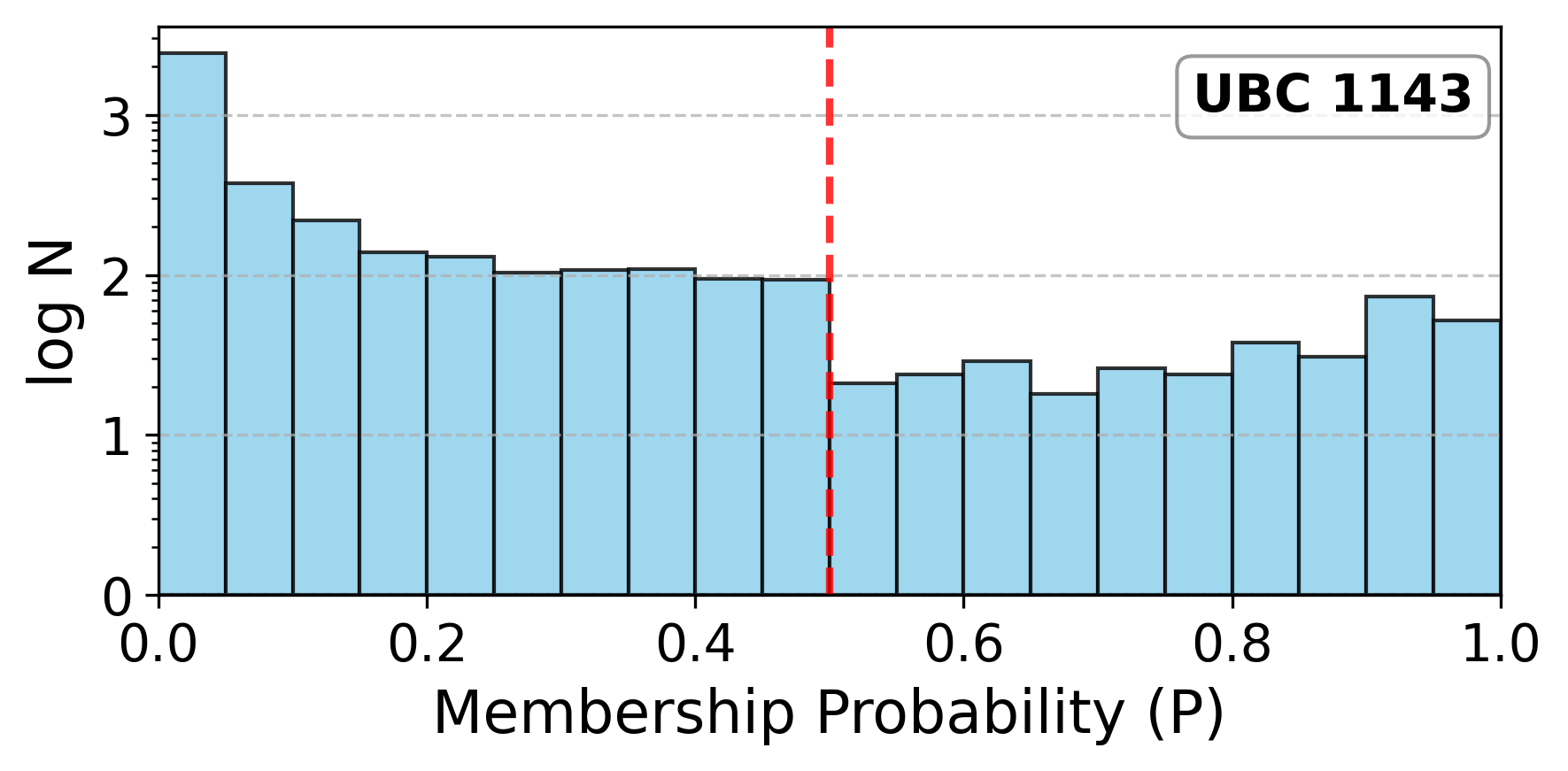}
\includegraphics[width=0.38\linewidth]{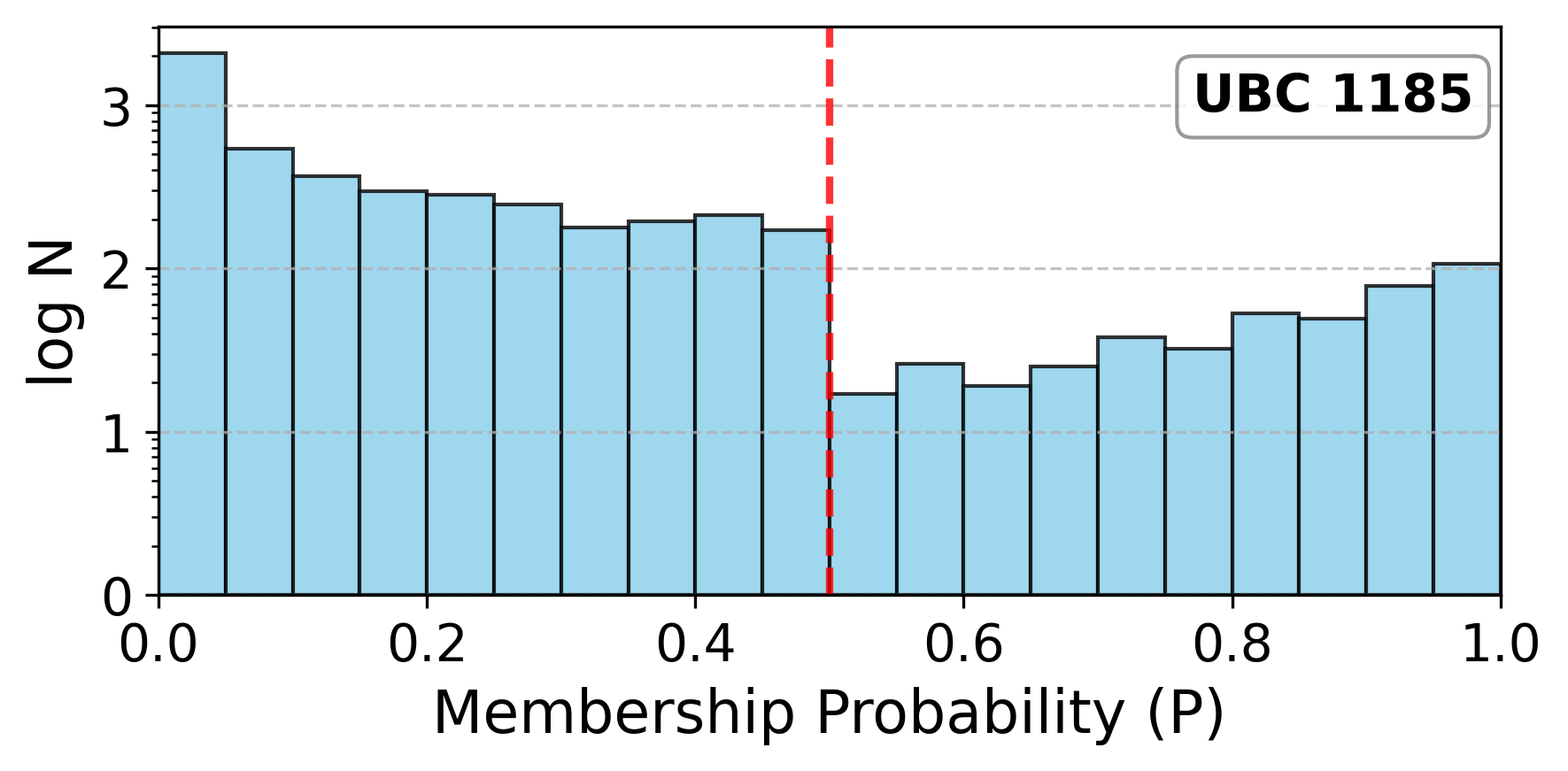}\\
\includegraphics[width=0.38\linewidth]{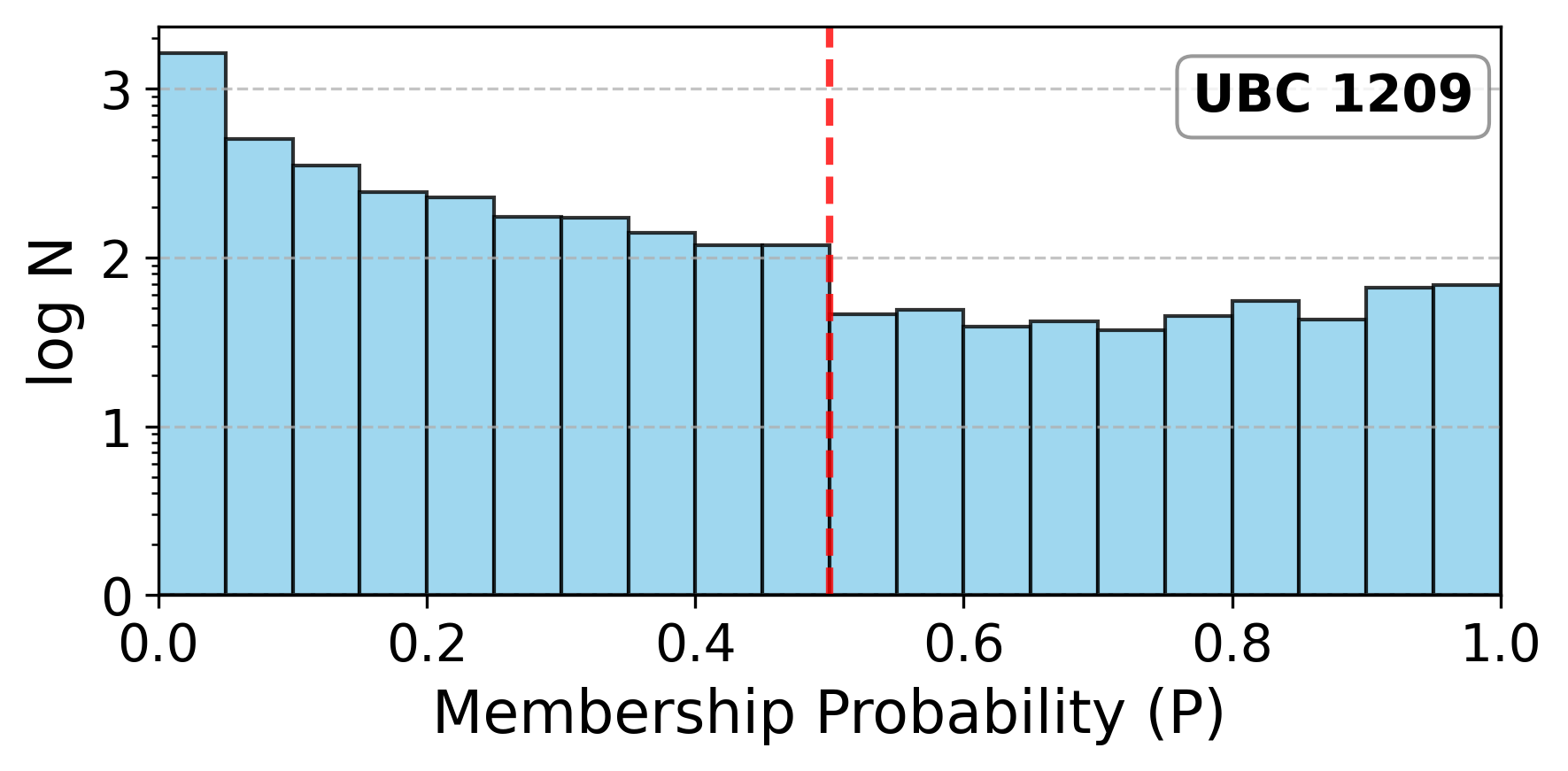}
\includegraphics[width=0.38\linewidth]{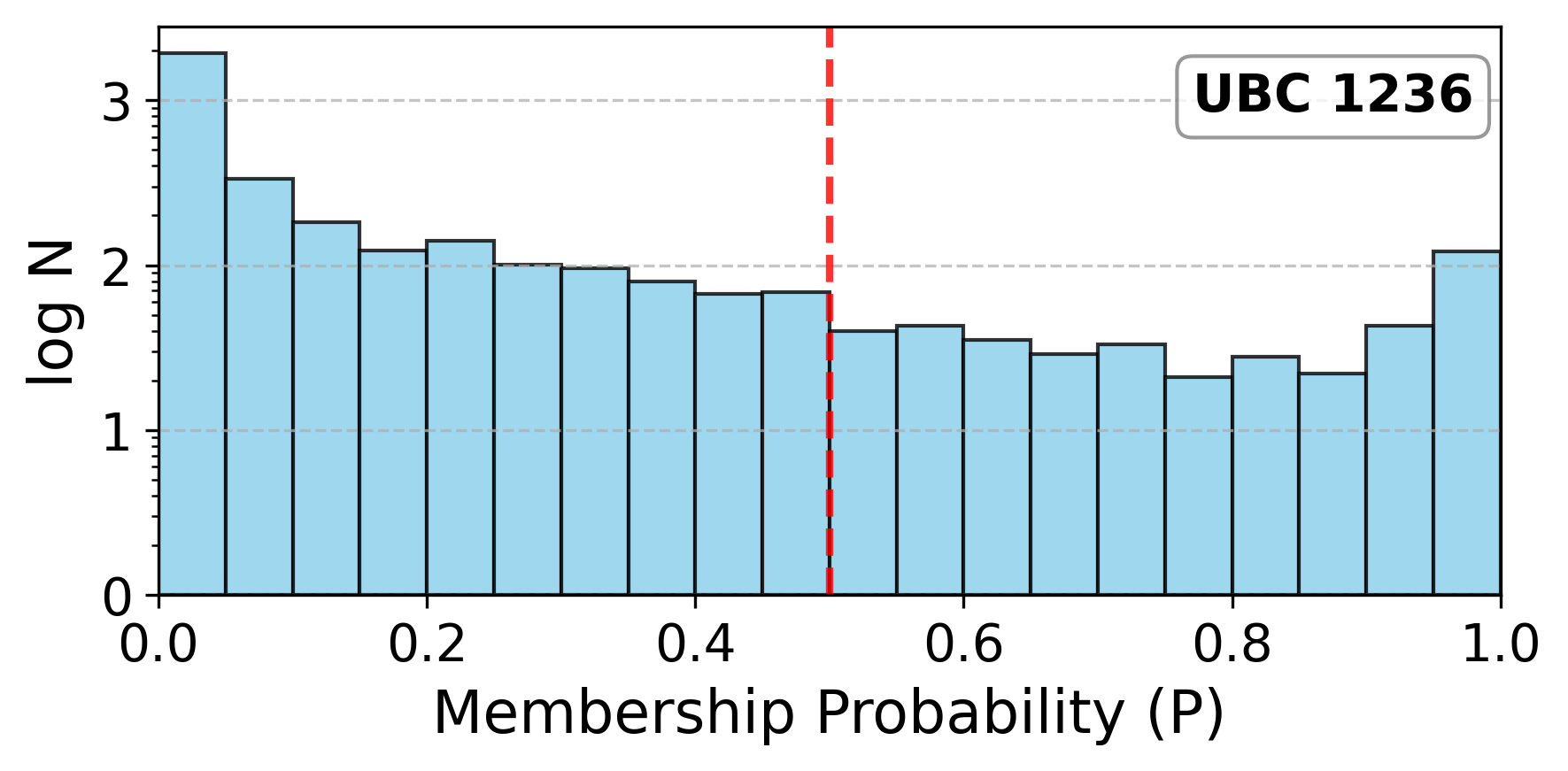}\\
\includegraphics[width=0.38\linewidth]{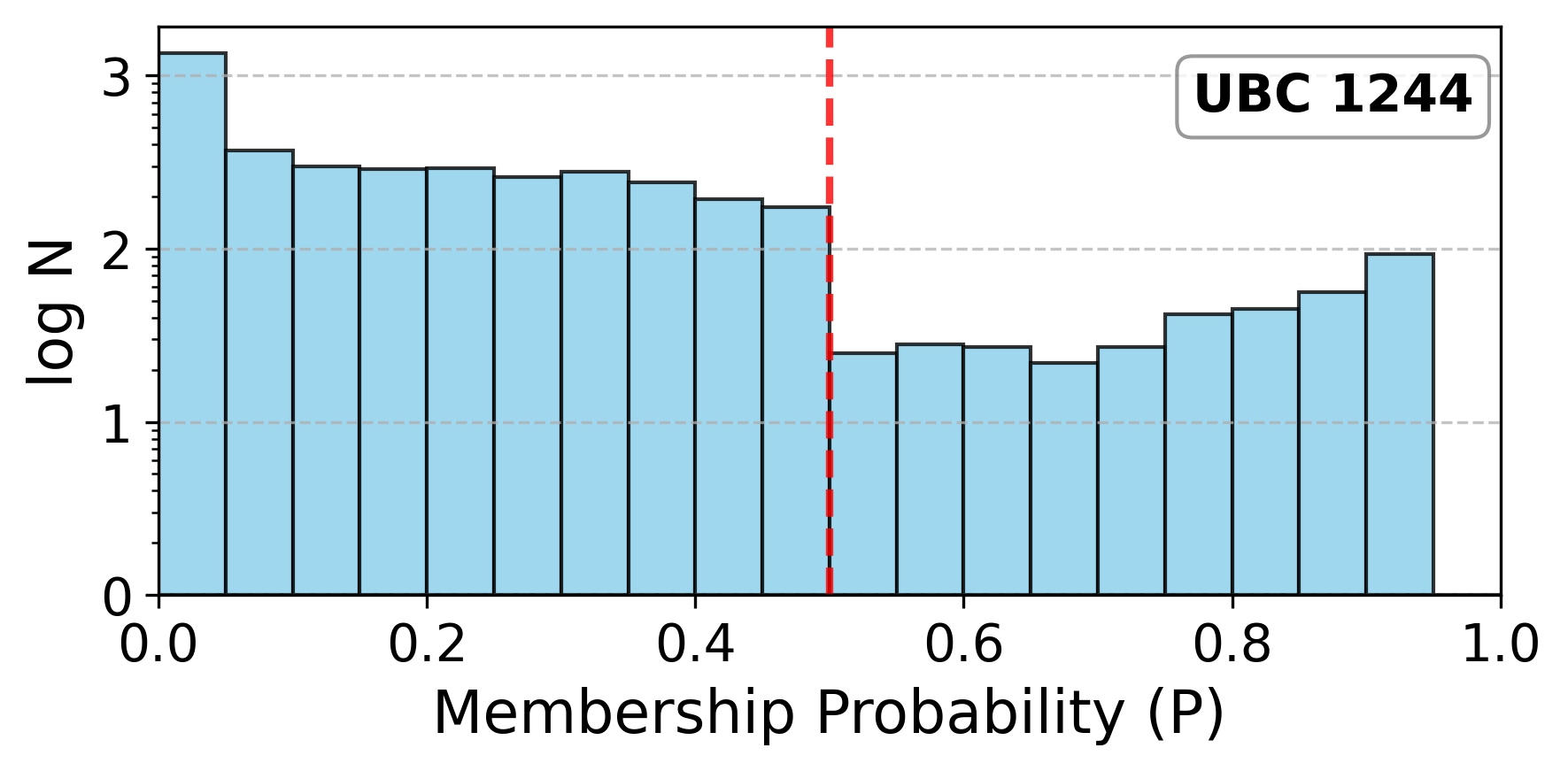}
\includegraphics[width=0.38\linewidth]{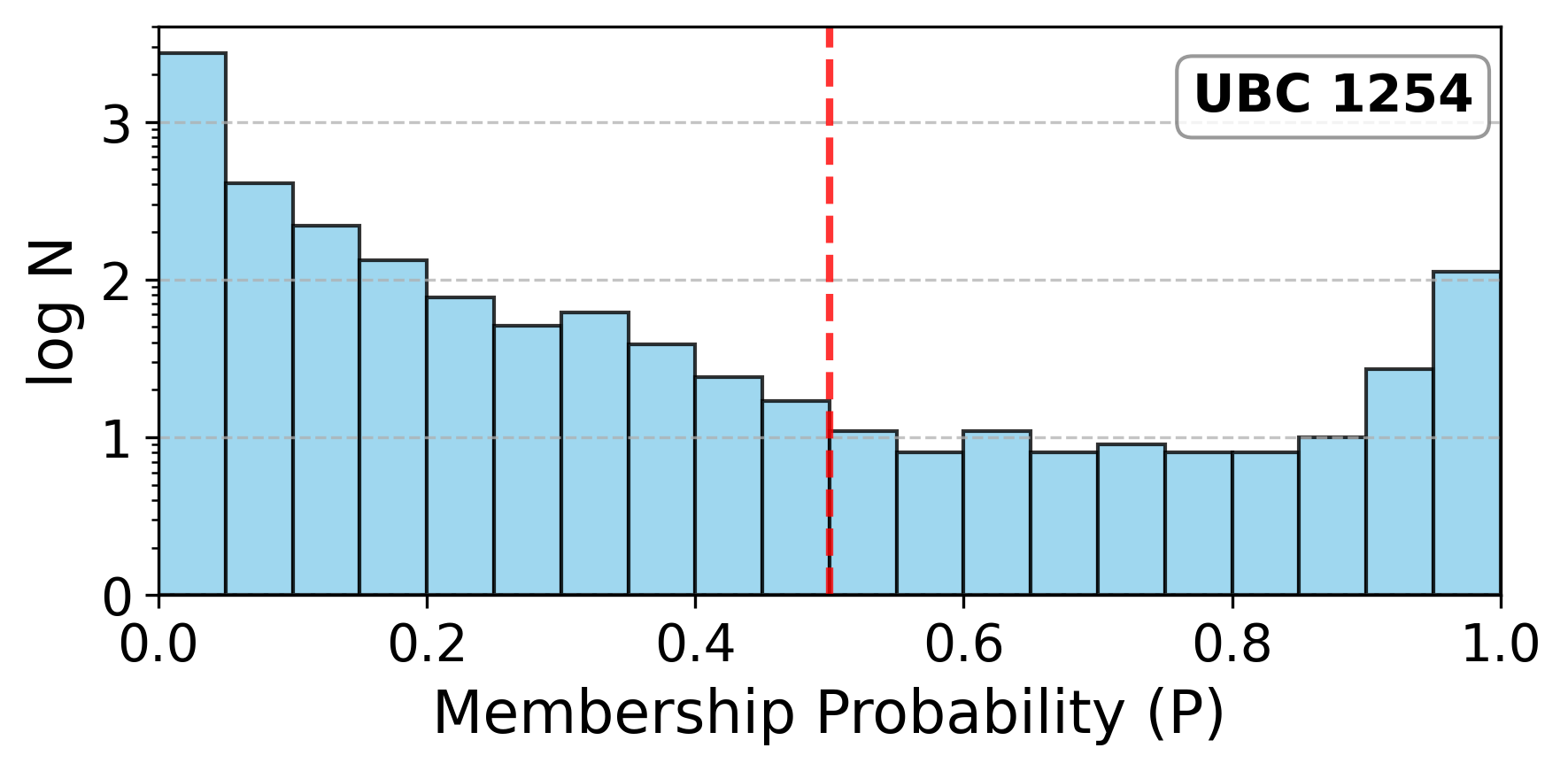}\\
\includegraphics[width=0.38\linewidth]{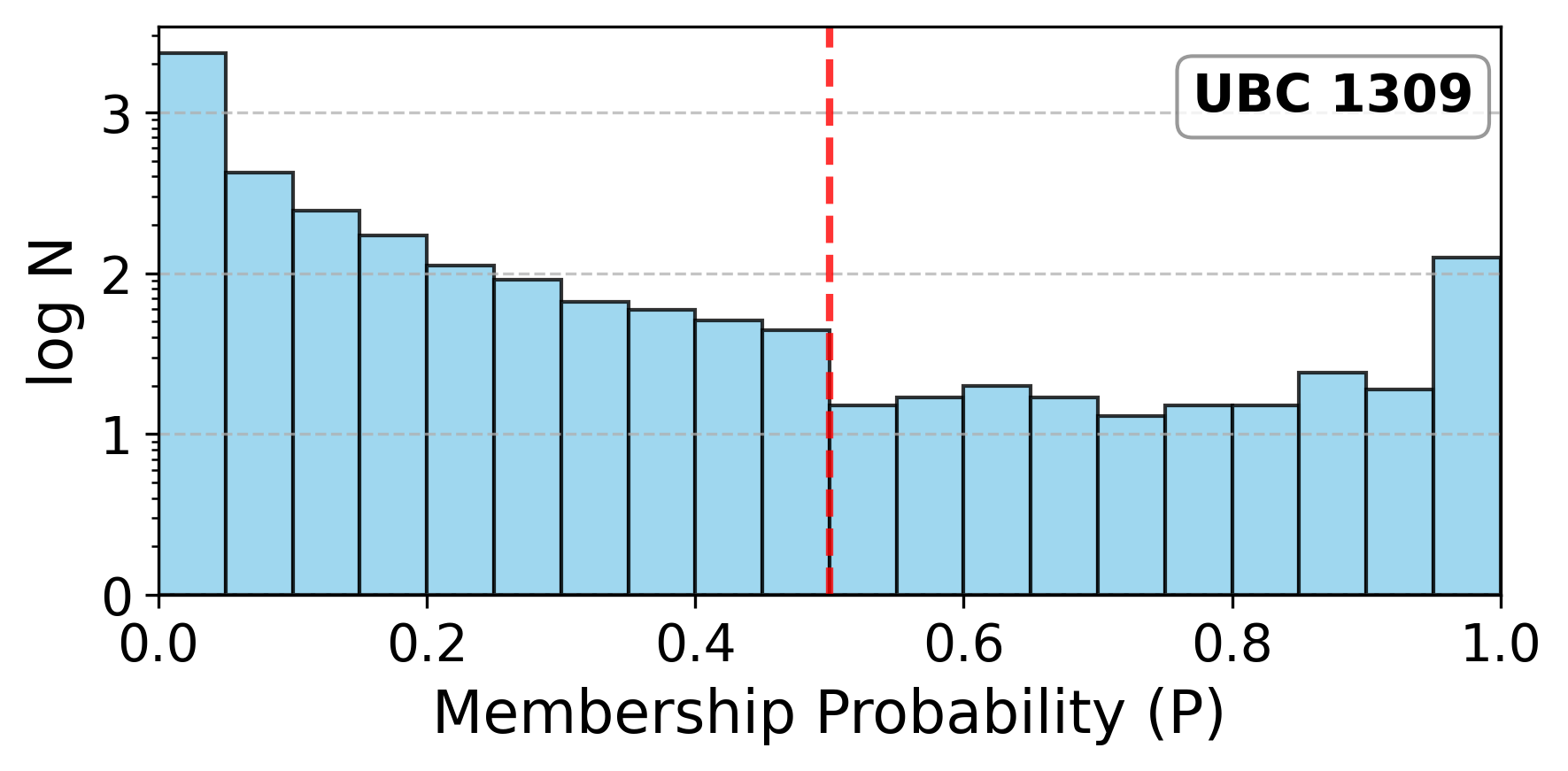}
\includegraphics[width=0.38\linewidth]{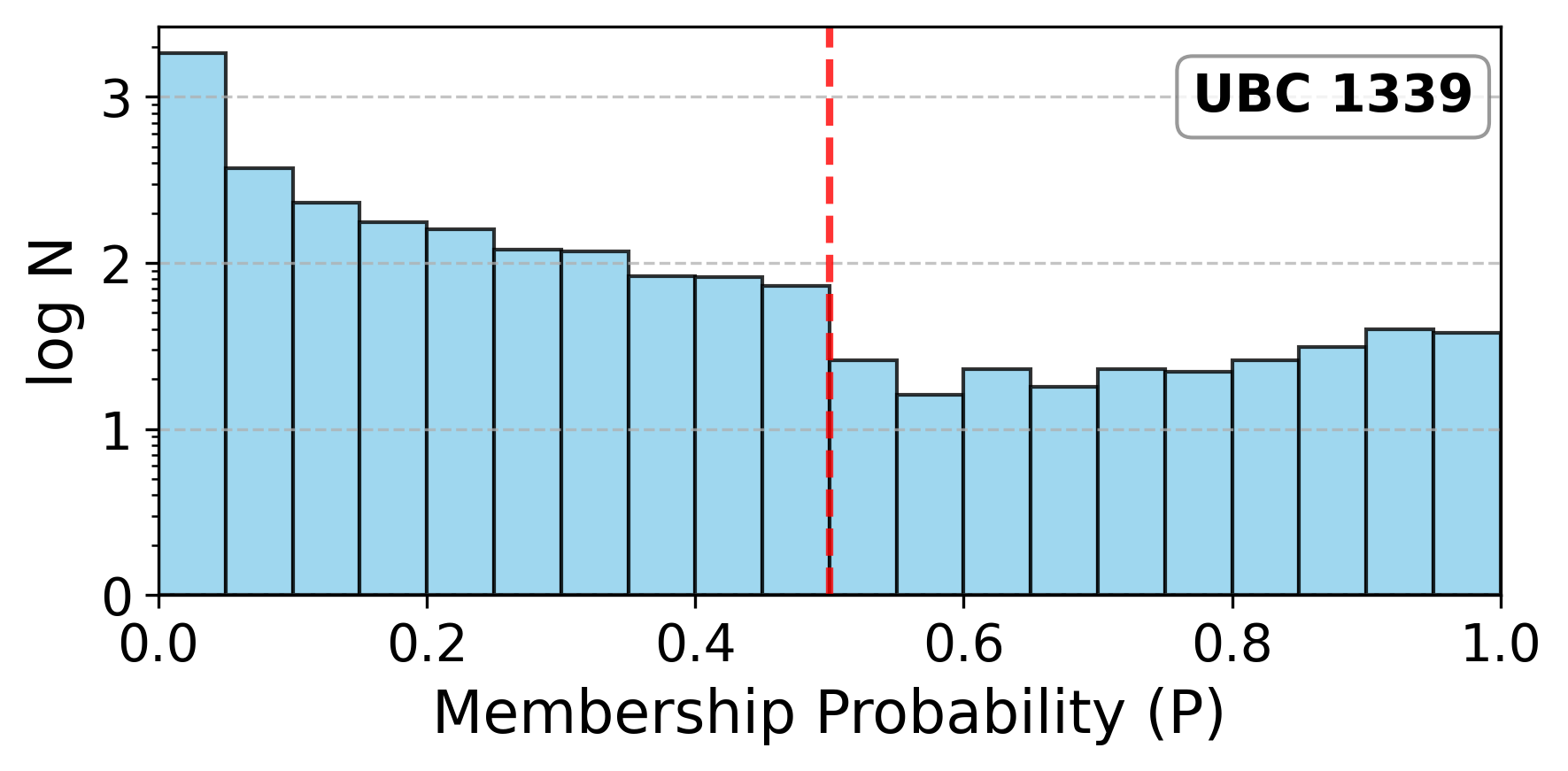}
\caption{The cluster membership probability of stars in the eight OC directions.}
\label{fig:P-histograms} 
\end{figure*}

Following the application of the \texttt{UPMASK} algorithm, stars with membership probabilities in the interval $0< P\leq 1$ were treated as candidate cluster members. This criterion was adopted to provide a balanced trade-off between sample completeness and membership reliability. The resulting probability distributions for the eight OCs are presented in Figure~\ref{fig:P-histograms}. The histograms show that the vast majority of stars fall into the low-probability regime ($P<0.5$), indicating a strong dominance of background-field stars. In the member region ($P \geq 0.5$), the star counts drop significantly but exhibit a distinct peak towards the highest probability bins. This suggests that, while clear, high-probability cluster members exist, the clusters are relatively sparse and heavily outnumbered by the surrounding field, consistent with observations of sparse or low-density OCs. For the final membership determination, a photometric completeness limit of $G\leq 20.5$ mag was imposed, and only stars with membership probabilities of $P\geq 0.5$ were considered as the most probable members of the OCs. The resulting numbers of probable cluster members for the eight OCs are given in Table~\ref{tab:astrometric}.

\begin{figure*}
\centering
\includegraphics[width=0.60\linewidth]{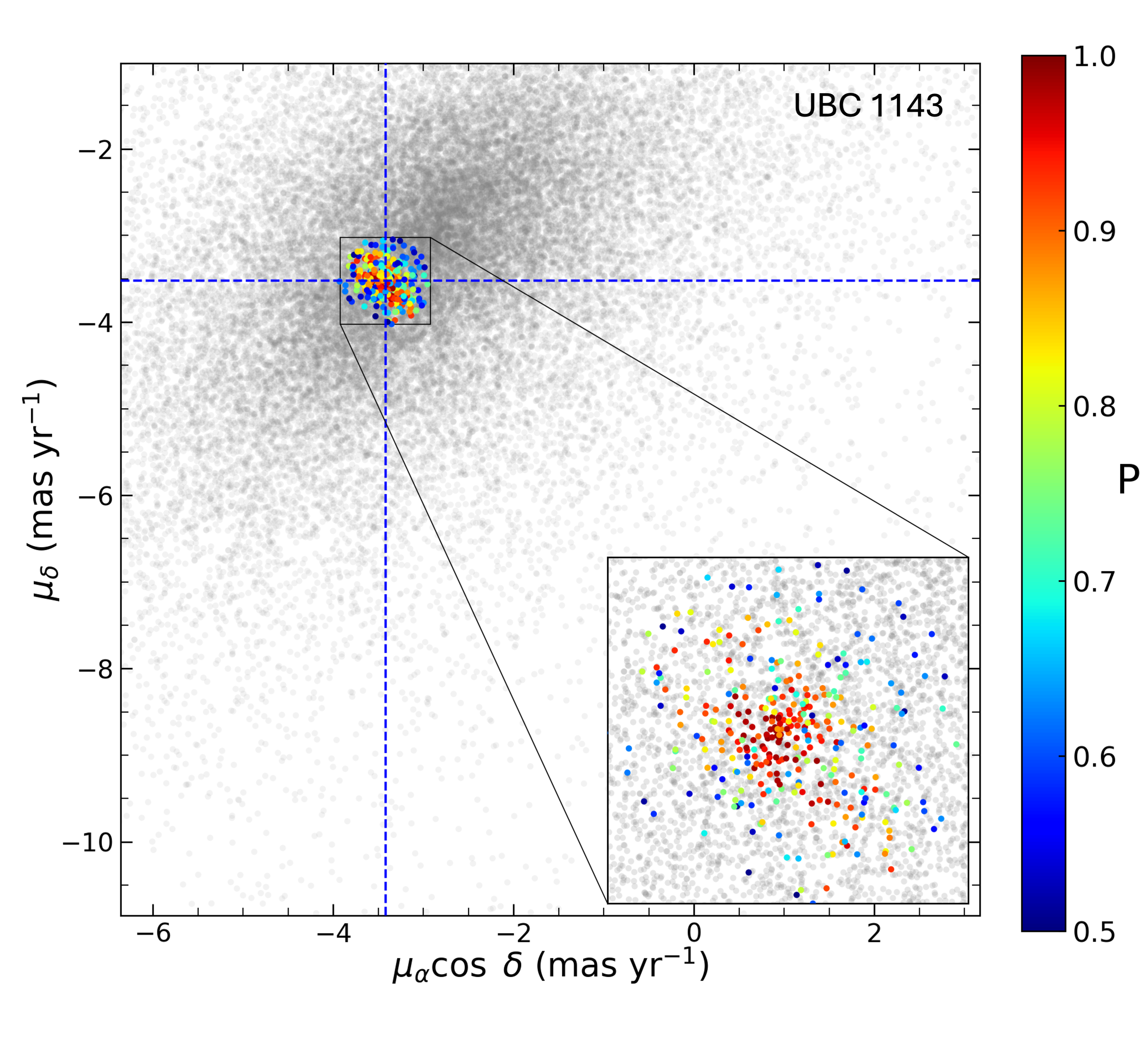}

\caption{VPD of the stellar population in the direction of UBC 1143. Gray dots represent the background field distribution, while color-coded points denote cluster member candidates. The color scale indicates the membership probability $P$, ranging from 0.5 to 1. Dashed blue lines mark the mean proper motion of the cluster. The inset provides a magnified view of the high-probability overdensity.}
\label{fig:vpd_1143} 
\end{figure*}

\section{Astrometric Parameters}
To characterize both the kinematic and dynamical properties of the selected OCs, the astrometric data of the most probable members, extracted from the Gaia DR3 catalog, were employed to estimate the median proper-motion components, trigonometric parallaxes, and to compute the clusters’ orbital parameters. The proper-motion distributions of stars in the selected cluster field UBC~1143 are shown in Figure~\ref{fig:vpd_1143} as an example of the applied procedure. This analysis was applied to all clusters in the study, and the vector point diagrams (VPDs) plotted for each cluster are presented in the Appendix~\ref{fig:vpss-append}.

These diagrams reveal that stars with membership probabilities $P \ge 0.5$ form compact groups that are clearly separated from the field population and display a coherent alignment of their proper-motion vectors. Based on these VPDs, high-probability members were identified for each cluster, and their median proper-motion components ($\langle \mu_{\alpha}\cos\delta, \mu_{\delta} \rangle$) are listed in Table~\ref{tab:astrometric}.

\begin{figure*}
\centering
 \includegraphics[width=1\linewidth]{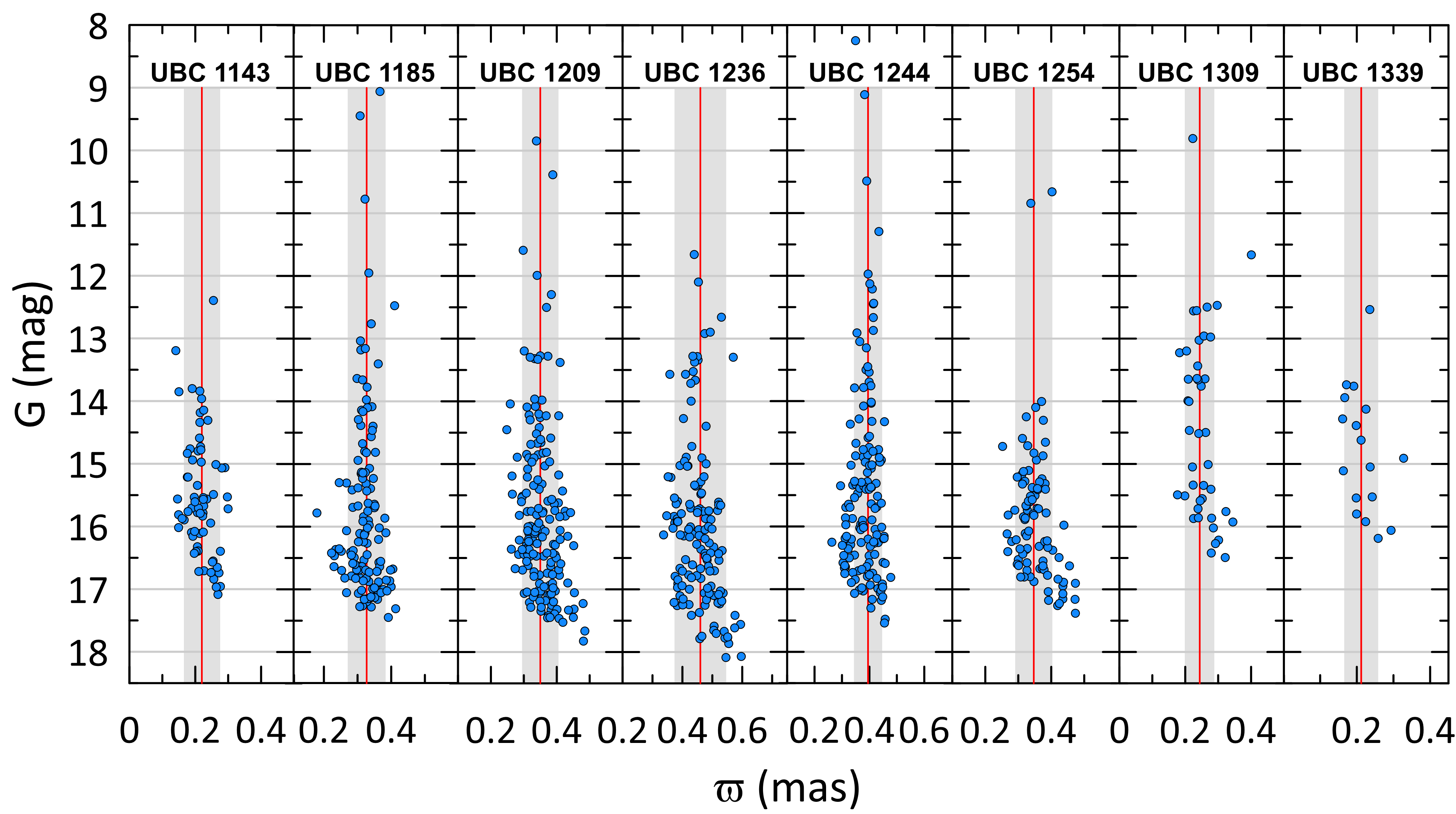}
\caption{Positions of OC member stars satisfying $\sigma_{\varpi}/\varpi \leq 0.2$ in the $G$ versus $\varpi$ diagrams. The red line shows the median trigonometric parallax of the sample, while the shaded regions denote the $\pm 1\sigma$ standard deviations.}
\label{fig:paralaks}
\end{figure*}

The median trigonometric parallaxes of the OCs were derived from the distribution of member stars with relative parallax uncertainties satisfying $\sigma_{\varpi}/\varpi \leq 0.2$ in the magnitude-trigonometric parallax ($G\times \varpi$) diagram shown in Figure~\ref{fig:paralaks}, and the number of stars applying this criterion, which have well-determined trigonometric parallaxes and extend to limiting magnitudes in the range $G \approx 16$-18, is provided in Table \ref{tab:astrometric}. As shown in Figure~\ref{fig:paralaks}, although the cluster members cover a broad interval in magnitude, they are tightly grouped around similar $\varpi$ values, which is consistent with their cluster membership. The resulting median trigonometric parallaxes and their standard deviations are calculated for each OC, while the uncertainties reported in Table \ref{tab:astrometric} are given as standard errors. Heliocentric distances were then estimated from these median trigonometric parallaxes using the relation $d({\rm pc}) = 1000 / \varpi~({\rm mas})$, and the resulting distances and their uncertainties are also listed in Table~\ref{tab:astrometric}.

\begin{table*}[ht]
\centering
\caption{Median astrometric parameters of eight UBC OCs. $N_{\mu}$ denotes the number of stars used in the proper motion estimates. $\langle\mu_{\alpha}\cos\delta\rangle$ and $\langle\mu_{\delta} \rangle$ represent the mean proper motion components of the clusters. $N_{\varpi}$ is the number of member stars considered in the trigonometric parallax analysis. $\langle \varpi\rangle$ indicates the median trigonometric parallax and its standard errors, while $d_{\varpi}$ corresponds to the distances derived from the trigonometric parallaxes of the OCs.}
\label{tab:astrometric}
\begin{tabular}{lcccccc}
\hline
Cluster & $N_{\mu}$ & $\langle\mu_{\alpha}\cos\delta\rangle$ & $\langle\mu_{\delta}\rangle$  & $N_{\varpi}$ & $\langle\varpi\rangle$ & $d_{\varpi}$\\
        &           & (mas yr$^{-1}$)                        & (mas yr$^{-1}$)               &              & (mas)                  &  (pc)    \\
\hline  
UBC~1143 & 326 & $-3.423\pm0.102$ & $-3.524\pm0.092$ & ~80 & $0.220\pm0.006$ & $4535\pm123$ \\
UBC~1185 & 433 & $-4.130\pm0.116$ & $-1.334\pm0.115$ & 145 & $0.327\pm0.005$ & $3056\pm47$ \\
UBC~1209 & 481 & $-2.420\pm0.088$ & $-0.369\pm0.105$ & 174 & $0.350\pm0.004$ & $2858\pm33$ \\
UBC~1236 & 405 & $-0.954\pm0.101$ & $-0.528\pm0.111$ & 160 & $0.460\pm0.007$ & $2175\pm33$ \\
UBC~1244 & 356 & $+0.379\pm0.088$ & $-1.246\pm0.108$ & 150 & $0.395\pm0.004$ & $2532\pm26$ \\
UBC~1254 & 202 & $+0.786\pm0.087$ & $-1.250\pm0.073$ & ~93 & $0.347\pm0.006$ & $2879\pm50$ \\
UBC~1309 & 270 & $-0.158\pm0.141$ & $-0.469\pm0.098$ & ~42 & $0.244\pm0.007$ & $4106\pm118$ \\
UBC~1339 & 253 & $-0.066\pm0.123$ & $+0.562\pm0.121$ & ~17 & $0.212\pm0.008$ & $4715\pm178$ \\\hline
\end{tabular}
\end{table*}

\begin{figure*}
\centering
\includegraphics[width=0.8\linewidth]{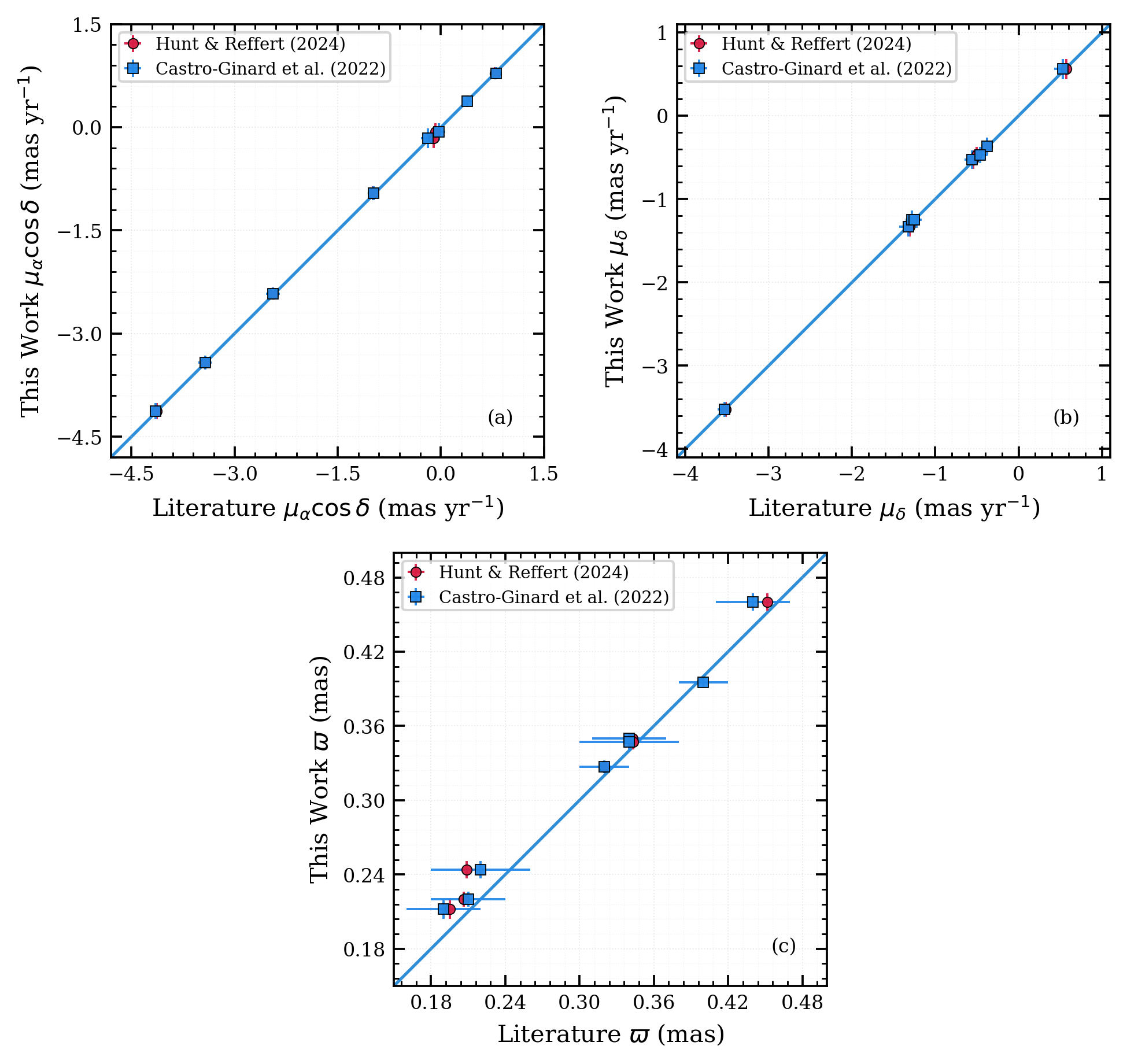}
\caption{Comparison of the clusters’ mean proper motion components (panels a and b) and trigonometric parallaxes (panel c) with values from the literature.}
\label{fig:literature-comparison} 
\end{figure*}

Since these clusters have only recently been introduced into the literature, there are not yet many studies focusing on their astrometric properties. Among the available works, \citet{Castro-Ginard2022}, who identified new OCs using Gaia DR2 data, and \citet{Hunt2024}, who investigated a large sample of OCs using Gaia DR3 data, are particularly noteworthy. The comparison between the astrometric parameters obtained in this study and those reported by these two research groups is presented in Figure \ref{fig:literature-comparison}. However, it should be noted that \citet{Hunt2024} did not include the  UBC 1244 OC in their catalog. Therefore, the comparisons were performed using eight clusters for \citet{Castro-Ginard2022}, while seven clusters were used for the comparison with \citet{Hunt2024}. In the parameter comparison, the mean differences between the parameters and their corresponding standard deviations were considered. For the proper motion comparison with \citet{Castro-Ginard2022}, the mean difference and standard deviation were found to be 0.02 and 0.01 mas yr$^{-1}$, respectively, while similar values were obtained in the comparison with \citet{Hunt2024}. Likewise, for the trigonometric parallax comparison, the mean difference and standard deviation were both determined to be 0.01 mas yr$^{-1}$ for the two studies. Overall, despite the use of different numbers of member stars, the results show good consistency with the literature, and no significant discrepancies were found.

\section{Astrophysical Parameters}
\label{sec: MCMC}
The CMD remains the most fundamental tool for deciphering the evolutionary history of OCs. Historically, the ``main-sequence fitting'' method has provided the classical backbone for distance and age determination by aligning the observed cluster sequence with standard empirical or theoretical templates. This approach leverages the distinct morphology of the CMD, specifically the turn-off point and the slope of the main sequence, to disentangle the physical degeneracy between age, metallicity, and extinction \citep[e.g.][]{Yontan2015, Ak2016, Cinar2024}. Despite the advent of sophisticated automated pipelines, precise morphological analysis of the CMD remains essential for validating the physical membership and evolutionary consistency of stellar populations in the Galactic disk. A detailed description of the methodology is provided in our previous studies \citep[e.g.,][]{Bilir2006b, Bilir2016, Karaali11, Yontan2019, Banks2020, 
Akbulut2021, Yontan2021, Yontan2022, Bostanci2015, Bostanci2018, Koc2022, Ak2024, Karagoz25}.

Robust determination of the fundamental astrophysical parameters, including age, chemical composition, distance, and interstellar extinction, plays a central role in constraining both the evolutionary stage and Galactic context of OCs. In this study, we derive the physical properties of eight UBC OCs through a probabilistic framework based on MCMC sampling. The analysis is carried out using the \texttt{PARSEC} stellar evolution models \citep{Bressan2012}, adopting an implementation tailored to ensure internal consistency between photometric observables and theoretical predictions.

Parameter inference is performed within a Bayesian framework, with posterior distributions sampled assuming Gaussian photometric uncertainties. The likelihood function quantifies the agreement between observed stellar magnitudes and model predictions in the Gaia photometric system ($G,~G_{\rm BP},~G_{\rm RP}$), and is expressed as in \citet{Tanik2025, Tasdemir2026, Bilir2026}:

\begin{equation}
\ln \mathcal{L}(\boldsymbol{\theta}) = - \frac{1}{2} \sum_{i=1}^{N_{\mathrm{stars}}} \sum_{X} \left[ \frac{(m_{X,i} - m^{\mathrm{mod}}_{X,i})^2}{\sigma_{X,i}^2} + \ln(2\pi \sigma_{X,i}^2) \right]
\end{equation}
Here, $m_{X,i}$ and $\sigma_{X,i}$ denote the observed magnitude and its uncertainty for the $i$-th star in passband $X$, while $m^{\mathrm{mod}}_{X,i}$ corresponds to the synthetic magnitude interpolated from the theoretical isochrone grid. The parameter vector $\boldsymbol{\theta}$ comprises $\log(t/\mathrm{yr})$, metallicity ($Z$), heliocentric distance ($d$), and $G$-band extinction ($A_{\rm G}$). The method incorporates the photometric measurements of the most probable member stars in each OC. To sample the posterior distribution, we employed the \texttt{emcee} algorithm \citep{Foreman-Mackey2013}, with the number of walkers equal to the number of high-probability members in each cluster. Each walker was evolved for 5,000 steps to ensure adequate sampling of the parameter space. Following convergence, posterior distributions of the cluster parameters were derived, and the best estimates along with their 1$\sigma$ uncertainties were determined from these distributions.

To minimize the effects of field star contamination, only sources with high membership probabilities were included in the fitting procedure. This selection ensures that the inferred parameters predominantly reflect the intrinsic properties of the clusters. The extinction values in the Gaia $G$ band were derived from the color excesses using the relation $A_{\rm G} = 1.8626 \times E(G_{\rm BP}-G_{\rm RP})$. The obtained Gaia-based reddening and extinction parameters were then transformed into the classical \textit{UBV} photometric system through the relations $E(G_{\rm BP}-G_{\rm RP}) = 1.41 \times E(B-V)$ and $A_{\rm G} = 0.83626 \times A_{\rm V}$ \citep{Canbay_2023}. 

\begin{table*}[ht]
\centering
\caption{Astrophysical parameters estimated for the eight UBC OCs based on the MCMC analysis.}
\label{tab:Final_Results_UBC}
\begin{tabular}{lcccccccc}
\hline
Cluster & $\log(t)$ & $Z$ & [Fe/H] & $d$  & $A_{\rm G}$ & $E(G_{\rm BP}-G_{\rm RP})$ & $A_{\rm V}$ & $E(B-V)$ \\
        & (yr)      &     & (dex)  & (pc) & (mag)       & (mag)                      & (mag)       & (mag)    \\
\hline
UBC~1143 & $8.26_{-0.36}^{+0.22}$ & $0.0070_{-0.0190}^{+0.0081}$ & $-0.34$ & $4394_{-404}^{+406}$ & $2.13_{-0.31}^{+0.26}$ & $1.14_{-0.17}^{+0.14}$ & $2.51_{-0.37}^{+0.31}$ & $0.81_{-0.12}^{+0.10}$ \\
UBC~1185 & $7.35_{-0.40}^{+0.22}$ & $0.0270_{-0.0400}^{+0.0146}$ & $+0.25$ & $3040_{-529}^{+541}$ & $2.43_{-0.40}^{+0.26}$ & $1.30_{-0.21}^{+0.14}$ & $2.87_{-0.47}^{+0.31}$ & $0.93_{-0.15}^{+0.10}$ \\
UBC~1209 & $7.35_{-0.56}^{+0.20}$ & $0.0152_{-0.0264}^{+0.0115}$ & $0.00$ & $3051_{-353}^{+655}$ & $1.98_{-0.45}^{+0.29}$ & $1.06_{-0.24}^{+0.16}$ & $2.34_{-0.53}^{+0.34}$ & $0.75_{-0.17}^{+0.11}$ \\
UBC~1236 & $7.99_{-0.35}^{+0.06}$ & $0.0126_{-0.0233}^{+0.0125}$ & $-0.08$ & $2153_{-423}^{+330}$ & $2.00_{-0.45}^{+0.38}$ & $1.07_{-0.24}^{+0.20}$ & $2.36_{-0.53}^{+0.45}$ & $0.76_{-0.17}^{+0.14}$ \\
UBC~1244 & $7.29_{-0.43}^{+0.04}$ & $0.0218_{-0.0296}^{+0.0220}$ & $+0.16$ & $2482_{-357}^{+208}$ & $2.18_{-0.48}^{+0.26}$ & $1.17_{-0.26}^{+0.14}$ & $2.57_{-0.57}^{+0.31}$ & $0.83_{-0.18}^{+0.10}$ \\
UBC~1254 & $8.67_{-0.42}^{+0.21}$ & $0.0171_{-0.0284}^{+0.0082}$ & $+0.05$ & $2993_{-489}^{+457}$ & $2.59_{-0.43}^{+0.29}$ & $1.39_{-0.23}^{+0.16}$ & $3.06_{-0.51}^{+0.34}$ & $0.99_{-0.16}^{+0.11}$ \\
UBC~1309 & $7.55_{-0.22}^{+0.16}$ & $0.0187_{-0.0280}^{+0.0228}$ & $+0.09$ & $4544_{-371}^{+140}$ & $0.91_{-0.27}^{+0.32}$ & $0.49_{-0.14}^{+0.17}$ & $1.07_{-0.32}^{+0.38}$ & $0.35_{-0.10}^{+0.12}$ \\
UBC~1339 & $8.04_{-0.22}^{+0.40}$ & $0.0149_{-0.0226}^{+0.0100}$ & $-0.01$ & $5215_{-127}^{+246}$ & $0.90_{-0.21}^{+0.16}$ & $0.48_{-0.11}^{+0.09}$ & $1.06_{-0.25}^{+0.19}$ & $0.34_{-0.08}^{+0.06}$ \\
\hline
\end{tabular}
\end{table*}

To account for the dependence of age estimates on heavy-element content, the metallicity ($Z$) values were transformed into iron abundances ([Fe/H]) using a calibrated relation consistent with the adopted stellar models. First, an intermediate parameter is defined as \citep[see also][]{Gokmen2023, Yontan2023a, Yontan2023b, Cakmak2024}

\begin{equation}
Z_{\rm x} = \frac{Z}{0.7515 - 2.78 \times Z}
\end{equation}

\begin{figure*}
\centering
  \includegraphics[width=0.48\linewidth]{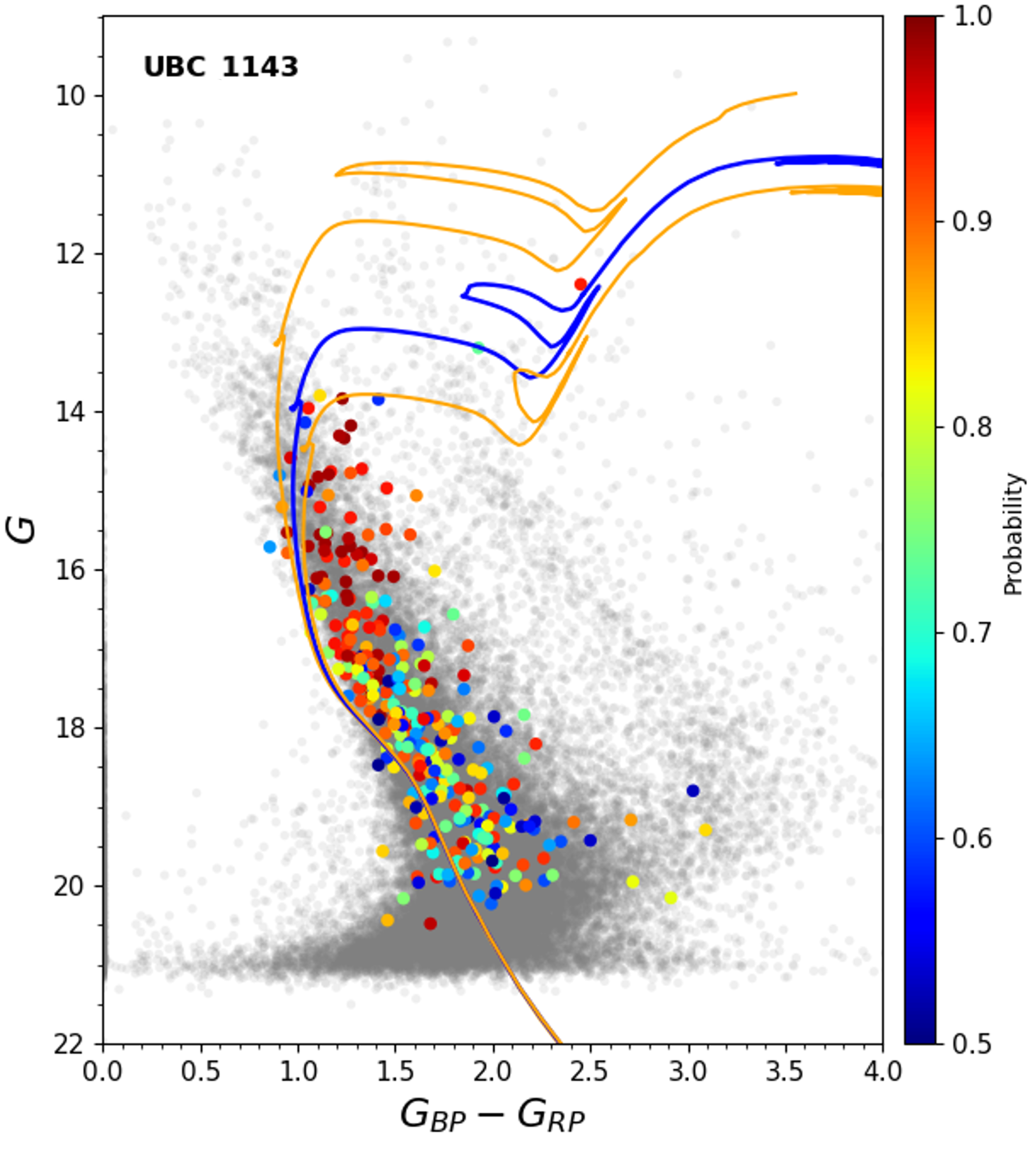}
 \includegraphics[width=0.48\linewidth]{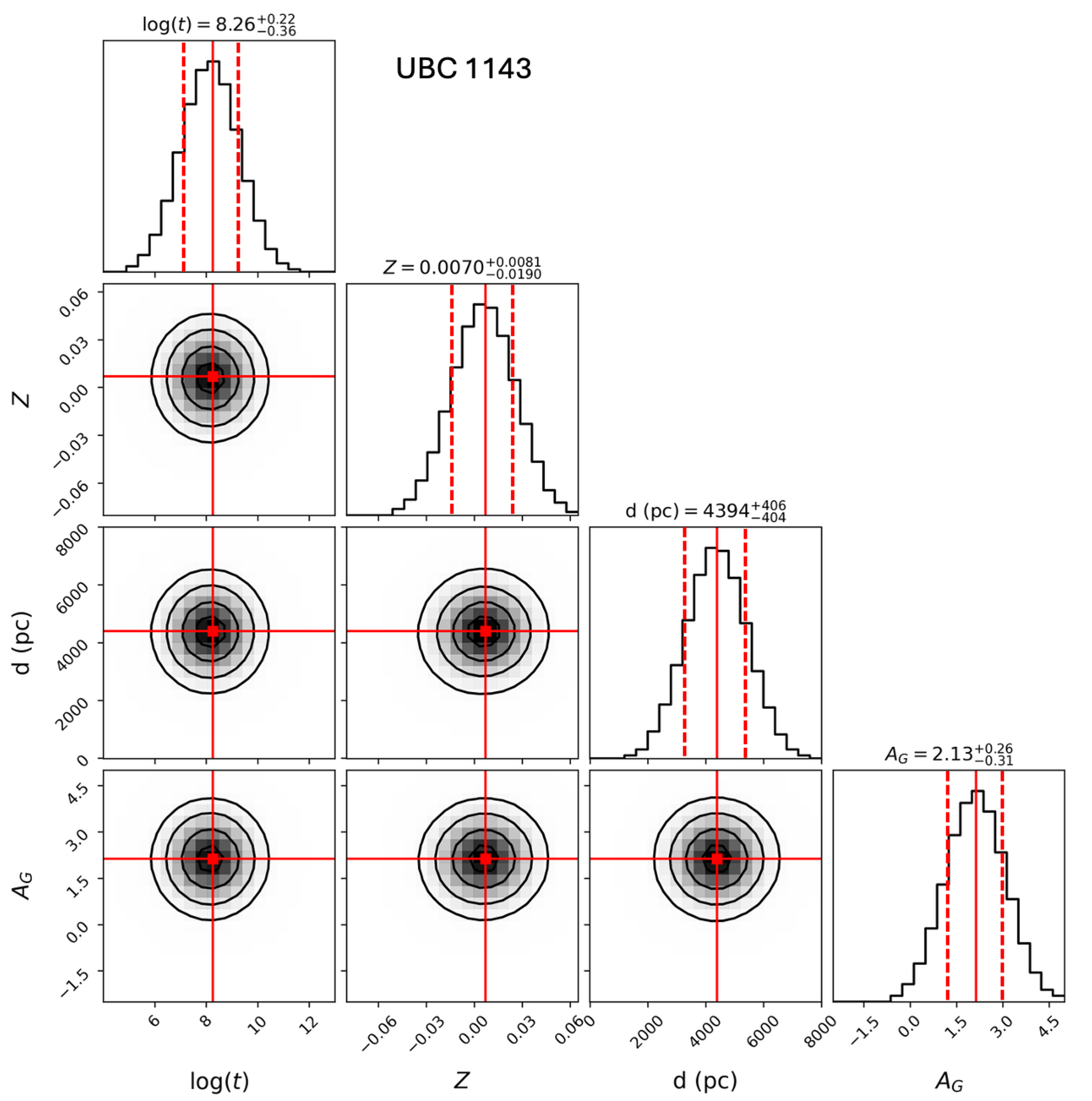}
\caption{CMD of the selected UBC~1143 OC, as a representative example, illustrating the member star distribution. Stars with $P \geq 0.5$ are displayed in different colors, while field stars are shown as filled gray circles. The best-fitting \texttt{PARSEC} isochrone (blue curve) and its uncertainty range (orange curves) are also presented (left panel). The corresponding corner plots illustrating the joint posterior distributions of the astrophysical parameters derived from the MCMC sampling are shown (right panel). The red solid lines indicate the best-fit parameter values, while the dashed red lines correspond to the 16th and 84th percentiles.}
\label{fig:mcmc_cmd}
\end{figure*}

Subsequently, the iron abundance is computed via:

\begin{equation}
\text{[Fe/H]} = \log (Z_{\rm x}) - \log \left( \frac{Z_{\odot}}{1 - 0.248 - 2.78 \times Z_{\odot}} \right)
\end{equation}
where the solar metallicity is adopted as $Z_\odot = 0.0152$. The resulting [Fe/H] values for the eight UBC OCs are listed in Table~\ref{tab:Final_Results_UBC}. The corresponding posterior distributions and best-fitting isochrones superimposed on the CMDs for UBC~1143, shown here as a representative example, are presented in Figure~\ref{fig:mcmc_cmd}, illustrating both the robustness of the fits and the intrinsic parameter correlations. The same analysis for all other clusters is provided in the Appendix~\ref{fig:cmd-mcmc-append}.

\begin{figure*}
\centering
\includegraphics[width=0.8\linewidth]{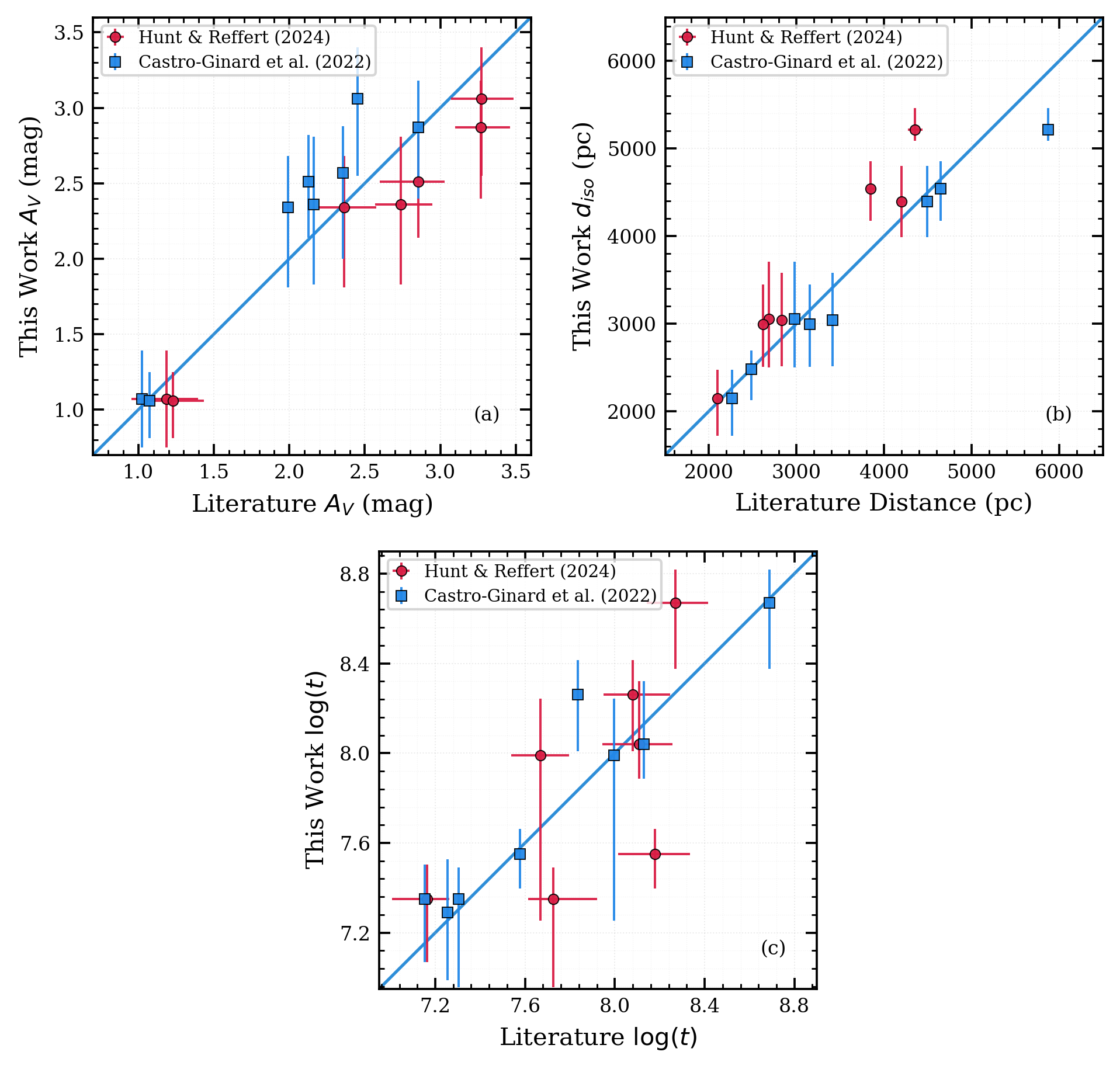}
\caption{Comparison of the clusters’ mean astrophysical parameters with those from the literature: $V$-band extinction (a), distance (b), and age (c).}
\label{fig:literature-comparison_astrophysics} 
\end{figure*}

The comparison of the fundamental parameters of the clusters obtained in this study with those reported by \citet{Castro-Ginard2022} and \citet{Hunt2024} was performed, and the results of these comparisons are shown in Figure~\ref{fig:literature-comparison_astrophysics}.
When considering the extinction values in the $V$ band, the differences and their corresponding standard deviations between this study and \citet{Castro-Ginard2022} were 0.26 and 0.22 mag, respectively. In contrast, \citet{Hunt2024} determined them to be 0.24 and 0.11 mag. Similar comparative analyses were carried out for the cluster distances. In this case, the results of \citet{Castro-Ginard2022} were found to differ by 197 pc and 213 pc, while those of \citet{Hunt2024} showed differences of 399 pc and 316 pc, respectively. Finally, when the logarithmic cluster ages were considered, the differences with \citet{Castro-Ginard2022} were 0.11 and 0.14 dex, whereas for \citet{Hunt2024} they were 0.30 and 0.20 dex, respectively. Considering these results, it is found that the parameters derived in this study are in better agreement with those of \citet{Castro-Ginard2022}, who used \textit{Gaia} DR2 data. Overall, it is concluded that the parameters derived in this study, which are based on a larger number of member stars, are consistent with those reported in the literature.

\section{Kinematic and Dynamical Orbital Analysis}

\subsection{Radial Velocity Determination}

A reliable estimate of the systemic radial velocity ($\langle V_{\rm R} \rangle$) is required to determine the space motion and Galactic orbit of the target clusters. For this purpose, radial velocity measurements were obtained from the Gaia DR3 catalog \citep{Gaia2023} and complemented with data from the Survey of Surveys (SoS) compilation \citep{Tsantaki2022}. The latter provides a homogenized set of spectroscopic measurements drawn from several large-scale surveys, thereby improving both the sample size and the robustness of the results. In total, the initial cross-matched sample comprises 26 stars with radial velocity measurements from Gaia DR3, two stars from LAMOST, and one star from APOGEE.

These kinematic data were cross-matched with our astrometrically selected cluster members. To ensure a clean and reliable sample, we imposed strict selection criteria: only stars with membership probability $P \ge 0.5$ and high-quality astrometric solutions ($\texttt{RUWE} \le 1.4$) were retained. In addition, sources identified as variables or members of binary systems were excluded to minimize potential kinematic contamination. After applying these constraints, the final radial velocity sample for the eight UBC clusters comprises 29 stars in total. The systemic radial velocities were computed as weighted means following the approach of \citet{Soubiran2018} and \citet{Carrera2022}. To make the calculation of the systemic radial velocities reproducible, we explicitly followed the weighted-mean procedure described by \citet{Carrera2022}. For each OC, the mean radial velocity was computed as
\begin{equation}
\langle V_R \rangle =
\frac{\sum_i V_{R,i} \times w_i}{\sum_i w_i},
\end{equation}
where \(V_{{\rm R},i}\) is the individual radial velocity of the \(i\)-th selected member star and \(w_i\) is the corresponding weight, defined as \(w_i=1/\sigma_{V_{\rm R},i}^{2}\). Here, \(\sigma_{V_{\rm R},i}\) denotes the uncertainty of the individual radial-velocity measurement. The internal radial-velocity dispersion was then calculated as
\begin{equation}
\sigma_{V_{\rm R}} =
\sqrt{\frac{\sum_i w_i}{\left(\sum_i w_i\right)^2+\sum_i w_i^2} \times
\sum_i w_i \left(V_{{\rm R},i}-\langle V_{\rm R} \rangle\right)^2}.
\end{equation}
Finally, the uncertainty of the mean radial velocity was adopted as the maximum between the standard error of the weighted dispersion and the contribution from the individual measurement uncertainties. This procedure ensures that radial-velocity measurements with smaller uncertainties contribute more strongly to the systemic velocity estimate, while stars with large radial-velocity errors have correspondingly lower weights. The resulting mean values for all OCs are listed in Table~\ref{tab:orbital_params}, together with the associated uncertainties computed using the procedure described above. Table~\ref{tab:RV_info} provides the properties of the stars used in this analysis, including their Gaia identifiers, equatorial coordinates, $G$-band magnitudes, individual radial velocities, and $\texttt{RUWE}$ values.

\begin{table*}
\footnotesize
\centering
\caption{Radial velocity measurements of stars used to determine the systemic velocities of the UBC clusters. Columns list: running number (Order), cluster name, equatorial coordinates ($\alpha$, $\delta$) in J2000, $G$-band magnitude, individual radial velocity ($V_{\rm R}$) and its uncertainty, \texttt{RUWE}, membership probability ($P$), and the origin of the measurement (Source).}
\begin{tabular}{c l l c c c c c c l}
\hline\hline
Order & Cluster & ID & $\alpha$ (J2000) & $\delta$ (J2000) & $G$ (mag) & $P$ & $V_{\rm R}$ (km s$^{-1}$) & RUWE & Source \\
\hline
01 & UBC 1143 & 1 & 21:38:49.93 & +52:13:45.8 & 12.39 & 0.94 & $-65.45 \pm 0.24$ & 0.98 & Gaia \\
02 & UBC 1143 & 2 & 21:39:06.48 & +52:17:30.5 & 15.07 & 0.88 & $-41.74 \pm 21.60$ & 1.08 & Gaia \\
03 & UBC 1143 & 3 & 21:39:40.54 & +52:15:02.1 & 13.20 & 0.73 & $-73.57 \pm 0.95$ & 1.38 & Gaia \\
\hline
04 & UBC 1185 & 1 & 23:30:49.10 & +61:53:53.2 & 12.76 & 0.89 & $-69.92 \pm 37.77$ & 1.12 & Gaia \\
05 & UBC 1185 & 2 & 23:32:16.43 & +61:58:08.4 & 9.45 & 0.97 & $-48.16 \pm 0.38$ & 0.89 & Gaia \\
06 & UBC 1185 & 3 & 23:32:19.71 & +62:10:22.9 & 10.77 & 0.98 & $-57.98 \pm 3.58$ & 0.96 & Gaia \\
07 & UBC 1185 & 4 & 23:32:30.11 & +62:20:12.6 & 11.95 & 0.94 & $-45.98 \pm 14.64$ & 0.93 & Gaia \\
08 & UBC 1185 & 5 & 23:32:59.03 & +62:12:55.1 & 9.06 & 0.97 & $-49.28 \pm 1.45$ & 0.92 & Gaia \\
\hline
09 & UBC 1209 & 1 & 00:49:53.27 & +64:38.16.05 & 8.23 & 0.84 & $-38.99 \pm 0.97$ & 0.97 & Gaia \\
\hline
10 & UBC 1236 & 1 & 02:23:15.30 & +60:14:00.31 & 12.80 & 0.89 & $-56.19 \pm 0.61$ & 1.09 & Gaia \\
11 & UBC 1236 & 2 & 02:28:15.51 & +59:52:29.32 & 12.05 & 0.88 & $-98.75 \pm 17.37$ & 0.95 & Gaia \\
\hline
12 & UBC 1244 & 1 & 02:51:50.61 & +57:00:12.2 &  9.11 & 0.89 & $-39.27 \pm 0.26$ & 1.01 & Gaia \\
\hline
13 & UBC 1254 & 1 & 03:44:44.59 & +50:50:43.4 & 15.12 & 0.99 & $+60.29 \pm 5.77$ & 1.03 & Gaia \\
14 & UBC 1254 & 2 & 03:45:09.88 & +50:45:52.2 & 14.65 & 0.99 & $-44.39 \pm 2.52$ & 1.00 & Gaia \\
15 & UBC 1254 & 3 & 03:45:10.82 & +50:45:01.8 & 14.10 & 1.00 & $-43.47 \pm 1.69$ & 1.02 & Gaia \\
16 & UBC 1254 & 4 & 03:45:39.05 & +50:44:40.8 & 14.00 & 0.99 & $-43.30 \pm 1.22$ & 0.96 & Gaia \\
17 & UBC 1254 & 5 & 03:45:56.64 & +50:37:17.0 & 14.71 & 0.99 & $+67.33 \pm 9.01$ & 1.06 & Gaia \\
18 & UBC 1254 & 6 & 03:46:08.65 & +50:37:50.1 & 10.84 & 0.99 & $-8.11 \pm 0.62$ & 1.17 & LAMOST \\
19 & UBC 1254 & 7 & 03:46:18.40 & +50:39:14.0 & 10.66 & 0.77 & $-17.59 \pm 0.14$ & 1.02 & Gaia \\
20 & UBC 1254 & 8 & 03:46:24.56 & +50:33:58.0 & 14.59 & 0.94 & $-5.65 \pm 5.59$ & 1.04 & Gaia \\
21 & UBC 1254 & 9 & 03:46:38.85 & +50:39:17.2 & 14.30 & 0.95 & $-52.08 \pm 7.39$ & 1.09 & Gaia \\
22 & UBC 1254 & 10 & 03:46:12.11& +50:34:01.38& 14.26 & 0.99 & $-51.99 \pm 3.79$ & 1.02 & LAMOST \\
\hline
23 & UBC 1309 & 1 & 06:25:01.30 & +19:50:56.12 & 9.80  & 0.99 & $+59.98 \pm 23.76$ & 1.25 & APOGEE \\
24 & UBC 1309 & 2 & 06:26:05.29 & +19:46:24.00 & 14.20 & 0.91 & $+39.87 \pm 3.18$  & 0.99 & Gaia \\
25 & UBC 1309 & 3 & 06:26:29.89 & +19:58:13.07 & 15.17 & 0.89 & $+33.56 \pm 7.75$  & 1.09 & Gaia \\
26 & UBC 1309 & 4 & 06:24:30.93 & +19:59:41.28 & 15.04 & 0.79 & $+20.20 \pm 6.59$  & 0.99 & Gaia \\
27 & UBC 1309 & 5 & 06:25:25.29 & +19:58:40.15 & 13.86 & 0.82 & $+10.27 \pm 1.93$  & 1.03 & Gaia \\
28 & UBC 1309 & 6 & 06:25:06.16 & +19:39:46.25 & 15.26 & 0.86 & $+7.80 \pm 9.18$   & 1.13 & Gaia \\
\hline
29 & UBC 1339 & 1 & 06:41:06.57 & -00:52:14.79 & 14.66 & 0.89 & $+63.55 \pm 4.44$ & 1.05 & Gaia \\
\hline
\end{tabular}

\label{tab:RV_info}
\end{table*}

\subsection{Space Velocities and Galactic Orbital Parameters}

To investigate the dynamical evolution and orbital properties of the eight studied OCs, backward orbital integrations were executed employing the galactic dynamics library \textsc{galpy} \citep{Bovy2015}. The Galactic gravitational field was approximated using the axisymmetric \textsc{MWPotential2014} model, which comprises a Hernquist bulge \citep{Hernquist1990}, a Miyamoto-Nagai stellar disk \citep{Miyamoto1975}, and a Navarro-Frenk-White dark matter halo \citep{Navarro1996}. For these computations, we adopted a solar galactocentric distance of $R_0 = 8.2\pm 0.1$ kpc and a local circular velocity of $V_0 = 238\pm 15$ km s$^{-1}$, in accordance with \citet{Bland2016}. The initial phase-space conditions required for the integration process were defined using the present-day equatorial coordinates, distances, proper motion components, and mean radial velocities ($V_{\mathrm{R}}$) derived in this study. Based on these fundamental observational inputs, we calculated a comprehensive set of kinematic and dynamical parameters for each cluster, including heliocentric Cartesian coordinates ($X, Y, Z$), space velocity components ($U, V, W$), apogalactic ($R_{\rm a}$) and perigalactic ($R_{\rm p}$) distances, orbital eccentricity ($e$), maximum distance from the Galactic plane ($Z_{\mathrm{max}}$), orbital period ($P_{\rm orb}$), estimated birth radii (i.e., birthplace), present-day Galactocentric distance ($R_{\mathrm{gc}}$), and guiding radius ($R_{\rm guiding}$) as listed in Table \ref{tab:orbital_params}). 

Since the initial phase-space conditions include the adopted systemic $V_{\mathrm{R}}$, this parameter represents an important input in the kinematic and orbital analysis of OCs. To evaluate its effect on the orbital solutions, we repeated the orbital integrations by considering the upper and lower \(1\sigma\) limits of the observational input parameters, including the uncertainties in the systemic $V_{\mathrm{R}}$. The corresponding uncertainty-based solutions are presented in Appendix~(\ref{fig:orbit-append-err}). These additional integrations are not intended to replace the nominal orbital parameters reported in the main analysis; instead, they provide an estimate of the sensitivity of the orbital solutions to the input uncertainties. This assessment also emphasizes that the reliability of the derived orbits depends on the number and quality of available stellar $V_{\mathrm{R}}$ measurements.

In the analysis, the heliocentric Cartesian coordinates of the OCs are defined as $X=d\cos l \cos b$, $Y=d \sin l \cos b$, and $Z=d\sin b$, where $d$ denotes the heliocentric distance given Table~\ref{tab:Final_Results_UBC}, and $(l, b)$ represent the Galactic longitude and latitude of the OC center, respectively \citep[see also][]{Canbay2025, Caliskan2025}. In this framework, $X$ corresponds to the axis directed toward the Galactic center, $Y$ indicates the axis aligned with the direction of Galactic rotation at the Sun, and $Z$ represents the axis pointing out of the Galactic plane toward the North Galactic Pole. The $R_{\rm gc}$ of each OC is then calculated using the expression $R_{\rm gc} = \sqrt{R_{0}^{2} + (d \cos b)^2 - 2 R_{0} d \cos l \cos b}$ \citep{Cinar2025, Cinar2026}. The $R_{\rm guiding}$ is defined as the radius of a circular orbit that has the same angular momentum as the actual orbit of the cluster. In axisymmetric potentials, it represents the mean orbital radius around which a star or cluster undergoes epicyclic oscillations and is therefore a key quantity for characterizing radial migration processes \citep[e.g.][]{Binney2008, Schonrich2009}.

To assess the Galactic population membership of the investigated clusters, their space velocity components were corrected to the local standard of rest (LSR). For this transformation, we adopted the solar motion parameters $(U, V, W)_{\odot} = (8.83 \pm 0.24, 14.19 \pm 0.34, 6.57 \pm 0.21)$ km s$^{-1}$, as determined by \citet{Coskunoglu2011}. Following the kinematic classification scheme outlined by \citet{Schuster2012}, objects belonging to the Galactic thin disk are typically characterized by rotational velocity components within the range of $-50 < V_{\mathrm{LSR}} \ (\mathrm{km \ s^{-1}})< 0$. Within our sample, the derived $V_{\mathrm{LSR}}$ values span from $-41.11 \pm 1.51$ km s$^{-1}$ (for UBC 1143) to $2.88 \pm 0.93$ km s$^{-1}$ (for UBC 1309). Although this upper boundary marginally exceeds the strict zero limit, the notably low total space velocities ($S_{\mathrm{LSR}}$) combined with the highly constrained vertical velocity components ($W_{\mathrm{LSR}}$) collectively substantiate the kinematic assignment of all eight clusters to the Galactic thin-disk population.

\begin{figure*}
    \centering
    \includegraphics[width=0.95\textwidth]{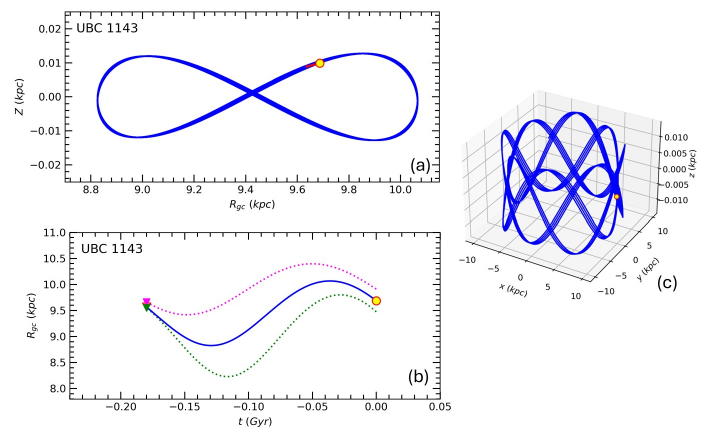}
    \caption{Galactic orbits of the UBC~1143 are shown in the $Z \times R_{\rm gc}$ plane (a) and the $R_{\rm gc} \times t$ plane (b). The three-dimensional spatial trajectory within the Galactic Cartesian reference frame (c). Filled yellow circle and triangle represented UBC~1143 present-day positions and birthplaces, respectively. The upper and lower sets of input parameters for the orbits and birthplaces are indicated by pink and green filled triangles, along with the corresponding dashed lines. Red arrows represent the motion vectors of UBC~1143.}
    \label{fig:ubc1143_orbit}
\end{figure*}

Based on the derived Galactic orbital parameters, the results of our backward orbital integrations robustly characterize the dynamical history and orbital configurations of the eight studied OCs, as listed in Table \ref{tab:orbital_params}. To rigorously analyze the spatial morphology of these orbits, the calculated orbital geometry for UBC 1143 is presented in Figure \ref{fig:ubc1143_orbit} as a representative case. This figure sequentially displays the meridional plane projection ($Z \times R_{\mathrm{gc}}$), the temporal evolution of the galactocentric distance ($R_{\mathrm{gc}} \times t$), and the three-dimensional spatial trajectory within the Galactic Cartesian reference frame ($X-Y-Z$). The kinematic orbital profiles for the remaining seven OCs are provided in the Appendix~(\ref{fig:orbit-append}). Our computations show that the orbital eccentricities are low, ranging from $e = 0.03$ to $0.09$, indicating that the clusters follow nearly circular orbits around the Galactic center. Furthermore, the maximum vertical distances from the Galactic plane are highly constrained, with $Z_{\mathrm{max}}$ amplitudes remaining below the 0.4 kpc threshold for the entire sample. Among the clusters, UBC 1236 exhibits the most tightly bound motion to the Galactic plane with a vertical deviation of merely $Z_{\mathrm{max}}=11\pm5$ pc; meanwhile, even UBC 1309, which possesses the highest vertical amplitude at $Z_{\mathrm{max}}=337\pm27$ pc, remains strictly confined within the geometric and kinematic boundaries of the Galactic thin disk. Ultimately, these dynamical descriptors strongly corroborate the preceding kinematic findings, thereby substantiating our classification that all eight clusters belong to the Galactic thin-disk population \citep{Tuncel-Guctekin2019, Plevne2015}.

In this study, we adopt the concept of the traceback early orbital radius ($R_{\rm teo}$), introduced by \citet{Akbaba2024}. This quantity characterizes the past orbital state of a stellar system by integrating its orbit backwards in time within a static (time-independent) Galactic potential. In contrast to the birth radius ($R_{\rm birth}$), which aims to directly identify the exact formation site, $R_{\rm teo}$ provides indirect yet dynamically consistent constraints on the earlier orbital locations of a cluster within the Galaxy. The estimation of $R_{\rm teo}$ is performed alongside other kinematic and dynamical parameters.

To infer the birth radii of the clusters, the ages derived in this study are traced back in the $R_{\rm gc}\times t$ plane. This approach enables us to combine the time evolution of cluster orbits with age information, thereby constraining their probable birth locations. The method has been successfully applied in several previous studies by our group \citep{Tasdemir2023, Yontan2023c, Cinar2024, Yucel2024, Tasdemir2025}. The lower panel of Figure~\ref{fig:ubc1143_orbit} shows the present-day and birth-time positions of the cluster UBC 1143. The same analysis has been carried out for the remaining clusters, and their birth radii have been determined accordingly. The results are listed in Table~\ref{tab:orbital_params}. The derived dynamical orbital solutions indicate that the eight clusters were born within a Galactocentric distance range of approximately 8.70-13.75 kpc. It should be noted, however, that such backward orbital integrations rely on the assumption of a static Galactic potential and therefore do not fully account for the effects of time-dependent non-axisymmetric structures, such as the Galactic bar and spiral arms \citep{Onal2018}.

\begin{table*}
\centering
\caption{Kinematic and dynamic orbital parameters for the eight UBC OCs. Parameters include Cartesian coordinates ($X, Y, Z$), velocities, and orbital elements derived from backward integration.}
\label{tab:orbital_params}
\renewcommand{\arraystretch}{1.1}
\small
\setlength{\tabcolsep}{5pt}
\begin{tabular}{lcccc}
\toprule
Parameter & UBC 1143 & UBC 1185 & UBC 1209 & UBC 1236 \\
\midrule
$X$ (pc) & -444 & -1229 & -1664 & -1507 \\
$Y$ (pc) & 4371 & 2780 & 2556 & 1538 \\
$Z$ (pc) & -12 & 36 & 93 & -31 \\
$R_{\mathrm{gc}}$ (kpc) & $9.69 \pm 0.22$ & $9.83 \pm 0.35$ & $10.19 \pm 0.38$ & $9.83 \pm 0.30$ \\
\hline
$V_{\mathrm{R}}$ (km s$^{-1}$)   & $-65.93 \pm 0.82$  & $-48.33 \pm 0.34$  & $-38.99 \pm 0.97$ & $-56.24 \pm 0.75$ \\
$U_{\mathrm{LSR}}$ (km s$^{-1}$) & $+117.02 \pm 6.43$ & $+85.57 \pm 8.02$  & $+59.31 \pm 3.22$ & $+53.42 \pm 0.24$ \\
$V_{\mathrm{LSR}}$ (km s$^{-1}$) & $-41.11 \pm 1.51$  & $-4.73 \pm 3.92$   & $+0.76 \pm 3.24$  & $-21.02 \pm 1.07$ \\
$W_{\mathrm{LSR}}$ (km s$^{-1}$) & $-0.92 \pm 0.72$   & $+5.57 \pm 1.03$   & $-0.13 \pm 0.75$  & $-1.11 \pm 0.23$ \\
$S_{\mathrm{LSR}}$ (km s$^{-1}$) & $+124.03 \pm 6.65$ & $+85.88 \pm 8.99$  & $+59.32 \pm 4.63$ & $+57.41 \pm 1.01$ \\
\hline
$R_{\mathrm{a}}$ (kpc)           & $10.07 \pm 0.30$  & $11.50 \pm 1.06$  & $11.50 \pm 1.01$  & $10.16 \pm 0.22$ \\
$R_{\mathrm{p}}$ (kpc)           & $8.83 \pm 0.55$   & $9.61 \pm 0.51$   & $10.19 \pm 0.38$  & $8.47 \pm 0.51$  \\
$e$                              & $0.07 \pm 0.02$   & $0.09 \pm 0.02$   & $0.06 \pm 0.02$   & $0.09 \pm 0.02$   \\
$Z_{\mathrm{max}}$ (pc)          & $13 \pm 5$        & $130 \pm 31$      & $129 \pm 14$      & $11 \pm 5$         \\
$P_{\mathrm{orb}}$ (Myr)         & $248 \pm 13$      & $281 \pm 24$      & $289 \pm 21$      & $244 \pm 1$      \\
$R_{\mathrm{teo}}$ (kpc)         & $9.57 \pm 0.06$   & $10.32\pm 0.45$   & $10.31 \pm 0.41$  & $8.70 \pm 0.61$   \\
$R_{\mathrm{guiding}}$ (kpc)     & $9.41 \pm 0.47$   & $10.47 \pm 0.75$  & $10.40 \pm 0.66$  & $8.95 \pm 0.40$ \\
\bottomrule
\end{tabular}

\vspace{0.2cm} 
\setlength{\tabcolsep}{5pt}
\begin{tabular}{lcccc}
\toprule
Parameter & UBC 1244 & UBC 1254 & UBC 1309 & UBC 1339 \\
\midrule
$X$ (pc) & -1867 & -2556 & -4431 & -4387 \\
$Y$ (pc) & 1632 & 1549 & -970 & -2810 \\
$Z$ (pc) & -101 & -166 & 268 & -230 \\
$R_{\mathrm{gc}}$ (kpc) & $10.20 \pm 0.24$ & $10.87 \pm 0.44$ & $12.67 \pm 0.25$ & $12.90 \pm 0.18$ \\
\hline
$V_{\mathrm{R}}$ (km s$^{-1}$) & $-39.27 \pm 0.26$ & $-17.68 \pm 1.17$ & $+18.83 \pm 4.39$ & $+63.55 \pm 4.44$ \\
$U_{\mathrm{LSR}}$ (km s$^{-1}$) & $+31.80 \pm 1.25$ & $+14.07 \pm 2.82$ & $-8.40 \pm 4.11$ & $-51.88 \pm 4.85$ \\
$V_{\mathrm{LSR}}$ (km s$^{-1}$) & $-19.85 \pm 1.05$ & $-12.01 \pm 2.36$ & $+2.88 \pm 0.93$ & $-9.12 \pm 1.07$ \\
$W_{\mathrm{LSR}}$ (km s$^{-1}$) & $-2.96 \pm 0.35$ & $+0.39 \pm 0.42$ & $-0.01 \pm 3.40$ & $+8.65 \pm 4.10$ \\
$S_{\mathrm{LSR}}$ (km s$^{-1}$) & $+37.60 \pm 1.66$ & $+18.49 \pm 3.70$ & $+8.88 \pm 5.42$ & $+53.38 \pm 6.44$ \\
\hline
$R_{\mathrm{a}}$ (kpc) & $10.20 \pm 0.24$ & $11.20 \pm 0.60$ & $14.56 \pm 0.20$ & $13.81 \pm 0.25$ \\
$R_{\mathrm{p}}$ (kpc) & $8.94 \pm 0.16$ & $9.93 \pm 0.09$ & $12.53 \pm 0.32$ & $12.90 \pm 0.20$ \\
$e$                    & $0.07 \pm 0.01$ & $0.06 \pm 0.02$ & $0.08 \pm 0.01$ & $0.03 \pm 0.01$ \\
$Z_{\mathrm{max}}$ (pc) & $85 \pm 9$ & $144 \pm 29$ & $337 \pm 27$ & $332 \pm 19$ \\
$P_{\mathrm{orb}}$ (Myr) & $251 \pm 6$ & $281 \pm 11$ & $371 \pm 8$ & $365 \pm 7$ \\
$R_{\mathrm{teo}}$ (kpc) & $10.06 \pm 0.29$ & $10.06 \pm 0.45$ & $13.44 \pm 0.99$ & $13.75 \pm 1.70$ \\
$R_{\mathrm{guiding}}$ (kpc) & $9.81 \pm 0.20$ & $10.52 \pm 0.32$ & $13.46 \pm 0.27$ & $13.32 \pm 0.21$ \\
\bottomrule
\end{tabular}
\end{table*}

\section{Radial Migration}\label{sec:Radial_migration}
The present-day Galactocentric positions of OCs do not necessarily reflect their birth locations, as stellar orbits can be redistributed over time by non-axisymmetric structures. This process, commonly referred to as radial migration, is expected to play a key role in shaping both the chemical and structural properties of the Galactic thin disk \citep{Sellwood2002, Schonrich2009, Minchev2010}. Therefore, the interpretation of radial abundance gradients depends on separating evolutionary effects from those imprinted at birth. OCs are reliable tracers for such analyses, since their ages, distances, and chemical abundances are relatively well constrained. In this context, the concept of birth radius has emerged as a powerful diagnostic, whereby the formation radius of a stellar population is inferred from its chemical properties in conjunction with an assumed or empirically constrained Galactic abundance gradient \citep[e.g.,][]{Minchev2018, Frankel2018, Frankel2020}. The comparison between present-day Galactocentric distances and inferred birth radii enables a direct, quantitative estimate of the extent of radial redistribution.

\begin{figure*}
\centering
\includegraphics[width=1\linewidth]{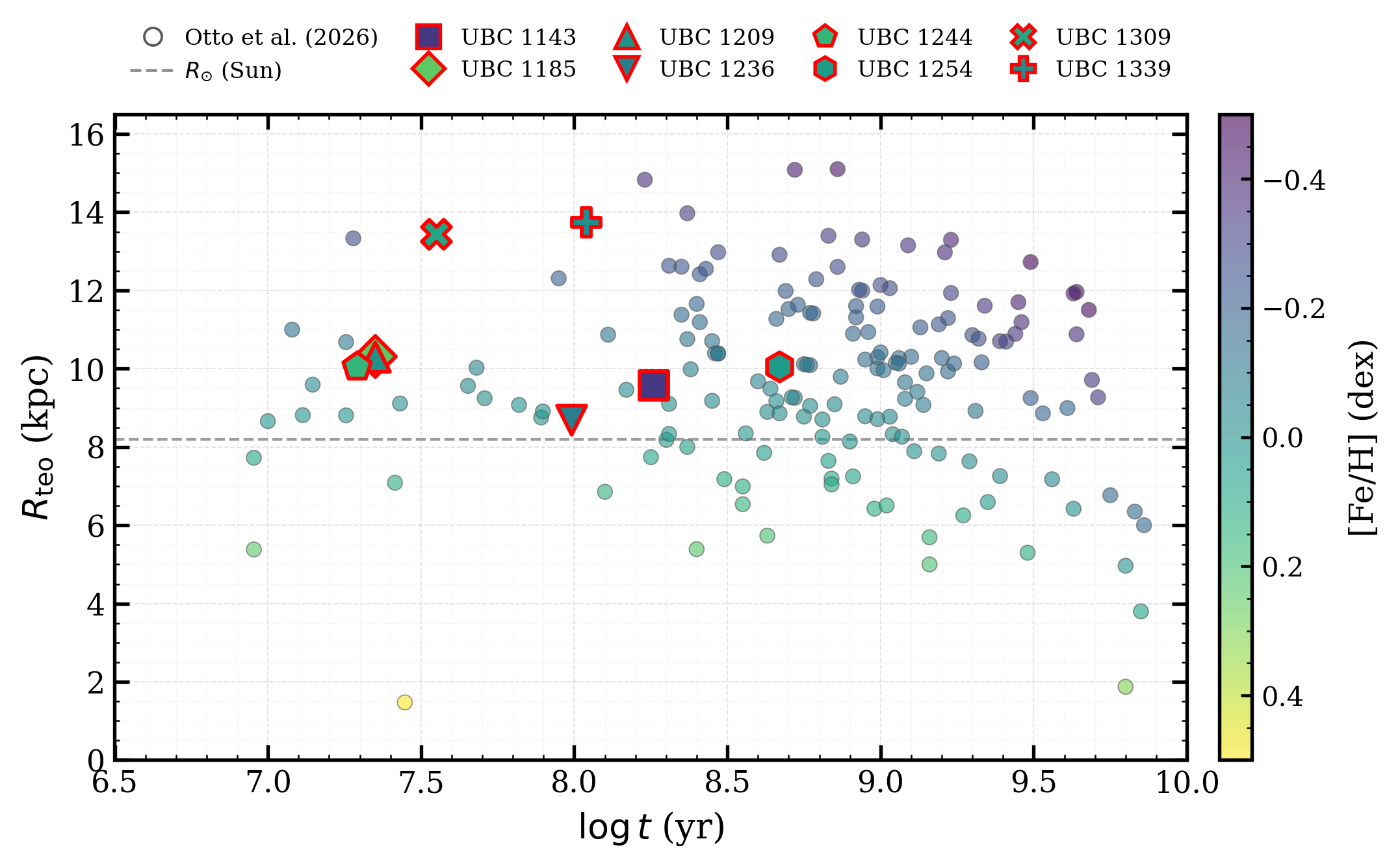}
\caption{$R_{\rm teo}\times \log t$ distribution of OCs. UBC OCs are shown with star symbols, while open circles represent the comparison sample of 158 OCs from \citet{Otto2026}. The color scale indicates the metallicity of each OC, and the horizontal dashed line marks the Sun’s Galactocentric distance.}
\label{fig:Rteo-logt}
\end{figure*}

To further explore the interplay between cluster age, chemical composition, and their formation sites, we examine the distribution of the sample in the $R_{\mathrm{teo}} \times \log t$ plane. The resulting birth radii are presented in this diagram in Figure~\ref{fig:Rteo-logt}, where clusters are color-coded according to their metallicities. This representation enables a direct visual comparison between age, chemical enrichment, and the inferred formation radius. Given the limited size of our sample, we also include, for comparison, a larger reference set of 158 OCs analyzed by \citet{Otto2026}, based on Sloan Digital Sky Survey Milky Way Mapper Data Release 19 (SDSS/MWM DR19) data \citep{SDSS2025}, overplotted on the same diagram.

To quantify the apparent relationship between cluster age and the inferred birth locations, we performed Pearson and Spearman correlation tests on the entire combined sample. The analysis reveals no statistically significant correlation between cluster age and birth radius  ($r = 0.12, p = 0.13$ and $\rho = 0.10, p = 0.21$, respectively), indicating the absence of a reliable systematic age-radius trend. In addition, the metallicities corresponding to the inferred birth radii show a clear decrease with increasing radius, consistent with the well-established negative radial metallicity gradient of the Galactic disk. A multiple linear regression analysis confirms that metallicity ([Fe/H]) is the dominant driver governing the distribution in this plane ($p < 0.001$), while the effect of age remains statistically insignificant ($p = 0.403$). This reinforces that the observed scatter in $R_{\mathrm{teo}}$ is primarily chemically driven rather than a temporal evolutionary effect.

To achieve a more robust characterization of the chemical and dynamical evolution of the Galactic disk, it is essential to consider not only the present-day positions of OCs but also their orbital histories. In this context, radial migration driven by both churning, i.e., changes in the guiding radius due to resonant interactions with transient spiral arms and/or the Galactic bar \citep{Sellwood2002, Roskar2008}, and blurring, i.e., epicyclic excursions around a nearly conserved guiding radius \citep{Schonrich2009}, plays a key role in shaping the observed spatial and chemical distributions of clusters. In this study, the radial displacements of the eight UBC clusters are quantified using the differences between their guiding radii ($R_{\rm guiding}$) and inferred birth radii ($R_{\rm teo}$), defined as $\Delta R = |R_{\rm guiding} - R_{\rm teo}|$. To place our results into a broader context, we extend the analysis by incorporating the OC sample presented by \citet{Otto2026}, enabling a homogeneous comparison across a wider range of ages and Galactocentric distances.

\begin{figure*}
\centering
\includegraphics[width=1\linewidth]{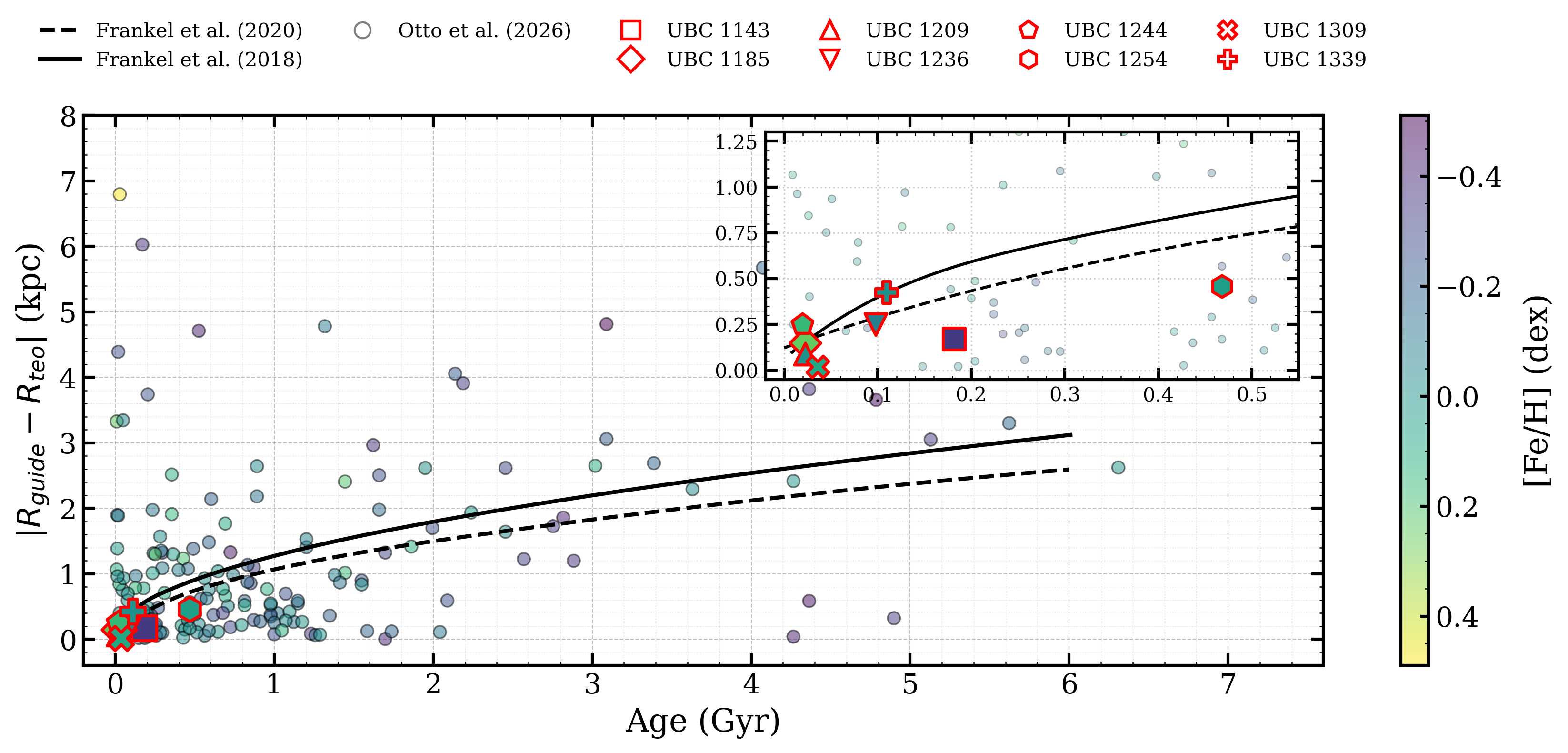}
\caption{Radial displacement–age diagram of the clusters. The clusters analyzed in this study are shown with different symbols, while those from \citet{Otto2026} are represented by circular symbols. The solid and dashed lines in the figure correspond to the theoretical radial migration predictions of \citet{Frankel2018, Frankel2020}. The inset panel shows a zoomed-in view of the region occupied by young clusters. The clusters are additionally color-coded according to their metallicities using the color scale shown on the right-hand side of the figure.}
\label{fig:Birth}
\end{figure*}

To assess the physical significance of the distance offsets inferred from the observations, we compare our results with the theoretical predictions of \citet{Frankel2018, Frankel2020}, as shown in Figure~\ref{fig:Birth}. Given that the ages of the eight clusters analyzed in this study are younger than 0.5 Gyr, an enlarged inset is also provided to better illustrate the expected level of radial migration in this age regime. As is evident from the figure, all eight OCs exhibit relatively small radial displacements from their inferred birth locations, remaining within $\sim$0.5 kpc. These values are consistent with the theoretical expectations for radial migration presented by \citet{Frankel2018, Frankel2020}, which predict limited migration amplitudes for such young stellar populations.

The fact that the clusters are both young and on nearly circular orbits further suggests that they have not experienced significant blurring within the Galactic disk. Taken together, these results indicate that the observed offsets are more likely dominated by mild dynamical evolution rather than strong radial migration processes. Overall, this combined approach provides a coherent framework for disentangling the relative roles of churning and blurring across different age regimes and offers a comprehensive view of the radial migration history and dynamical evolution of the UBC cluster sample within the Galactic disk.

\section{Investigating the Dynamical Behavior of the UBC OCs}
\subsection{Luminosity and Mass Functions}
The luminosity function (LF) is a fundamental diagnostic tool for understanding the stellar content and evolutionary state of star clusters, as it provides a direct link between the observed photometric distribution and the underlying mass distribution. In this study, we derived the LFs for the eight selected UBC OCs based on catalogs constructed from Gaia DR3 astrometric and photometric data \citep{Gaia2023}, converting the $G$-band apparent magnitudes into $G$-band absolute magnitudes ($M_{\rm G}$) using the standard distance modulus equation with correction for interstellar extinction. Using the distance ($d$) and color excess ($E(G_{\rm BP}-G_{\rm RP})$) obtained from the isochrone fitting procedure, we computed $M_{\rm G}$ for each member star using the relation $M_{\rm G}$ expressed as $G-5 \log d+5-A_{\rm G}$. All parameters required for the calculation of the absolute magnitude have been obtained in this study and are listed in Table~\ref{tab:Final_Results_UBC}.

The LF histograms of the eight UBC OCs are presented in Figure~\ref{fig:luminosity_functions}. The CMDs provided in the Appendix~\ref{fig:cmd-mcmc-append} for each cluster were carefully inspected to classify member stars into different evolutionary stages, including evolved stars, main-sequence stars (MS), and pre-main-sequence stars (pre-MS), and to determine the corresponding limits in apparent $G$-band magnitudes. These apparent magnitude limits were then converted into absolute $M_{\rm G}$ magnitude limits using the distance moduli of the clusters, and the corresponding regions in the CMDs are indicated with red dashed lines in Figure~\ref{fig:luminosity_functions}. The classification of stars into luminosity-based groups was subsequently used in the construction of the present-day mass functions (PDMFs), which were derived starting from the main-sequence stellar population of each cluster. Accordingly, the absolute magnitude ranges of the MS stars and the corresponding number of stars within these intervals were determined for each cluster. For UBC~1143, the MS stars span the range $-1.2 < M_{\rm G}~ {\rm (mag)}< 5.1$, comprising 321 stars. In UBC~1185, the MS is defined within $-2.9 < M_{\rm G}~ {\rm (mag)} < 3.6$, including 256 stars, while for UBC~1209, the range $-3.2 < M_{\rm G}~ {\rm (mag)} < 4.1$ contains 294 stars. Similarly, the MS stars in UBC~1236 lie between $-2.0 < M_{\rm G}~ {\rm (mag)} < 6.8$ with a total of 405 stars, and in UBC~1244, the interval $-3.7 < M_{\rm G}~ {\rm (mag)} < 3.8$ includes 202 stars. For UBC~1254, the MS extends over $-0.4 < M_{\rm G} < 5.5$, comprising 196 stars, whereas in UBC~1309, the range $-2.5 < M_{\rm G}~ {\rm (mag)} < 6.2$ contains 269 stars. Finally, for UBC~1339, the MS stars are distributed within $-2.0 < M_{\rm G}~ {\rm (mag)} < 5.9$, with a total of 254 stars. The absolute magnitude distributions of main-sequence stars suggest that they span early to intermediate spectral types, indicating that the clusters are generally young.

\begin{figure*}
\centering
\includegraphics[width=1\linewidth]{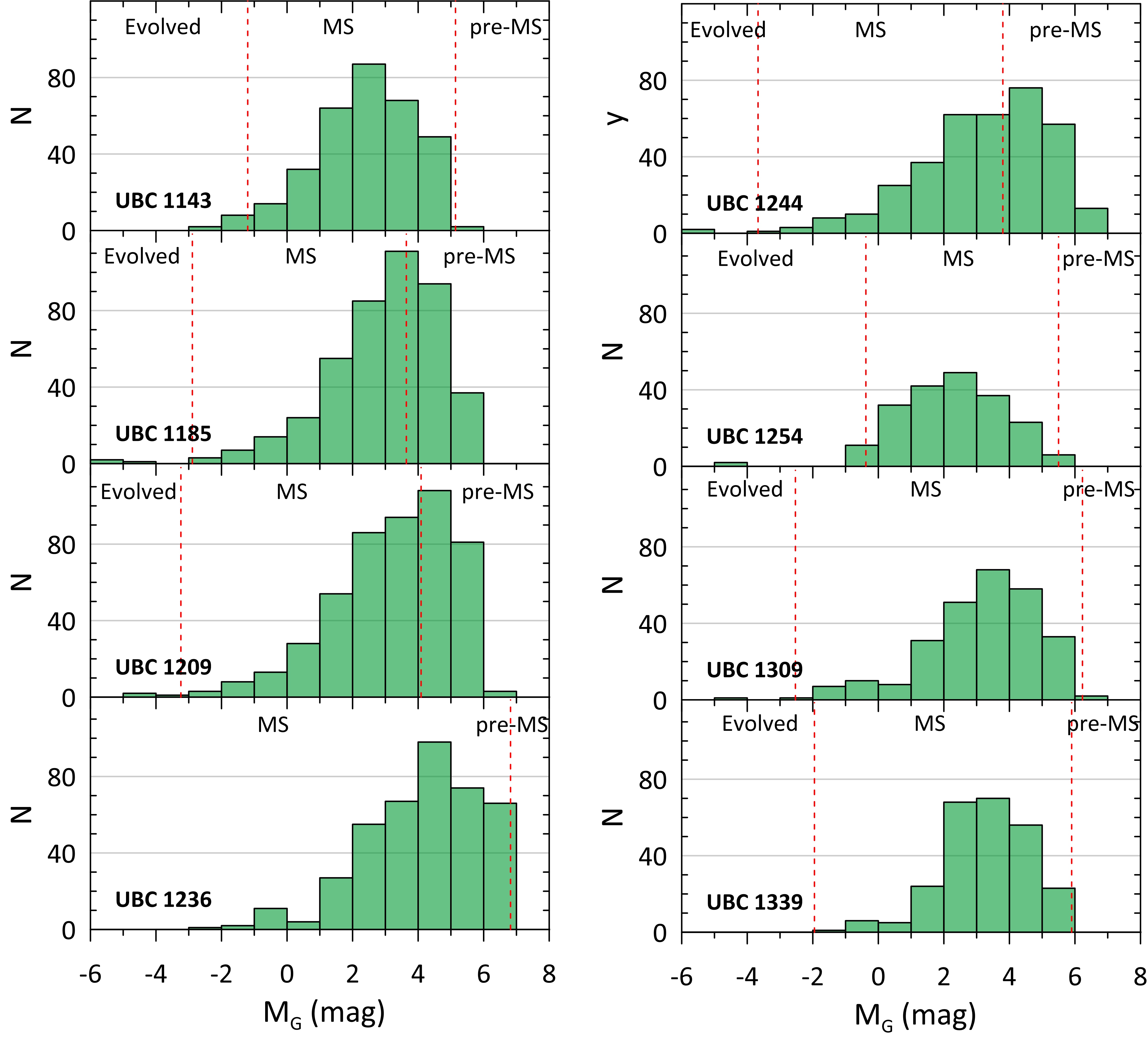}
\caption{The histograms of LFs for eight UBC OCs. The red dashed lines in the panels indicate the regions corresponding to stars of different luminosity classes in the clusters, based on their absolute magnitudes.}
\label{fig:luminosity_functions}
\end{figure*}

\begin{figure*}
\centering
\includegraphics[width=1\linewidth]{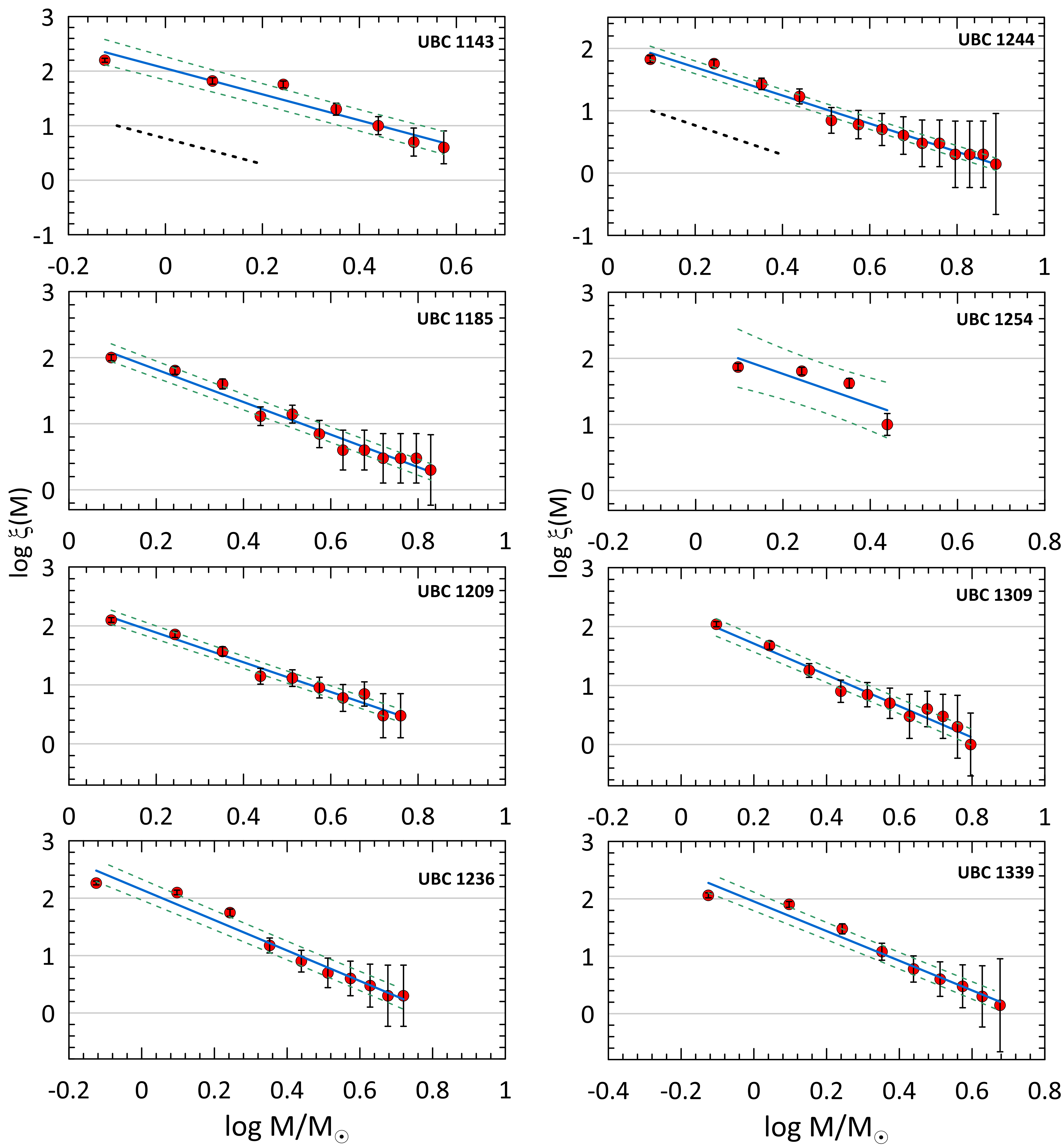}
\caption{PDMFs of the eight UBC OCs. Blue solid lines show the cluster PDMFs, while green dashed lines indicate the $\pm 1\sigma$ uncertainties. Black dashed lines in the upper panels show the \citet{Salpeter1955} slope.}
\label{fig:PDMFs}
\end{figure*}

The PDMFs of the OCs were constructed by deriving the stellar mass distributions for each system. For this purpose, stars were grouped into appropriate mass intervals, and the logarithmic number of stars within each bin was computed. The resulting mass distributions are presented in Figure~\ref{fig:PDMFs}. The uncertainties in the number of stars per mass bin were estimated under the assumption of Poisson statistics, adopting $1/\sqrt{N}$ as the corresponding error. To quantify the slopes of the PDMFs, we adopted the logarithmic representation of the classical power-law formulation originally proposed by \citet{Salpeter1955}. In this context, the mass function can be expressed as:
\begin{equation}
\log \left( \frac{dN}{dM} \right) = -(1 + \Gamma)\times\log M + C,
\label{eq:salpeter}
\end{equation}
where $dN/dM$ denotes the number of stars per unit mass interval, $M$ represents the central mass of each bin, $C$ is a normalization constant, and $\Gamma$ corresponds to the slope of the mass function. A linear least-squares fitting procedure was then applied to the logarithmic mass distributions to determine the PDMF slopes. The best-fit relations are shown by the blue solid lines in Figure~\ref{fig:PDMFs}. The number of MS stars in the OCs examined in this study, the slope $\Gamma$ derived from the linear fit along with its uncertainty, the correlation coefficient of the relation and the mass range of the MS stars are listed in Table~\ref{tab:PDMFs}. The number of MS stars used in the determination of the PDMFs of the eight UBCs studied in this study ranges from 196 to 405. The $\Gamma$ derived from linear fits to the mass distributions lie in the range 1.26-1.65, and the correlation coefficients of the fits are all above 0.7. These results indicate that the derived mass-function slopes are determined reliably. In addition, this study presents, for the first time, the PDMFs of eight UBC OCs in the literature. The total masses of the OCs and the mean stellar masses were derived by summing the masses of all member stars ($P\geq 0.5$) and are listed in Table~\ref{tab:Dynamic}. According to these results, the most massive OC is UBC~1185 with a total mass of 976 $M/M_{\odot}$, while the least massive OC is UBC~1254 with a total mass of 337 $M/M_{\odot}$. Furthermore, considering the mean stellar masses in the OCs, the highest value is found for UBC~1185 with 2.25 $M/M_{\odot}$, whereas the lowest mean stellar mass is obtained for UBC~1236 with 1.29 $M/M_{\odot}$.

\begin{table*}[ht]
\centering
\footnotesize
\caption{Parameters used to derive the PDMFs of the eight UBC OCs.}
\label{tab:PDMFs}
\renewcommand{\arraystretch}{1}
\begin{tabular}{ccccc} 
\hline
Cluster  & $N_{\rm MS}$ & $\Gamma$  & $R^2$ & Mass Range\\
         &     &   &       & ($M/M_{\odot}$)\\
\hline
UBC 1143 & 319 & 1.38$\pm$0.27 & 0.939 & (0.5-3.8] \\
UBC 1185 & 256 & 1.46$\pm$0.14 & 0.969 & (1.3-6.9] \\
UBC 1209 & 294 & 1.51$\pm$0.14 & 0.975 & (1.1-5.9] \\
UBC 1236 & 405 & 1.65$\pm$0.18 & 0.960 & (0.7-5.1] \\
UBC 1244 & 202 & 1.26$\pm$0.10 & 0.977 & (1.1-7.9] \\
UBC 1254 & 196 & 1.30$\pm$0.98 & 0.732 & (0.9-2.8] \\
UBC 1309 & 264 & 1.65$\pm$0.16 & 0.966 & (0.9-6.5] \\
UBC 1339 & 253 & 1.58$\pm$0.17 & 0.970 & (0.8-4.9] \\
\hline
\end{tabular}
\end{table*}

\subsection{Dynamical State and Relaxation Time}

The characteristic timescale associated with this dynamical evolution is the half-mass relaxation time ($T_{\rm E}$), which represents the time required for stellar encounters to significantly alter the velocity distribution of the system. This parameter depends on fundamental cluster properties, including the total number of stars ($N$), the half-mass radius ($R_{\rm h}$), and the mean stellar mass ($\langle m \rangle$). Following the classical formulation of \citet{Spitzer1971}, the relaxation time is expressed as:
\begin{equation}
T_{\rm E} = \frac{8.9 \times 10^{5} N^{1/2} R_{\rm h}^{3/2}}{\langle m\rangle^{1/2}\ln(0.4N)}.
\end{equation}
A useful diagnostic of the dynamical state of a cluster is given by the ratio of its age ($\tau_{\text{age}}$) to the relaxation time. Clusters with $\tau_{\text{age}} / T_{\rm E} \gg 1$ are considered dynamically evolved, indicating that the observed degree of mass segregation is most likely the result of long-term internal dynamical processes. In contrast, systems with $\tau_{\text{age}} / T_{\rm E} \lesssim 1$ are dynamically young, and any observed central concentration of massive stars may primarily reflect the initial conditions of star formation, often referred to as primordial mass segregation \citep{Allison2009, Haroon2025, Cinar2026, Cinar2026c}.

It should also be noted that the interpretation of mass segregation is not always straightforward. External influences, such as the Galactic tidal field, as well as internal mechanisms like stellar feedback and early gas expulsion, can significantly modify the dynamical evolution of clusters, potentially accelerating or even mimicking the effects of relaxation-driven segregation \citep{Kruijssen2012}. Therefore, a comprehensive assessment of a cluster’s dynamical state requires consideration of both internal relaxation processes and environmental effects.

\begin{table*}[htbp]
  \centering
  \caption{Structural and dynamical parameters of the eight UBC OCs.}
    \begin{tabular}{cccccccc}
    \hline
Cluster & $N$ & $r_{\rm t}$ & $r_{\rm c}$ & $R_{\rm h}$ & Total Mass & Mean Mass & $T_{\rm E}$ \\
&   &     (pc)  &     (pc)  &    (pc)   & ($M/M_{\odot}$) & ($M/M_{\odot}$) &  (Myr)  \\
    \hline
    UBC 1143 & 326 & 17.52  & 2.36 & 3.42 & 455 & 1.40 & 40.61 \\
    UBC 1185 & 433 & 14.82  & 2.85 & 3.47 & 976 & 2.25 & 35.65 \\
    UBC 1209 & 481 & 14.88  & 2.22 & 3.06 & 824 & 1.71 & 34.98 \\
    UBC 1236 & 405 & ~~9.53 & 4.45 & 3.52 & 523 & 1.29 & 47.13 \\
    UBC 1244 & 356 & ~~5.74 & 2.30 & 1.96 & 652 & 1.83 & 15.82 \\
    UBC 1254 & 202 & 13.36  & 1.31 & 2.22 & 337 & 1.67 & 16.97 \\
    UBC 1309 & 270 & ~~8.71 & 6.49 & 4.10 & 462 & 1.71 & 45.66 \\
    UBC 1339 & 253 & 17.32  & 8.71 & 6.65 & 378 & 1.49 & 99.18 \\
        \hline
    \end{tabular}%
  \label{tab:Dynamic}%
\end{table*}%

The dynamical relaxation times of the eight UBC OCs analyzed in this study were computed using the half-mass radii, total number of members, and mean stellar masses derived for each cluster. The resulting structural and dynamical parameters are listed in Table~\ref{tab:Dynamic}, which provides a compact overview of the key quantities governing the internal evolution of the OCs. The calculated relaxation times span a broad interval, ranging from $\sim 15.8$ to $\sim 99.2$ Myr, indicating a substantial diversity in the dynamical states of the sample.

A clear trend emerges when considering the dependence of $T_{\rm E}$ on OC structure. Systems with relatively small half-mass radii and higher central concentrations, such as UBC~1244 and UBC~1254, exhibit the shortest relaxation times, consistent with enhanced stellar encounter rates and more efficient energy redistribution. In contrast, more spatially extended clusters, particularly UBC~1339 with its large $R_{\rm h}$, yield significantly longer relaxation times, reflecting a slower approach toward energy equipartition. OCs with intermediate structural properties (e.g., UBC~1143, UBC~1185, and UBC~1209) naturally occupy an intermediate regime in $T_{\rm E}$, illustrating the combined influence of both the number of member stars and the spatial scale of the system. Although the dependence on $N$ is logarithmic, its contribution remains non-negligible when coupled with variations in $R_{\rm h}$ and $\langle m \rangle$.

From a dynamical perspective, the range of relaxation times listed in Table~\ref{tab:Dynamic} suggests that the efficiency of two-body relaxation, and hence the degree of mass segregation, varies across the sample. When compared with the cluster ages (see Table~\ref{tab:Final_Results_UBC}), a significant fraction of these systems likely satisfies $\tau_{\text{age}} / T_{\rm E} \gg 1$, implying that their PDMF distributions have been largely shaped by long-term internal dynamical evolution. However, clusters with relatively large $T_{\rm E}$ values may still be in a transitional regime, where both primordial conditions and dynamical evolution contribute to the observed level of mass segregation. Furthermore, the spread in structural parameters, particularly in $R_{\rm h}$ and core radii, suggests differences in the clusters’ interaction histories with the Galactic environment. External effects such as tidal stripping or disk shocking may have altered their density profiles, thereby indirectly influencing their relaxation timescales. Overall, the results presented in Table~\ref{tab:Dynamic} reinforce the interpretation that mass segregation in these clusters is closely linked to their dynamical evolution, while also highlighting that this process does not proceed uniformly across the sample. Instead, it reflects a complex interplay between internal relaxation mechanisms and environmental conditions, which must be considered jointly when assessing the evolutionary state of OCs.

\section{Summary and Conclusion}\label{sec:conclusion}
In this study, we present a homogeneous chemo-dynamical analysis of eight recently identified UBC OCs using Gaia DR3 data. These systems are located at relatively large heliocentric distances ($\sim$2-5 kpc), sampling the outer regions of the Galactic disk. Owing to the high precision of Gaia astrometry and the use of modern statistical techniques, we derive robust and self-consistent constraints on their structural, astrophysical, and dynamical properties. Our results provide new insights into the evolution of sparse OCs and their role as tracers of Galactic disk processes, particularly radial migration. The key results of this study are as follows:

\begin{itemize}
\item
RDP analysis based on King model fitting reveals tidal radii and overall sizes broadly consistent with expectations for OCs ($\sim$5-15 pc). However, core radii exhibit significant variation, indicating a diversity of internal configurations ranging from compact to extended systems. The derived concentration parameters suggest that the sample includes dynamically evolved as well as relatively diffuse clusters. These results highlight the structural heterogeneity of eight UBC OCs, particularly in low-density regimes.

\item
Bayesian isochrone fitting yields ages spanning from $\sim$20 Myr to $\sim$5 Gyr, covering a wide range of $-0.34 < {\rm [Fe/H] (dex)} <+0.25$. The OCs are located at heliocentric distances between $\sim 2$ and 5 kpc and are affected by moderate extinction ($E(B-V)<1$ mag). The derived parameters are in good agreement with previous studies, but achieve improved precision owing to the high quality of Gaia DR3 data and the use of a homogeneous analysis framework; importantly, the explicit determination of OC metallicities enables more reliable and better-constrained age estimates.

\item
All clusters follow nearly circular Galactic orbits with low eccentricities ($e\approx 0.03-0.09$) and low vertical distances from the Galactic plane ($Z_{\rm max}<0.4$ kpc). These properties firmly place the sample within the Galactic thin disk and indicate dynamically cold kinematics typical of OCs \citep{Tuncel-Guctekin2019}.

\item
The combination of structural and Galactic orbital parameters suggests that the clusters are dynamically stable systems evolving under the influence of the Galactic potential. The diversity in concentration and core radii, together with their nearly circular orbits, implies different evolutionary stages, including possible signatures of internal relaxation and external tidal effects.

\item
A comparison between the present-day guiding radii and the inferred traceback early orbital radii indicates modest radial displacements for the UBC sample. The derived $R_{\rm teo}$ values span $\sim 8.70-13.75$ kpc, while the corresponding offsets remain small ($\Delta R<0.5$ kpc). These results are consistent with mild radial redistribution rather than strong radial migration, in agreement with the dynamically cold and nearly circular orbits of the clusters.

\item
Given their relatively large distances from the Sun and locations in the outer disk, these clusters provide valuable constraints on Galactic structure beyond the solar neighborhood. Our results show that even sparse and distant OCs can be analyzed with high precision using Gaia data, and that mild radial redistribution should be considered when interpreting the spatial and chemical evolution of the Galactic disk.

\end{itemize}
In summary, this study demonstrates that newly discovered UBC OCs, despite their low densities and large distances, are powerful tracers of the chemo-dynamical evolution of the Galactic thin disk when analyzed with modern high-precision datasets and statistical methods.


\section*{Acknowledgements}
We sincerely thank the anonymous referee for a thorough review and constructive suggestions that greatly improved the clarity and quality of the manuscript.
We made use of SIMBAD and VizieR databases at CDS, Strasbourg, France. This research has made use of the Astrophysics Data System, funded by NASA under Cooperative Agreement 80NSSC21M0056. We made use of data from the European Space Agency (ESA) mission \emph{Gaia}\footnote{https://www.cosmos.esa.int/gaia}, processed by the \emph{Gaia} Data Processing and Analysis Consortium (DPAC)\footnote{https://www.cosmos.esa.int/web/gaia/dpac/consortium}. Funding for DPAC has been provided by national institutions, in particular the institutions participating in the \emph{Gaia} Multilateral Agreement. 

\appendix
\section{RDPs}\label{fig:rdps-append}

\begin{figure}
    \centering
    \includegraphics[width=0.45\linewidth]{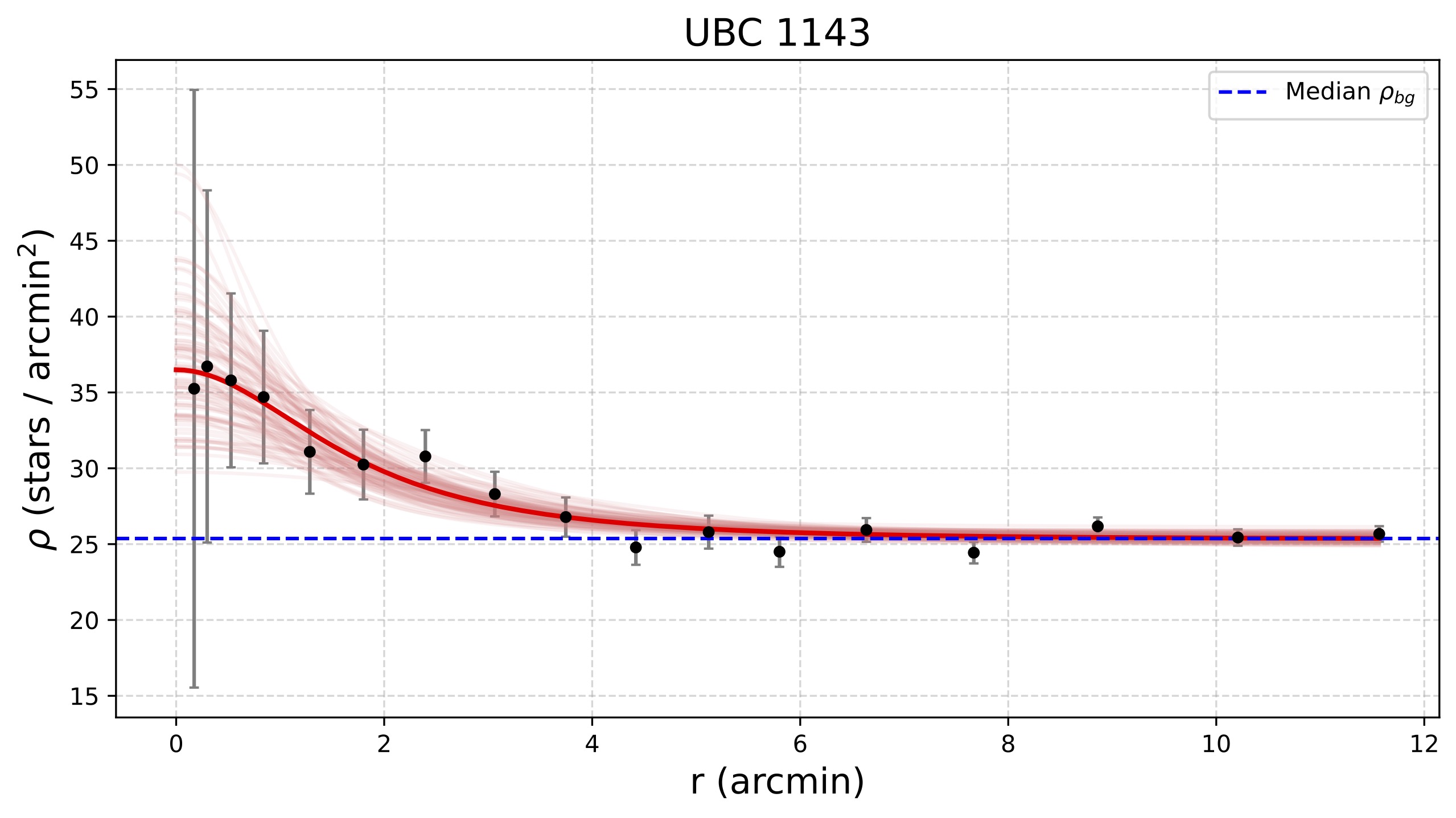}
    \includegraphics[width=0.45\linewidth]{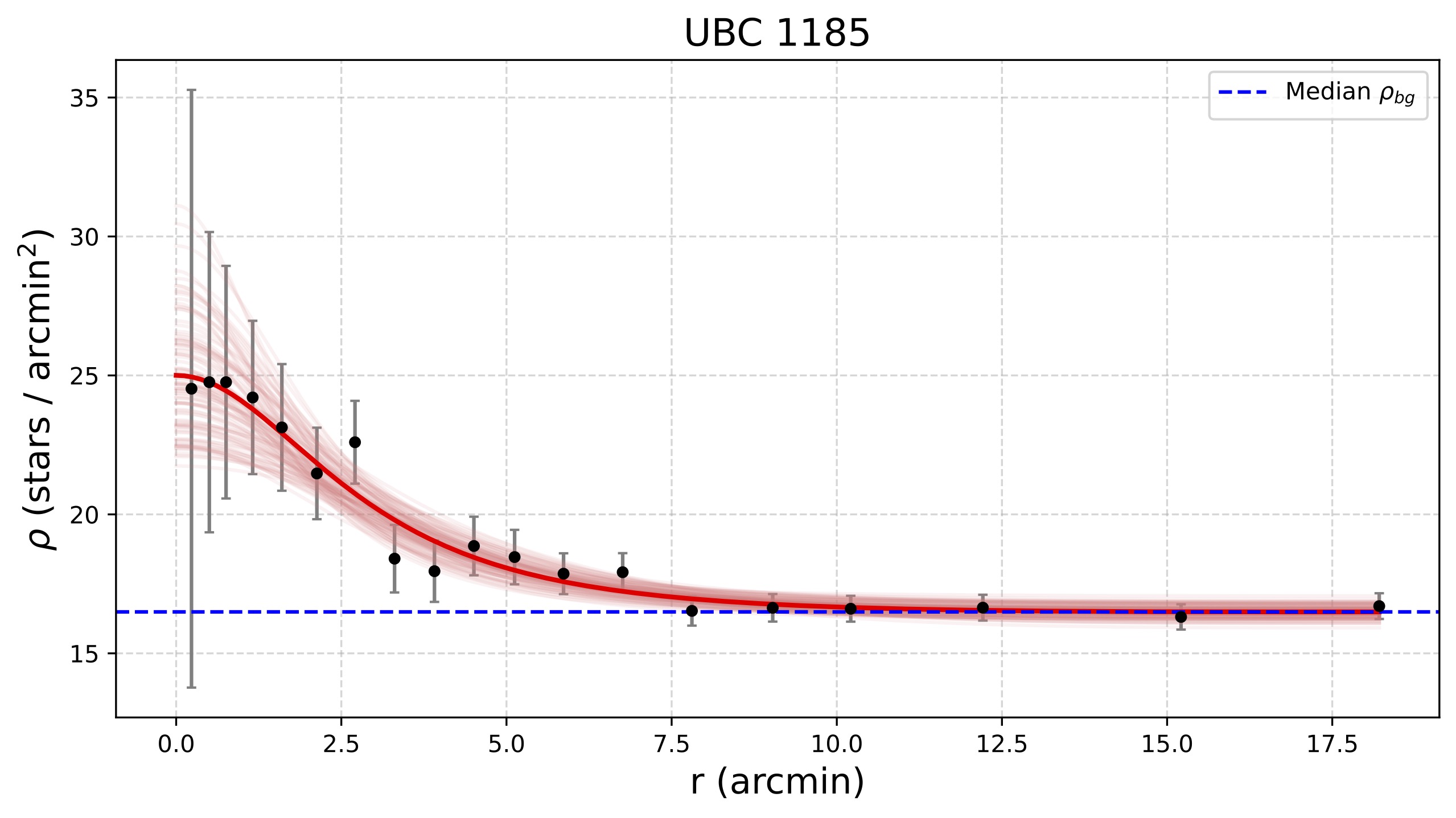}
    \includegraphics[width=0.45\linewidth]{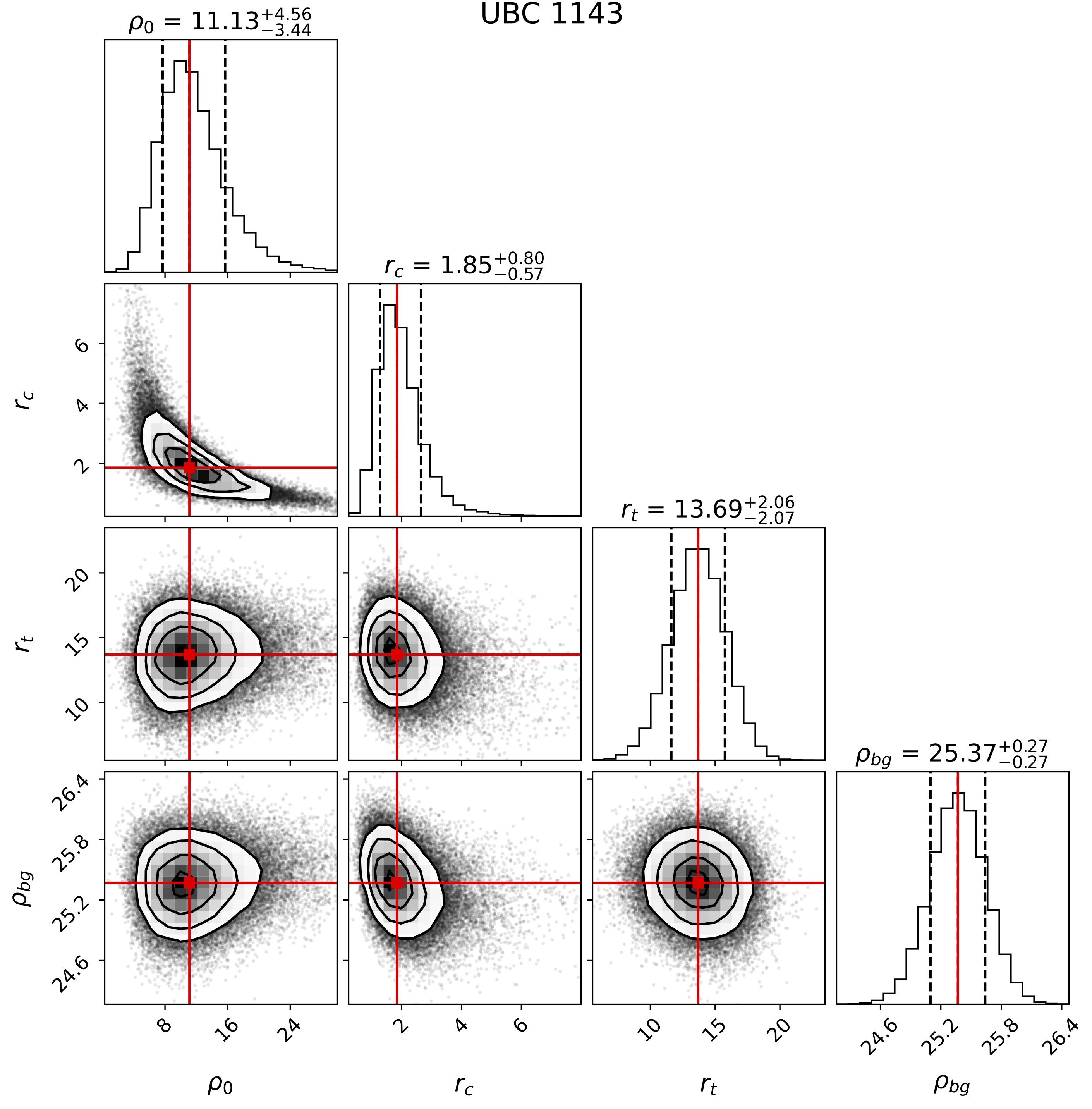}
    \includegraphics[width=0.45\linewidth]{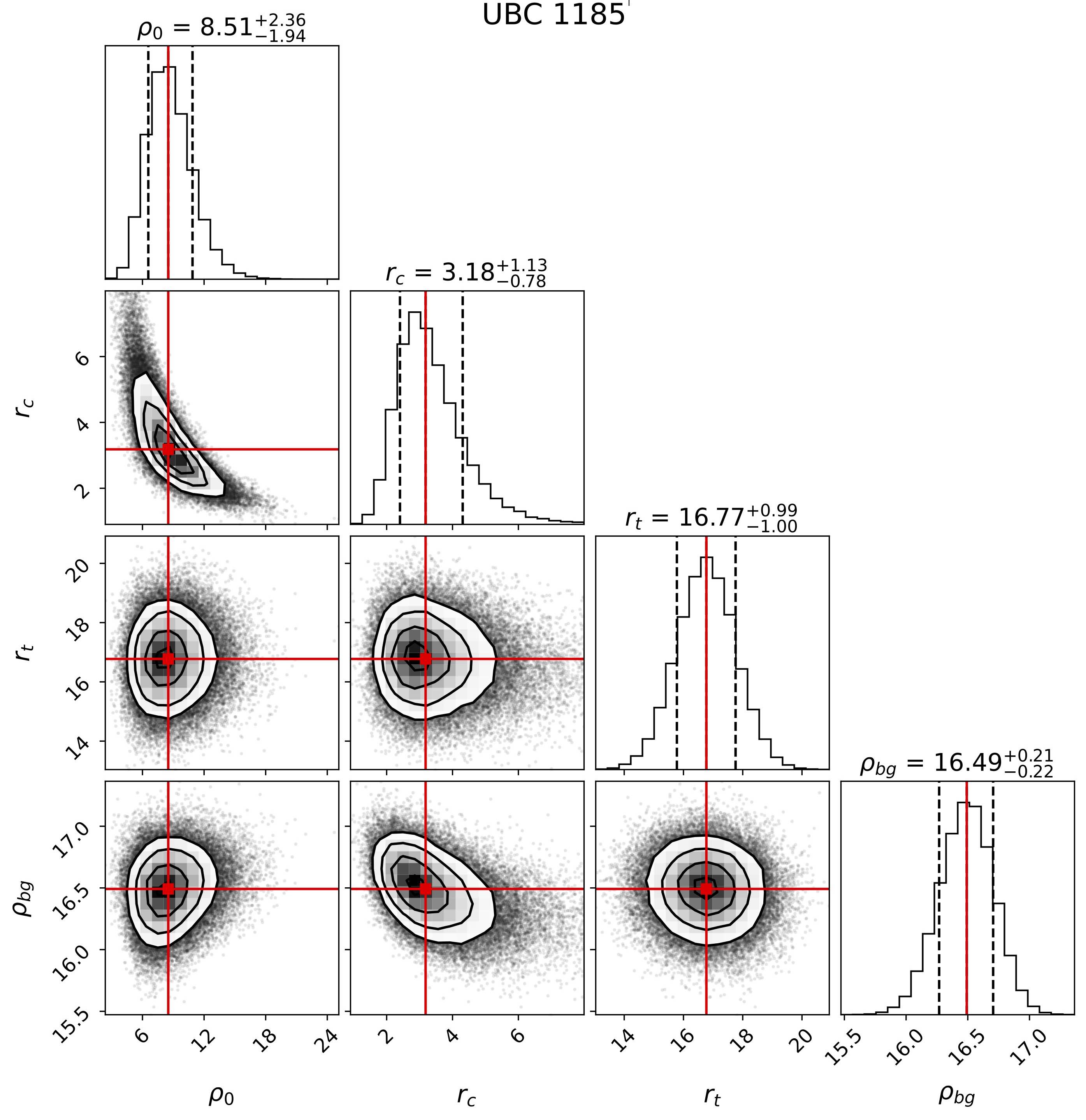}\\
    \includegraphics[width=0.45\linewidth]{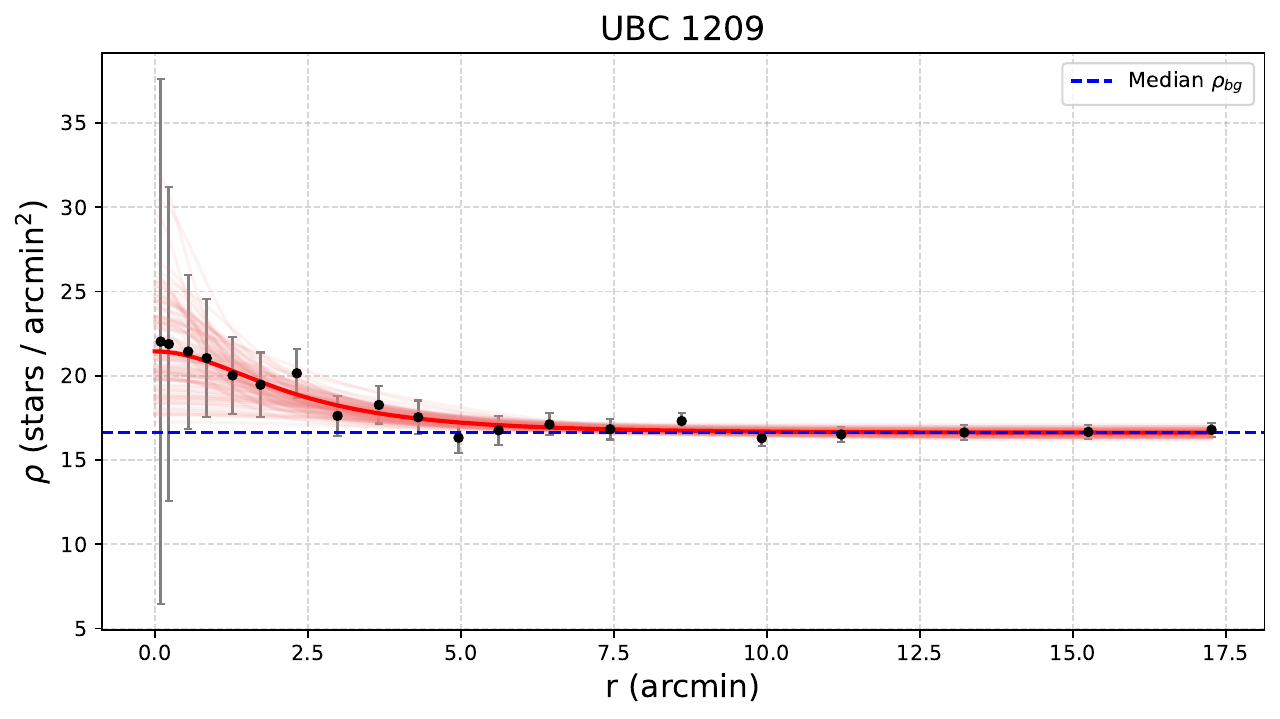}
    \includegraphics[width=0.45\linewidth]{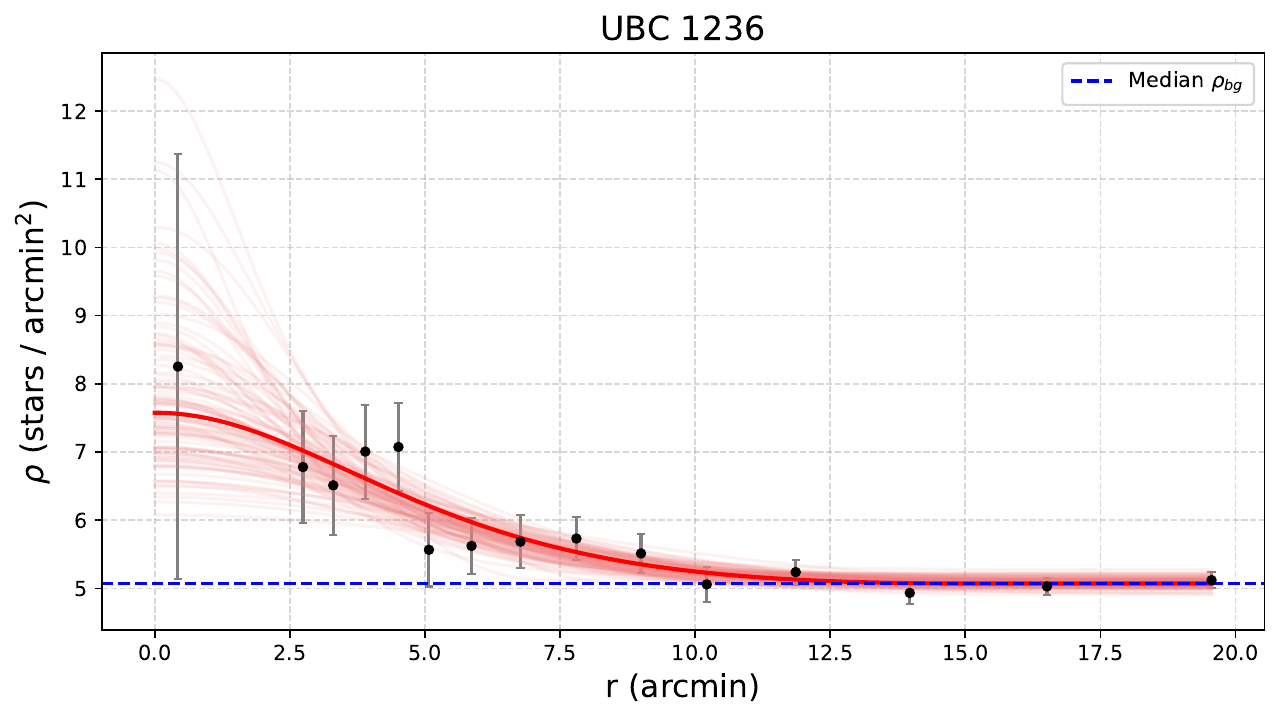}
    \includegraphics[width=0.45\linewidth]{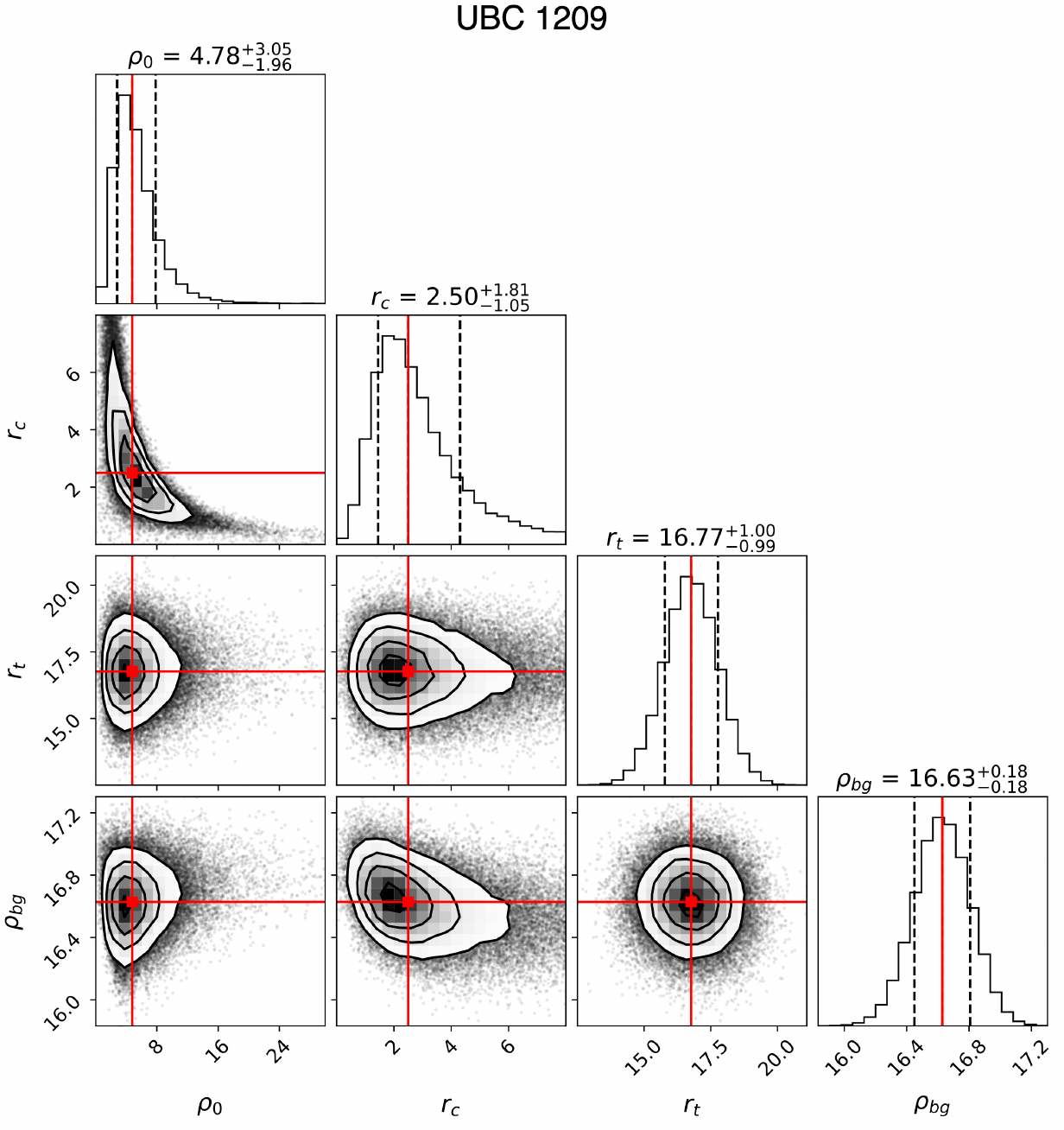}
    \includegraphics[width=0.45\linewidth]{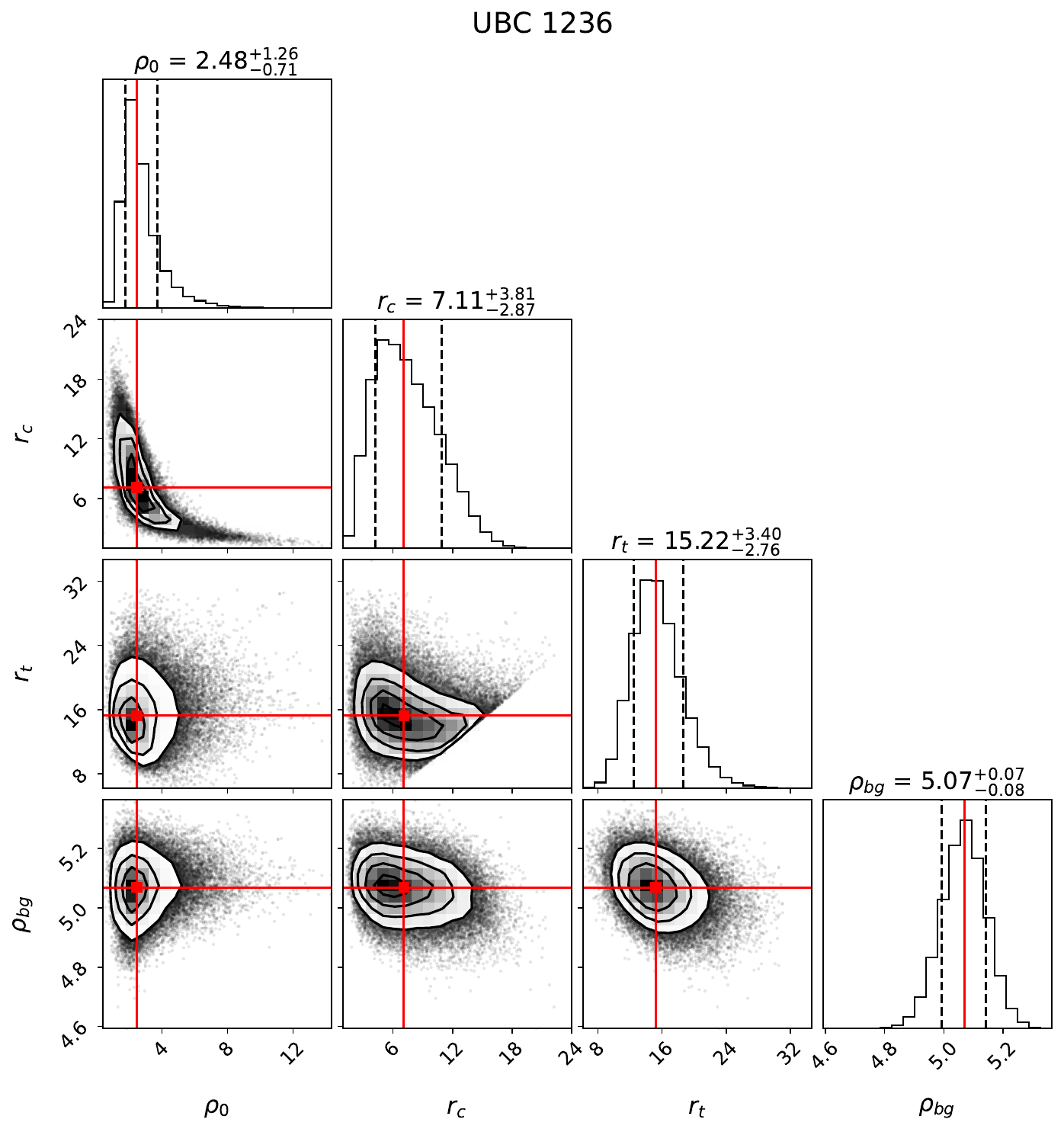}\\

    \caption{RDPs and the corner plots for the clusters UBC 1143, UBC 1185, UBC 1209, and UBC 1236.}
\end{figure}

\begin{figure}
    \centering
        \includegraphics[width=0.45\linewidth]{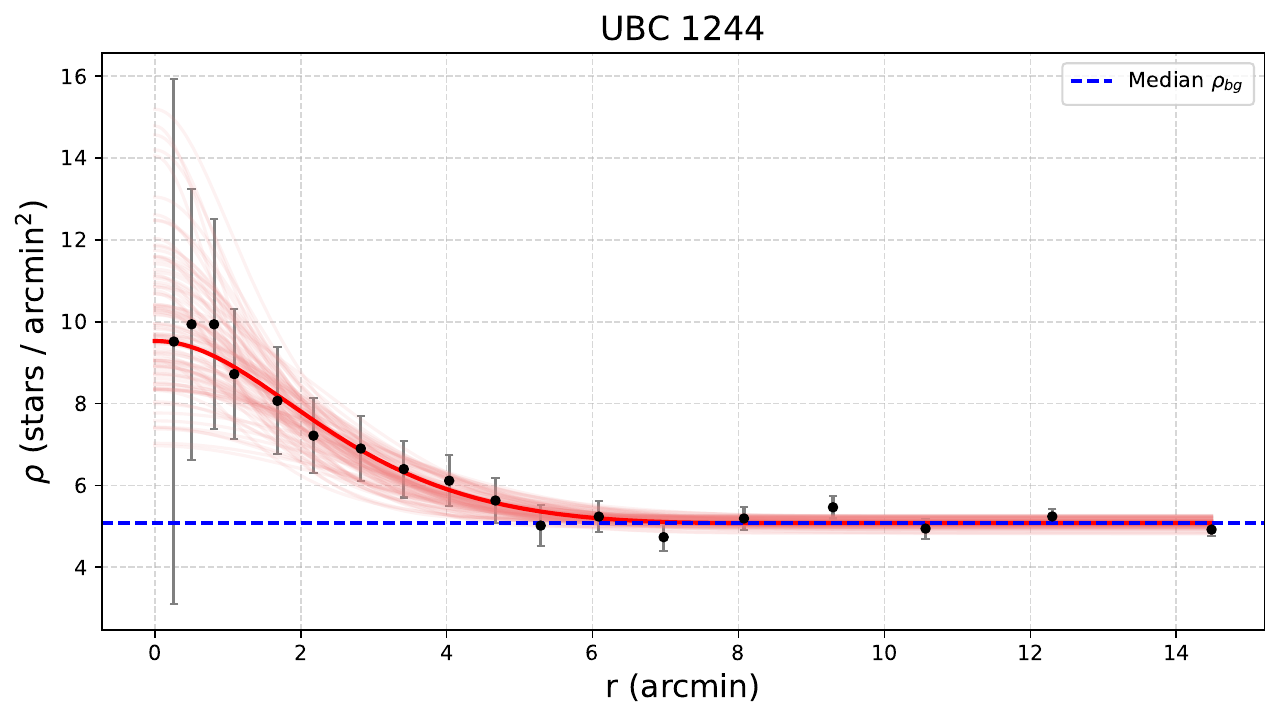}
    \includegraphics[width=0.45\linewidth]{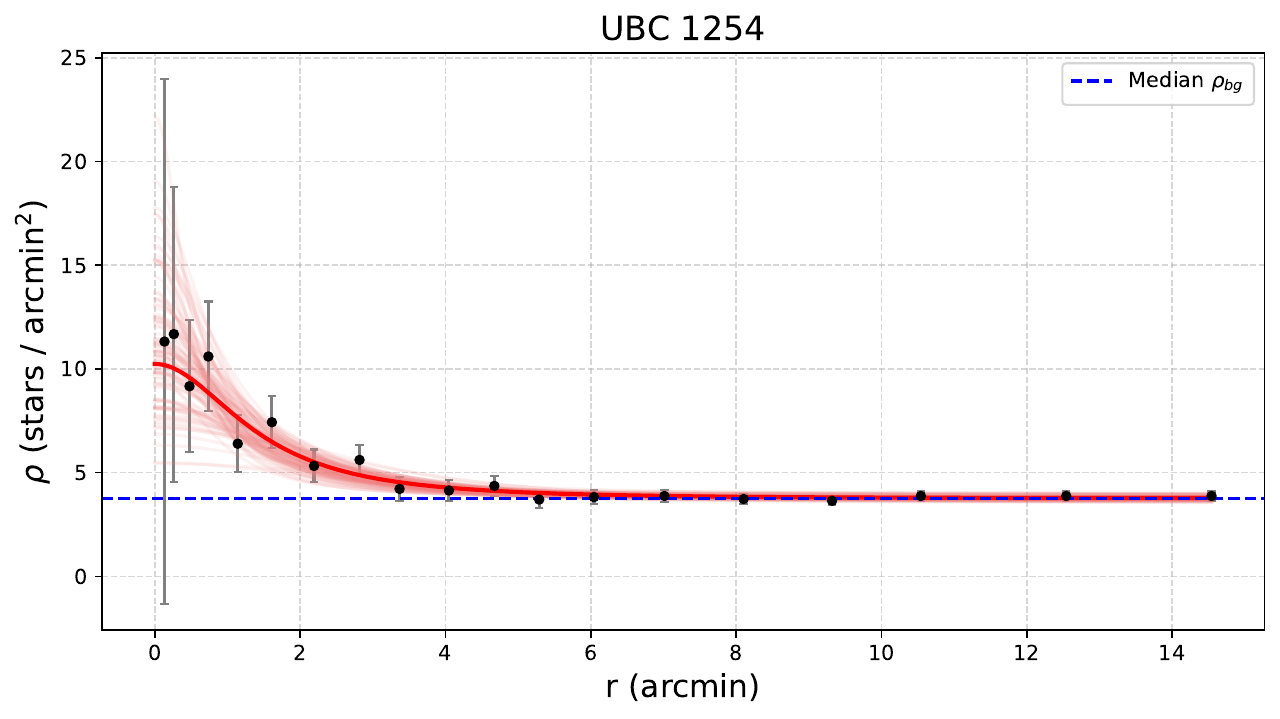}\\
    \includegraphics[width=0.45\linewidth]{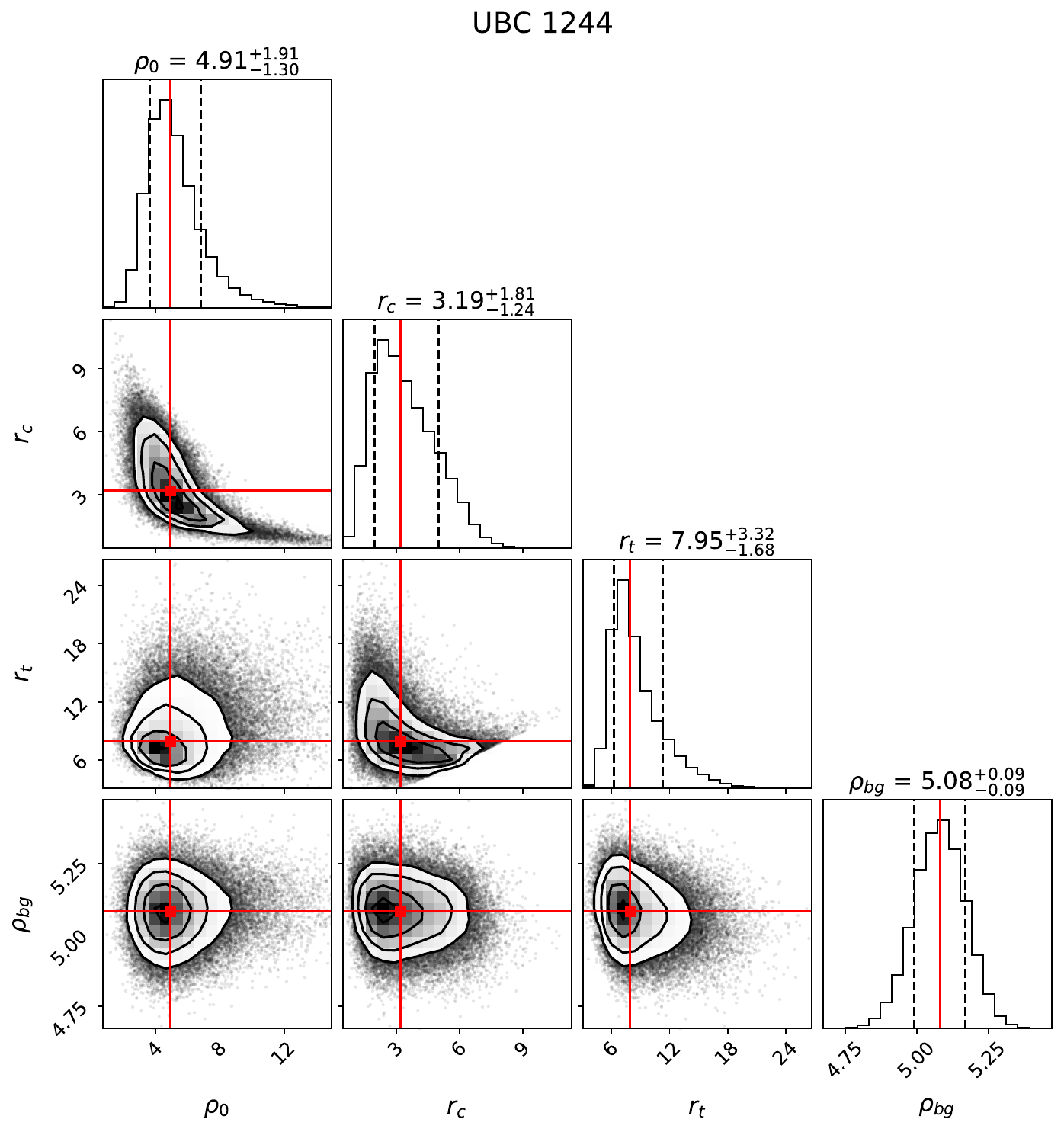}
    \includegraphics[width=0.45\linewidth]{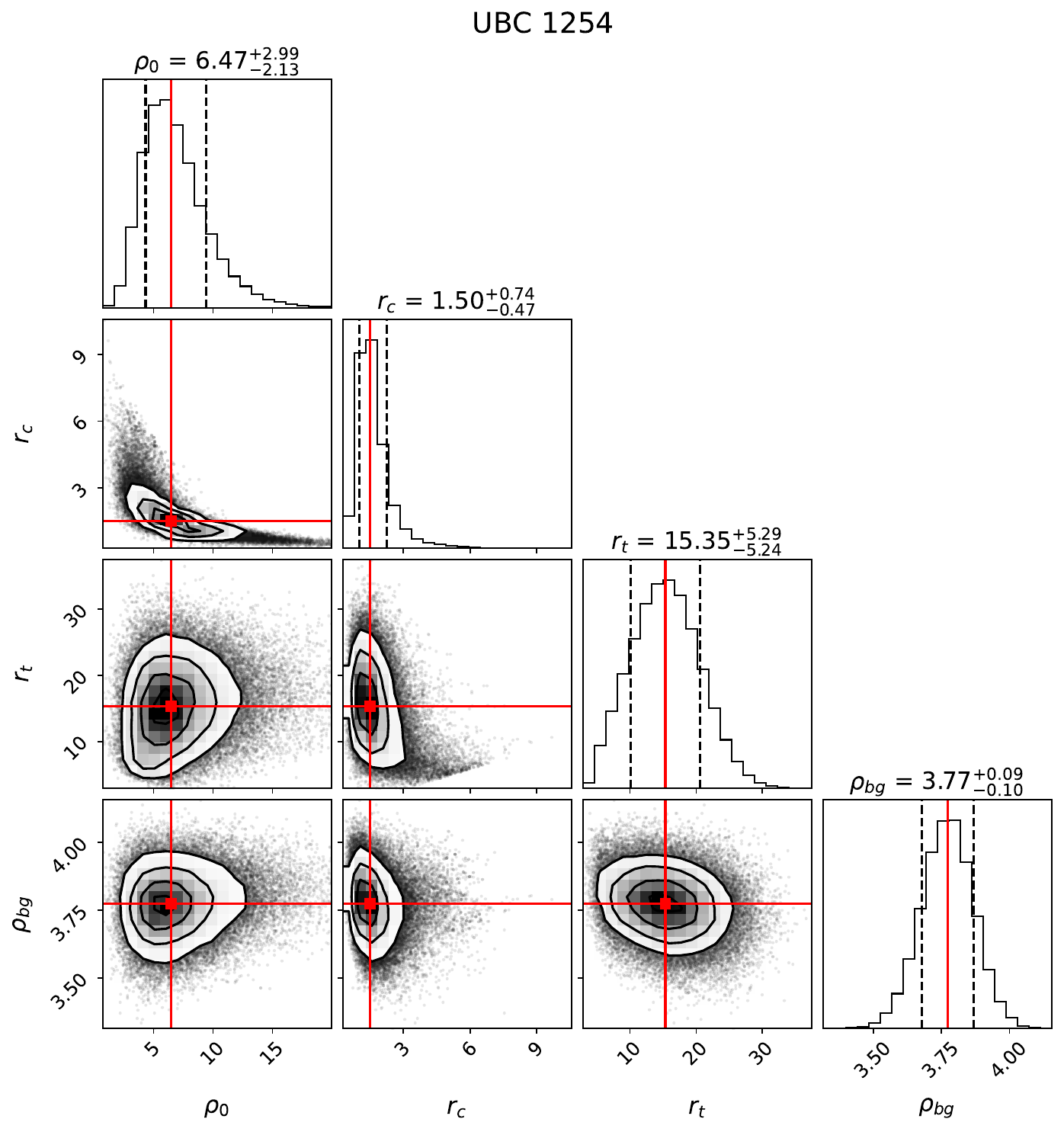}
    \includegraphics[width=0.45\linewidth]{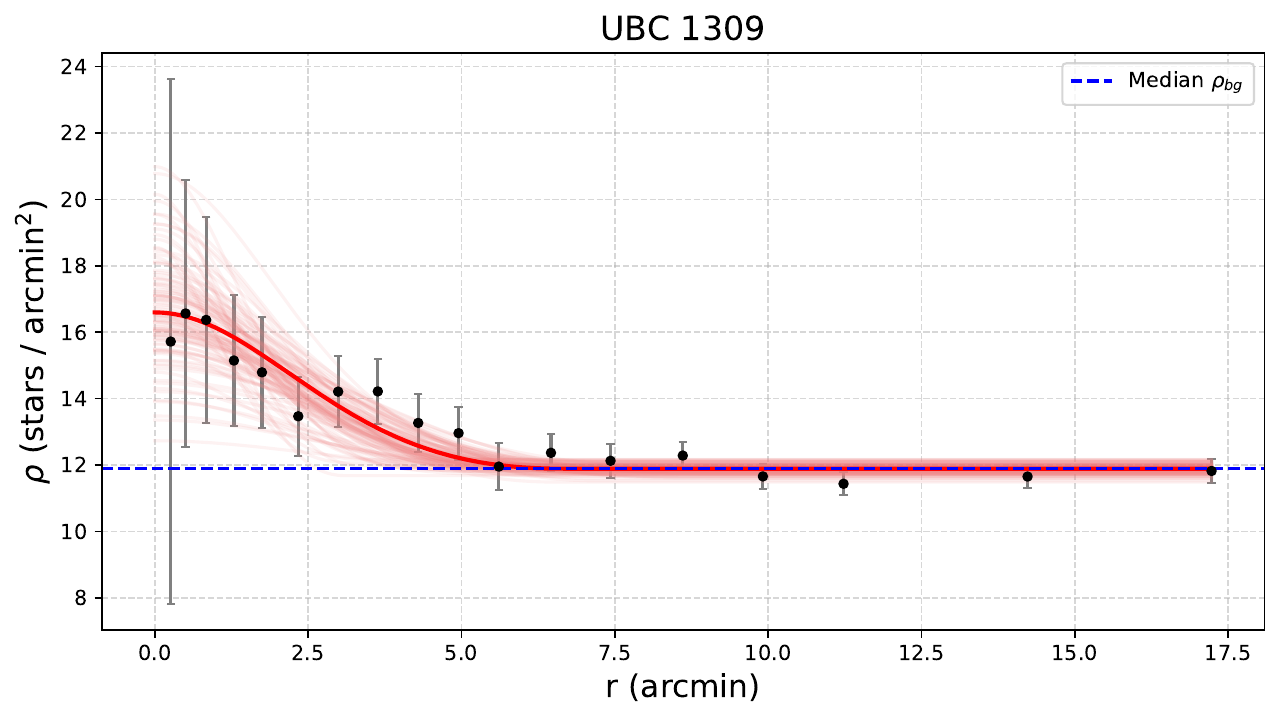}
    \includegraphics[width=0.45\linewidth]{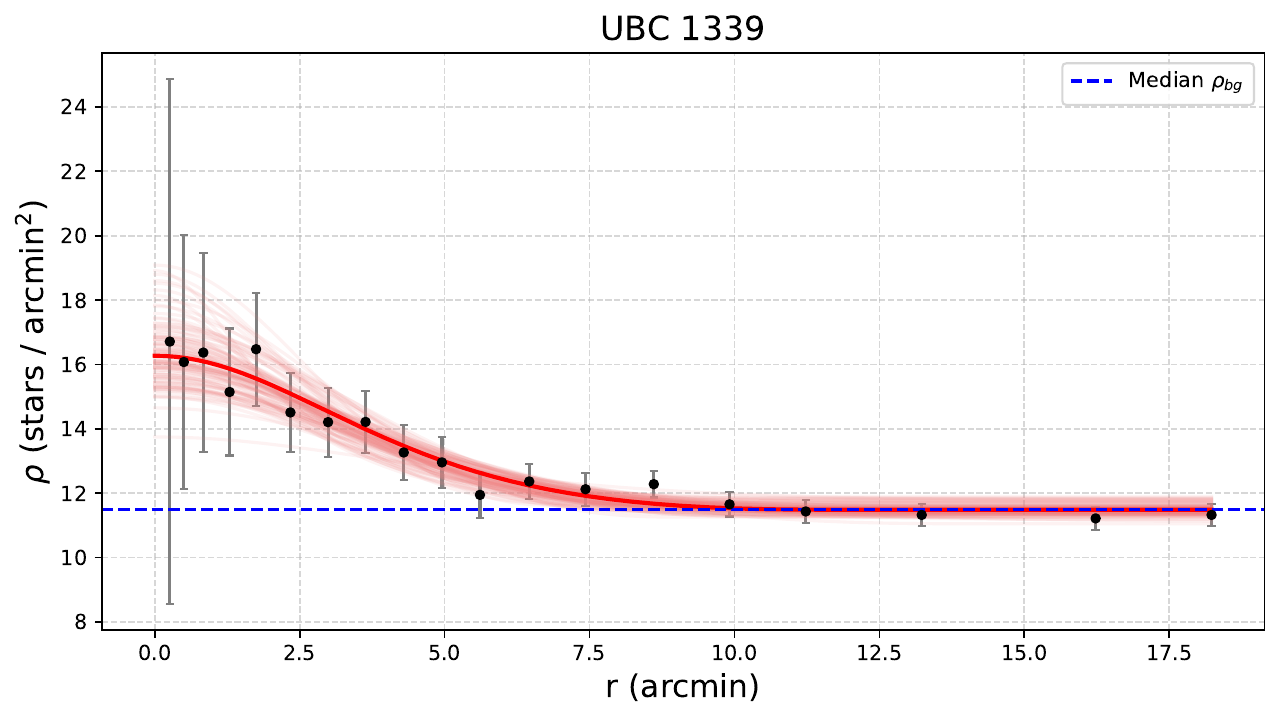}\\
    \includegraphics[width=0.45\linewidth]{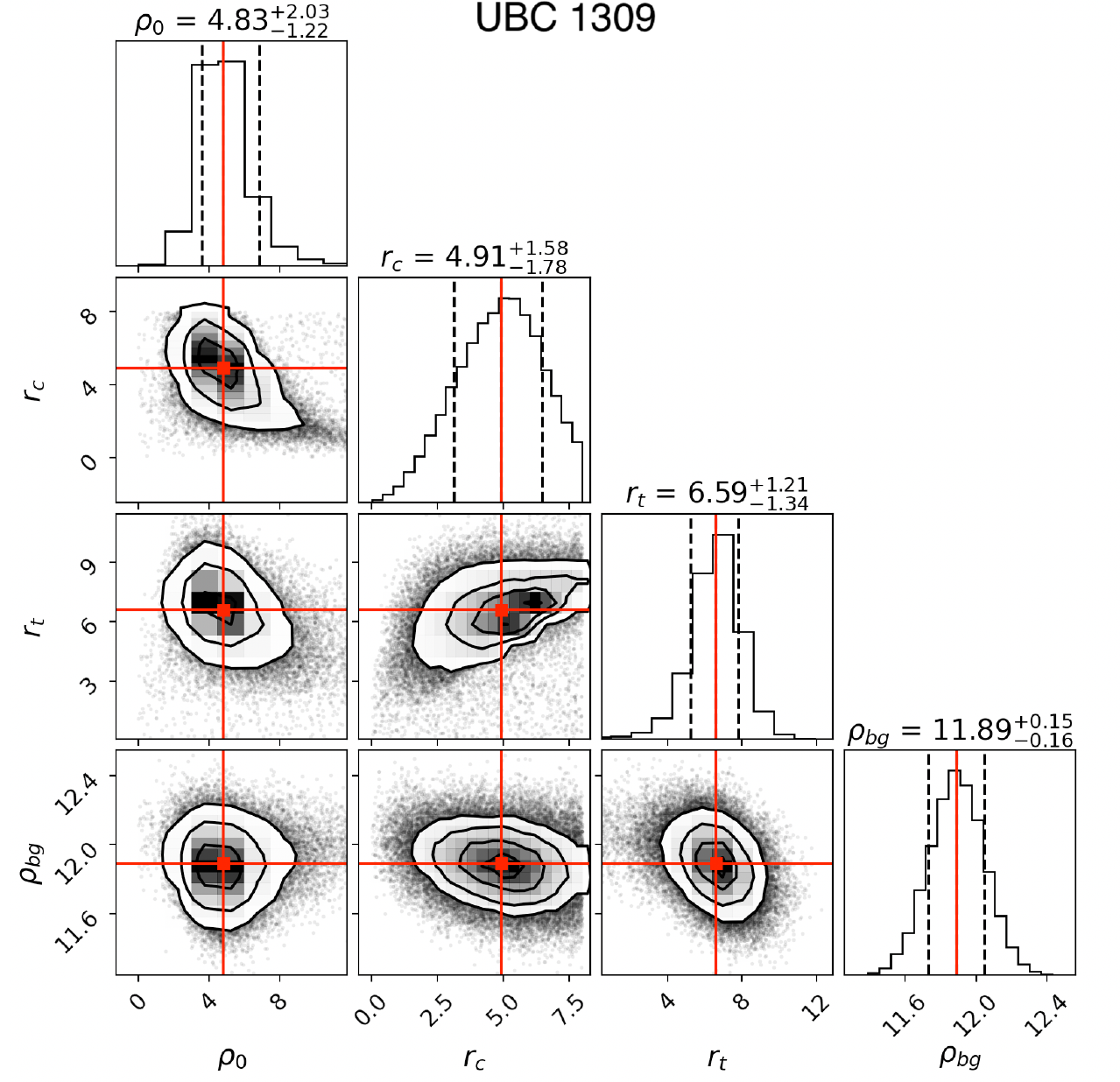}
    \includegraphics[width=0.45\linewidth]{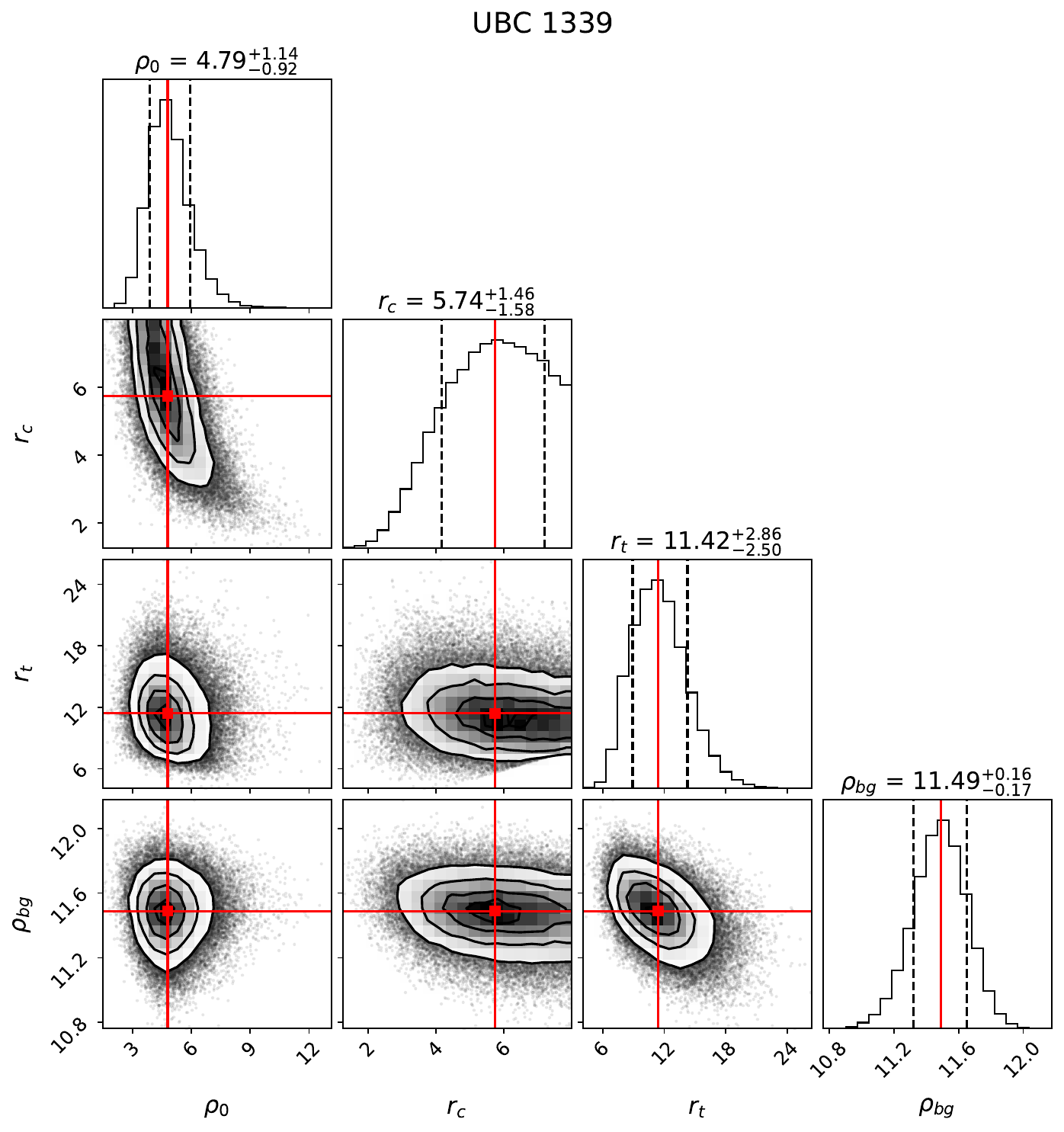}
    \caption{RDPs and the corner plots for the clusters UBC 1244, UBC 1254, UBC 1309, and UBC 1339.}
\end{figure}

\section{VPDs}\label{fig:vpss-append}

\begin{figure*}
\centering
\includegraphics[width=0.38\linewidth]{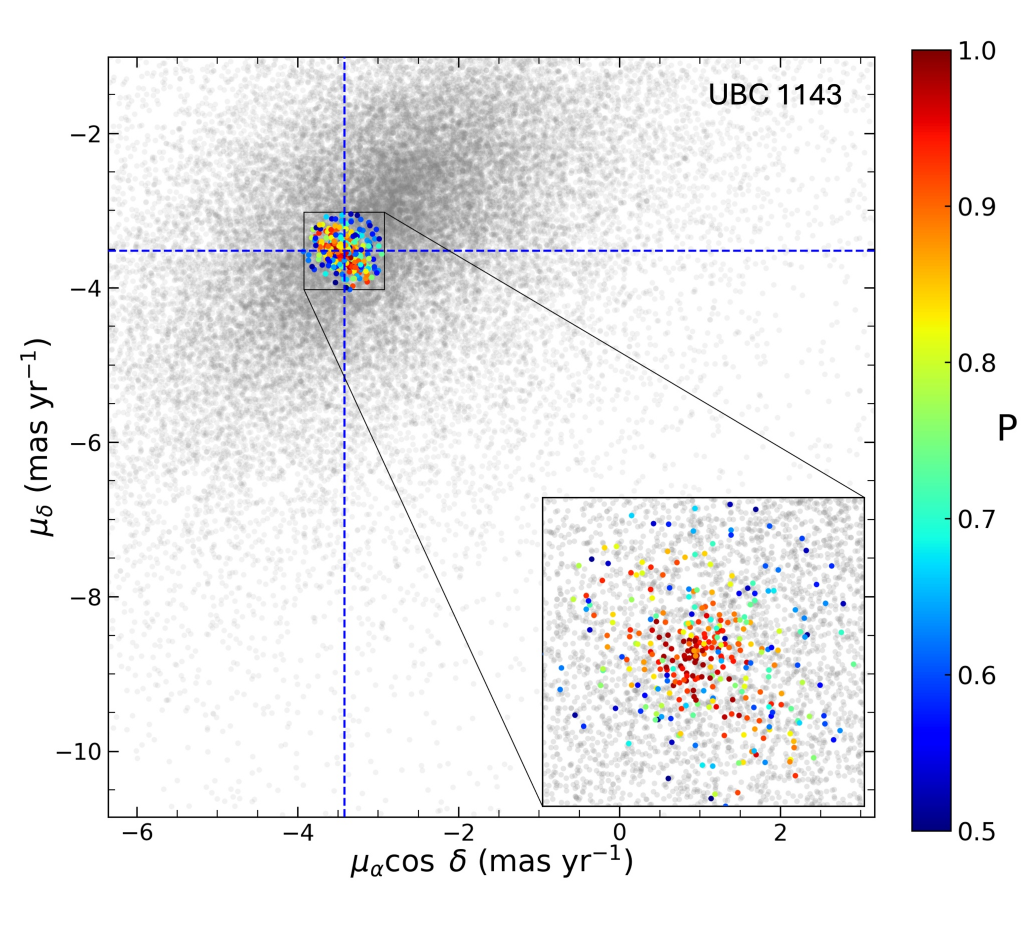}
\includegraphics[width=0.38\linewidth]{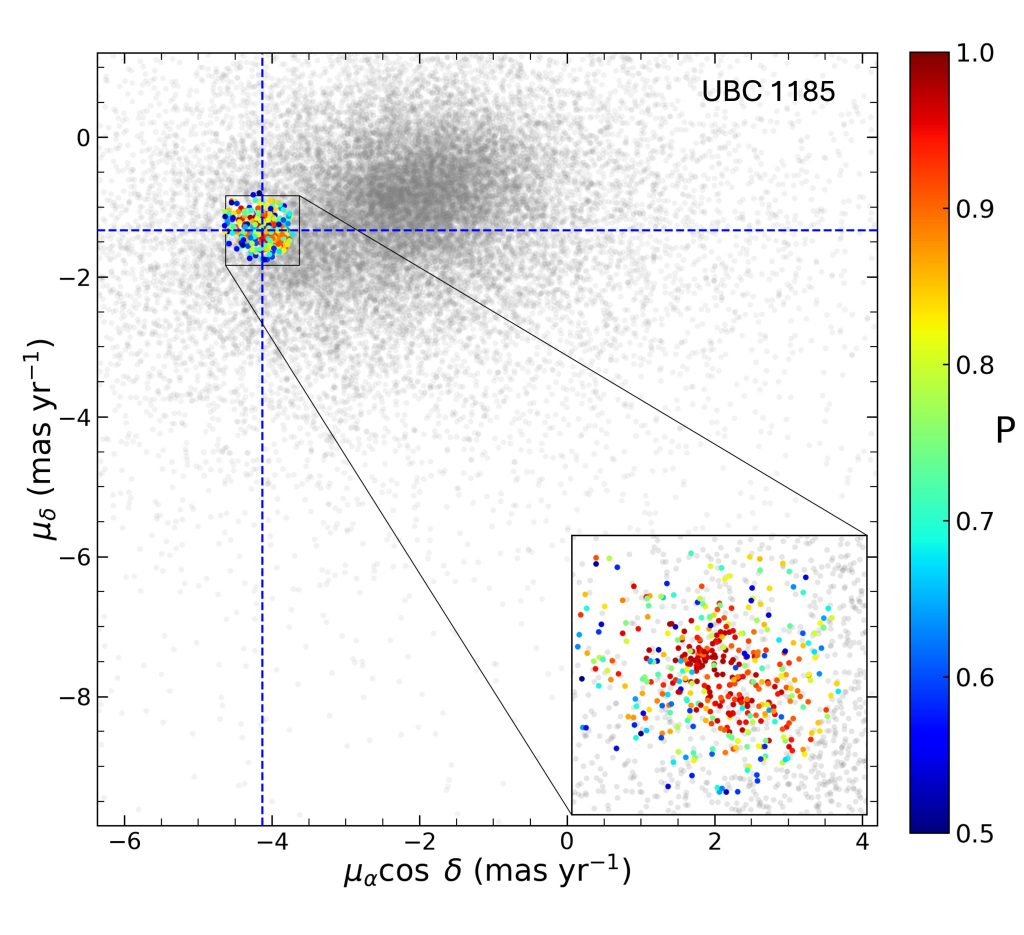}\\
\includegraphics[width=0.38\linewidth]{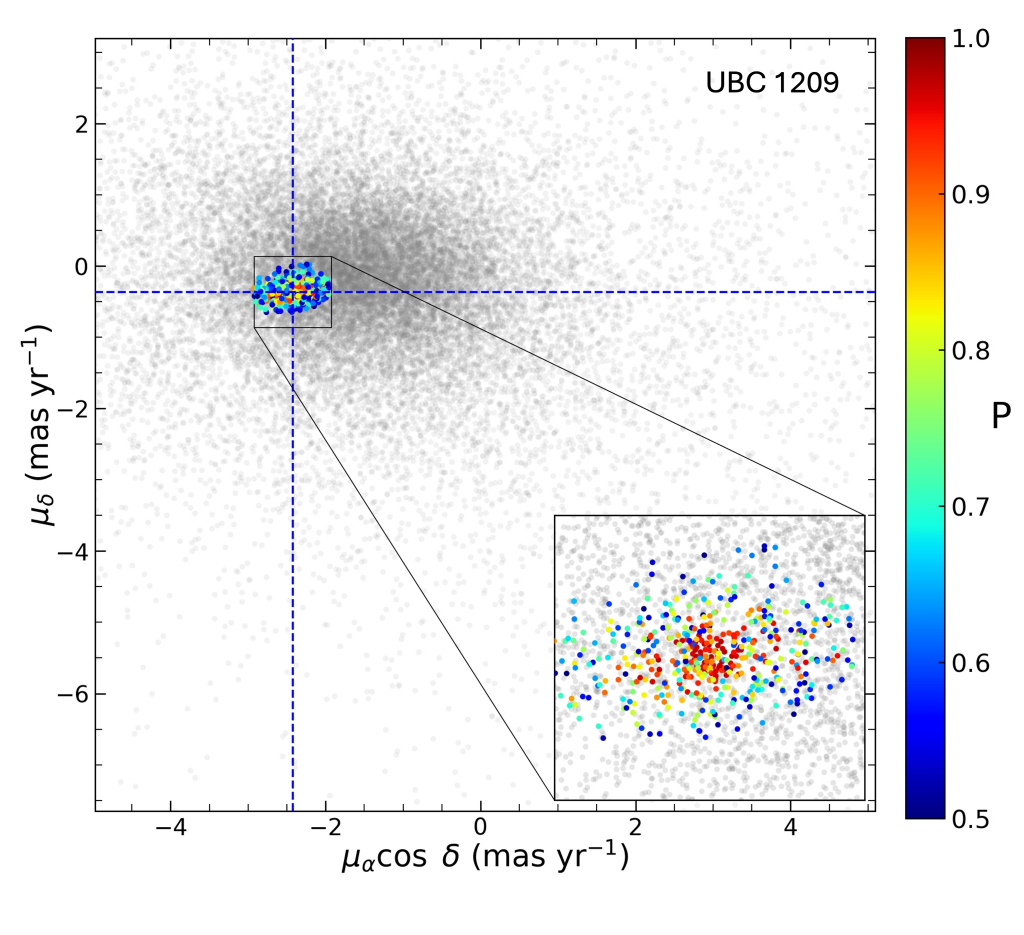}
\includegraphics[width=0.38\linewidth]{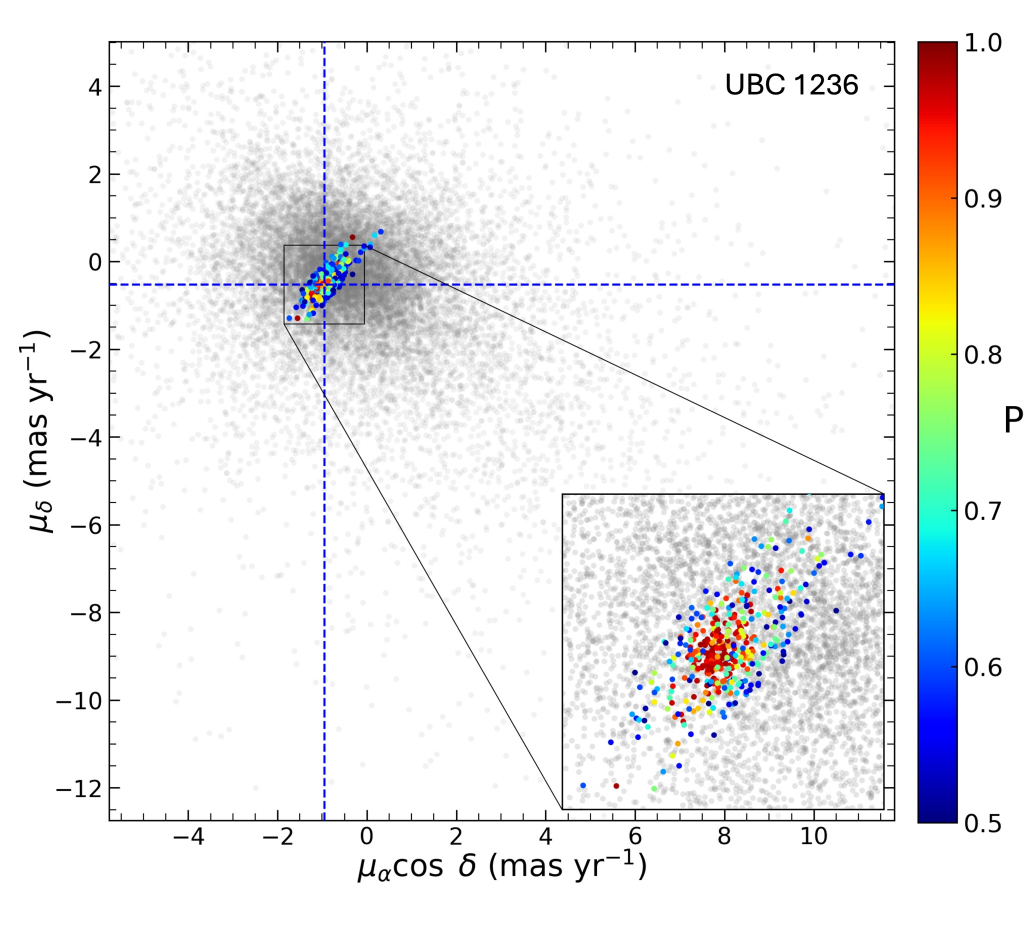}\\
\includegraphics[width=0.38\linewidth]{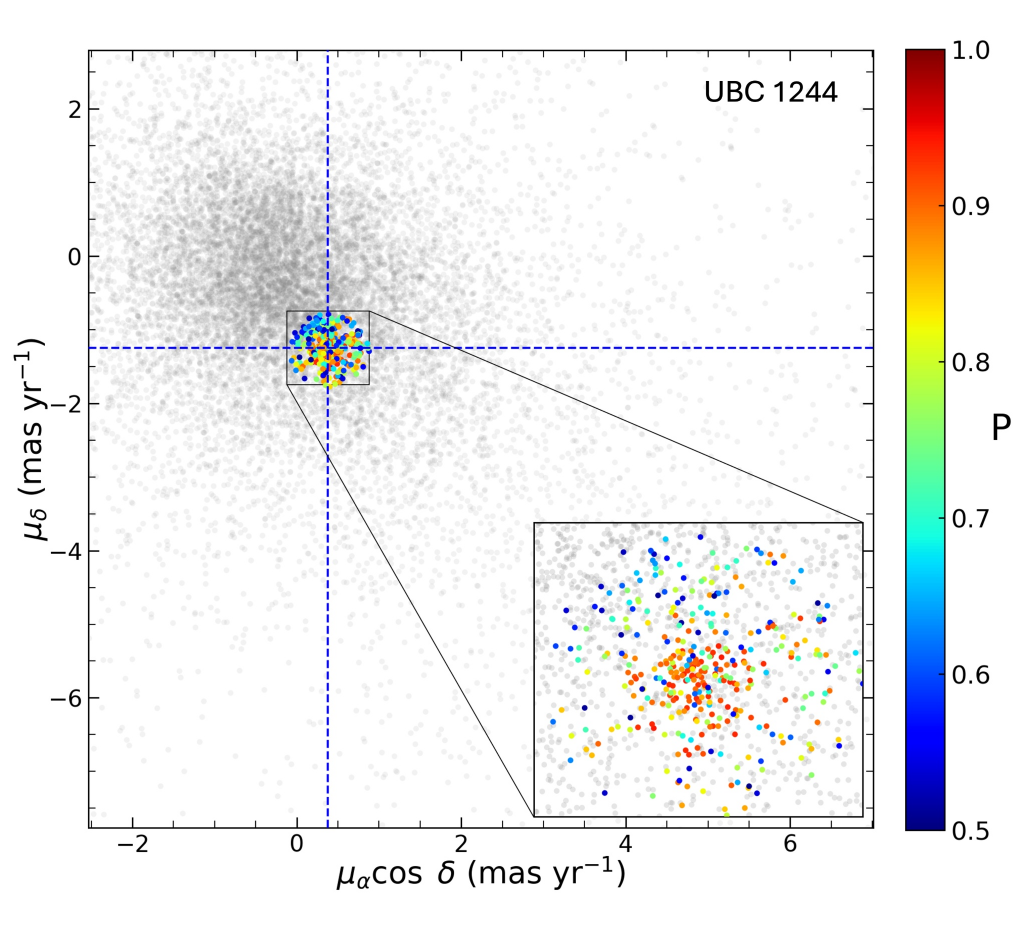}
\includegraphics[width=0.38\linewidth]{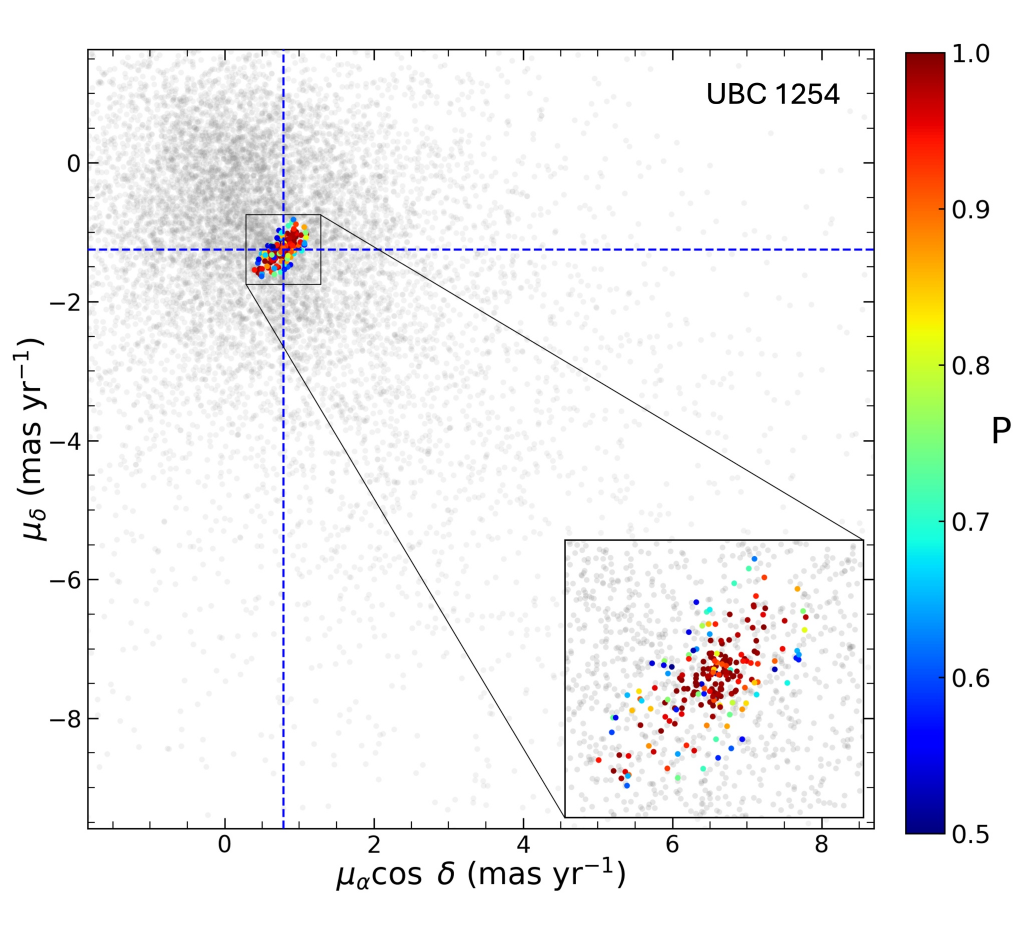}\\
\includegraphics[width=0.38\linewidth]{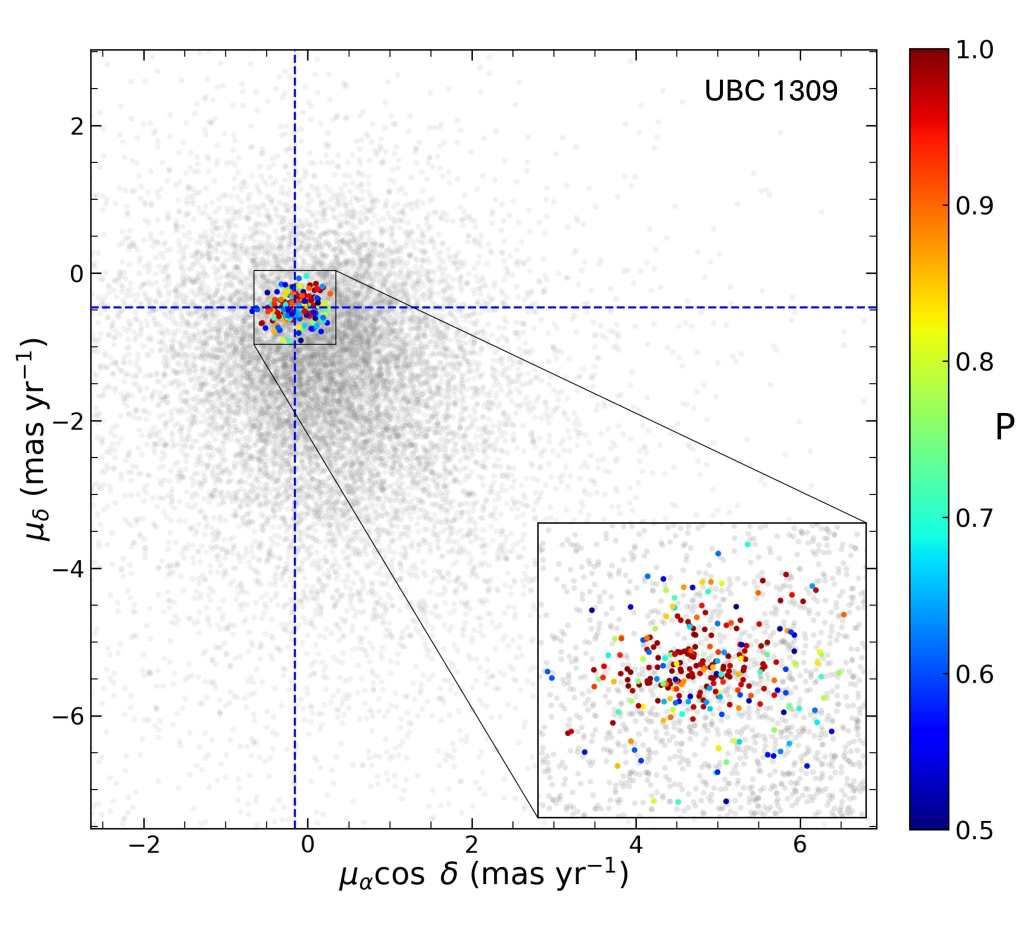}
\includegraphics[width=0.38\linewidth]{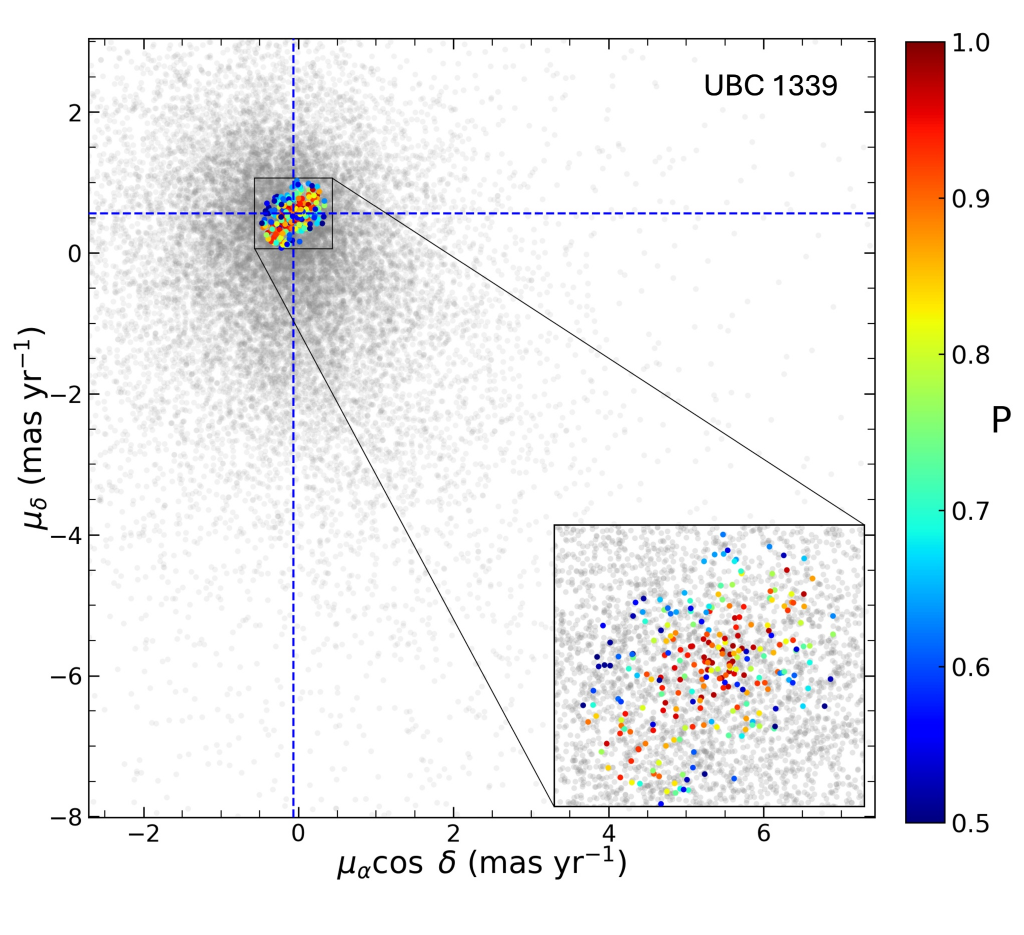}
\caption{Representative VPDs of the studied eight UBC OCs in this study.}
\label{fig:vpd_all} 
\end{figure*}

\section{CMD-MCMC}\label{fig:cmd-mcmc-append}

\begin{figure}
\centering
\includegraphics[width=0.35\linewidth]{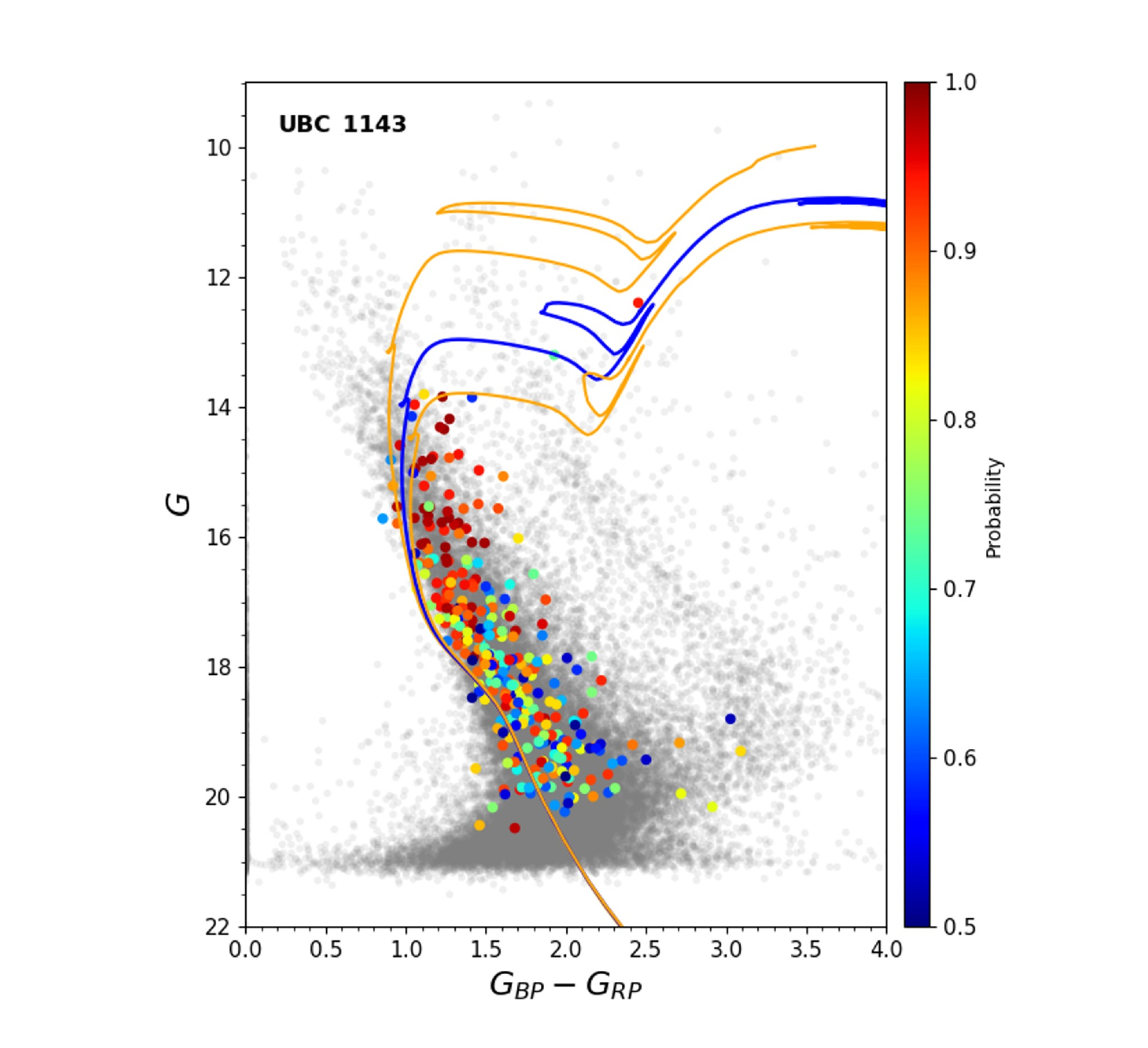}
\includegraphics[width=0.35\linewidth]{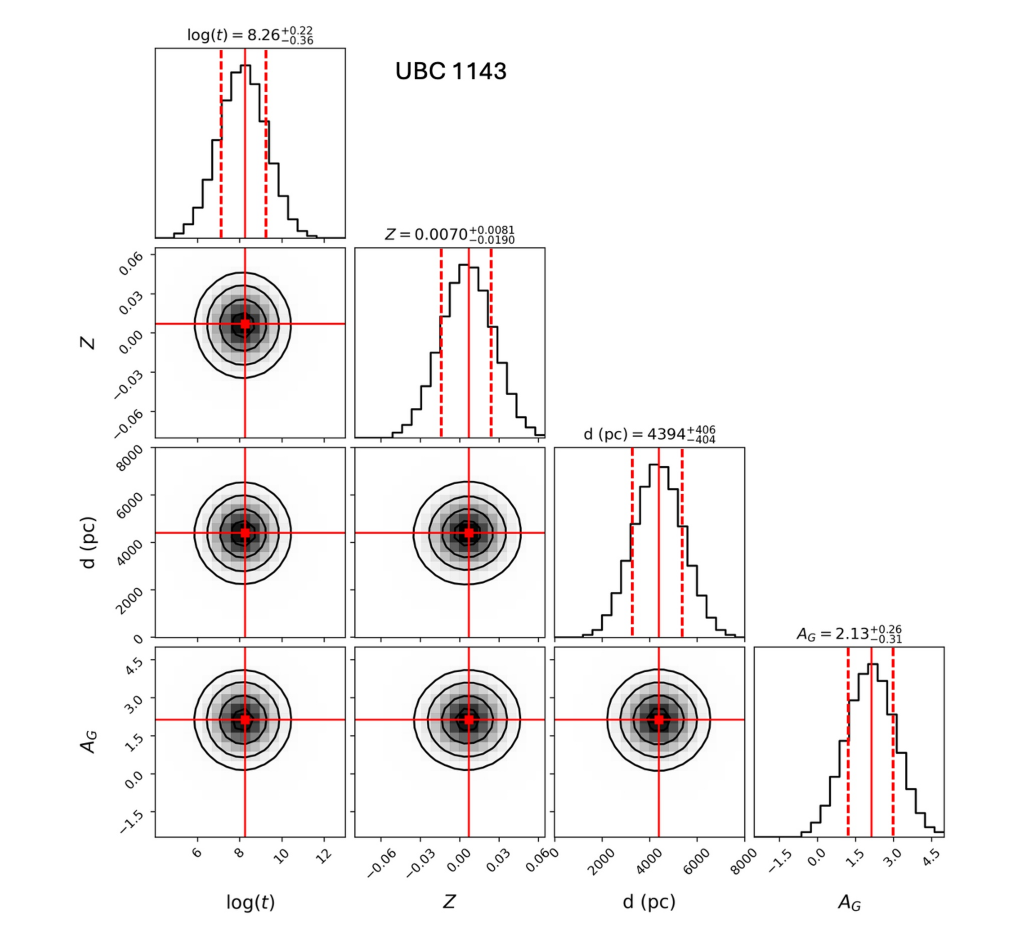}
\includegraphics[width=0.35\linewidth]{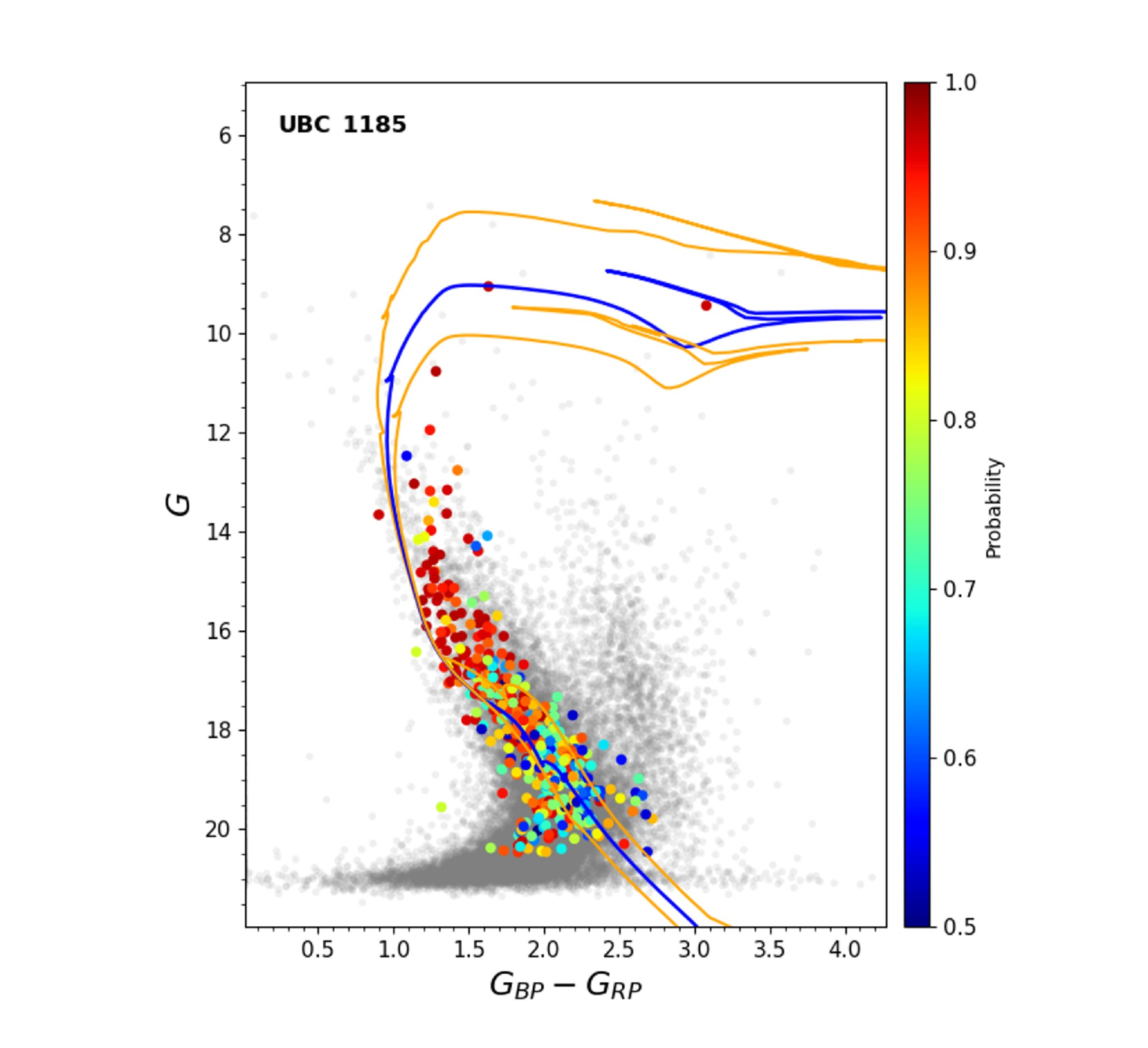}
\includegraphics[width=0.35\linewidth]{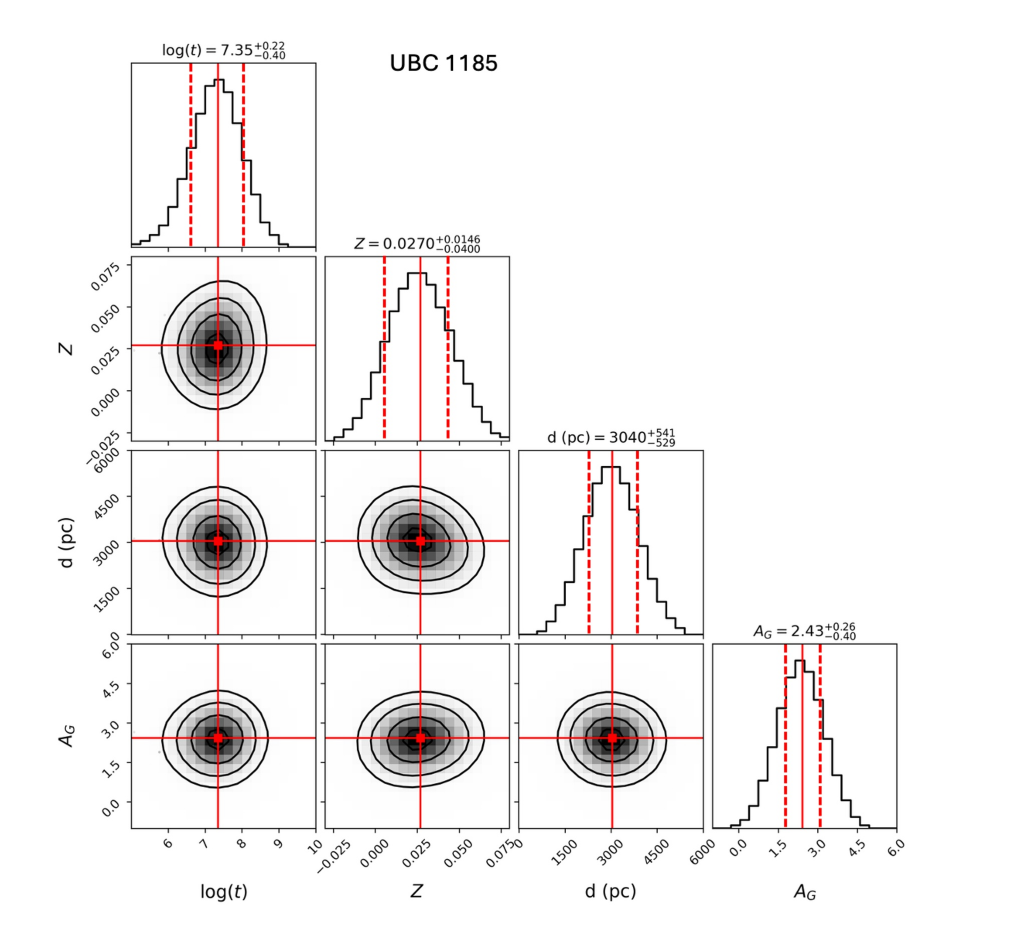}
\includegraphics[width=0.35\linewidth]{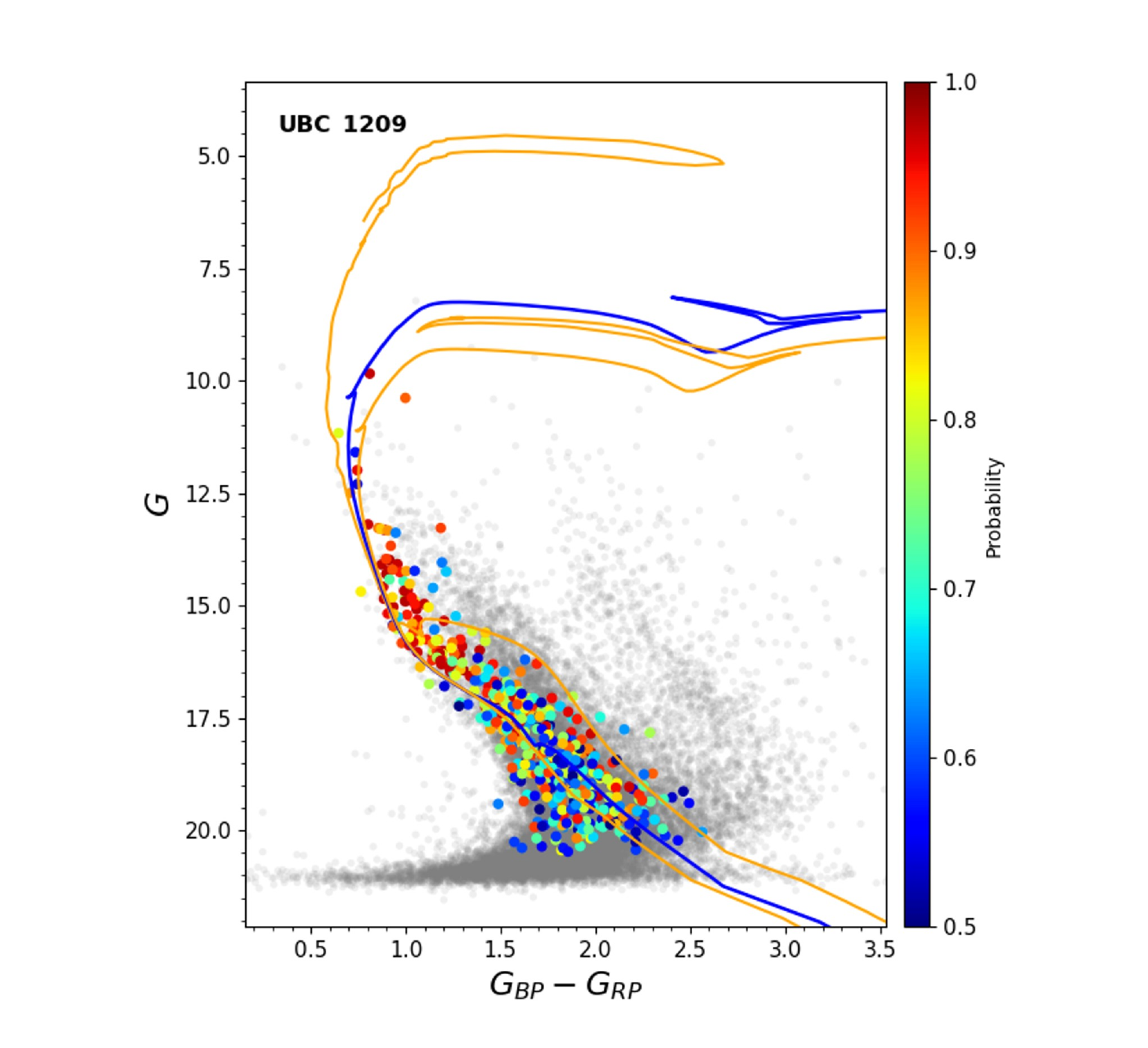}
\includegraphics[width=0.35\linewidth]{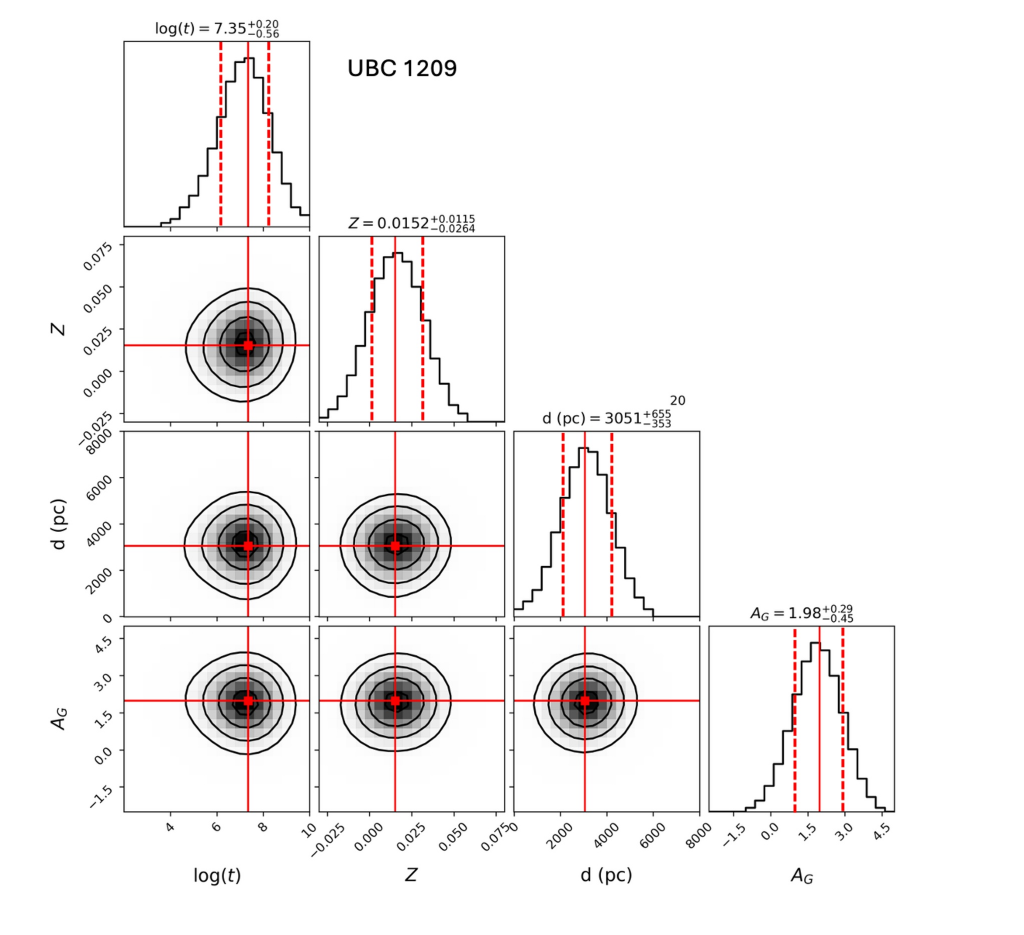}
\includegraphics[width=0.35\linewidth]{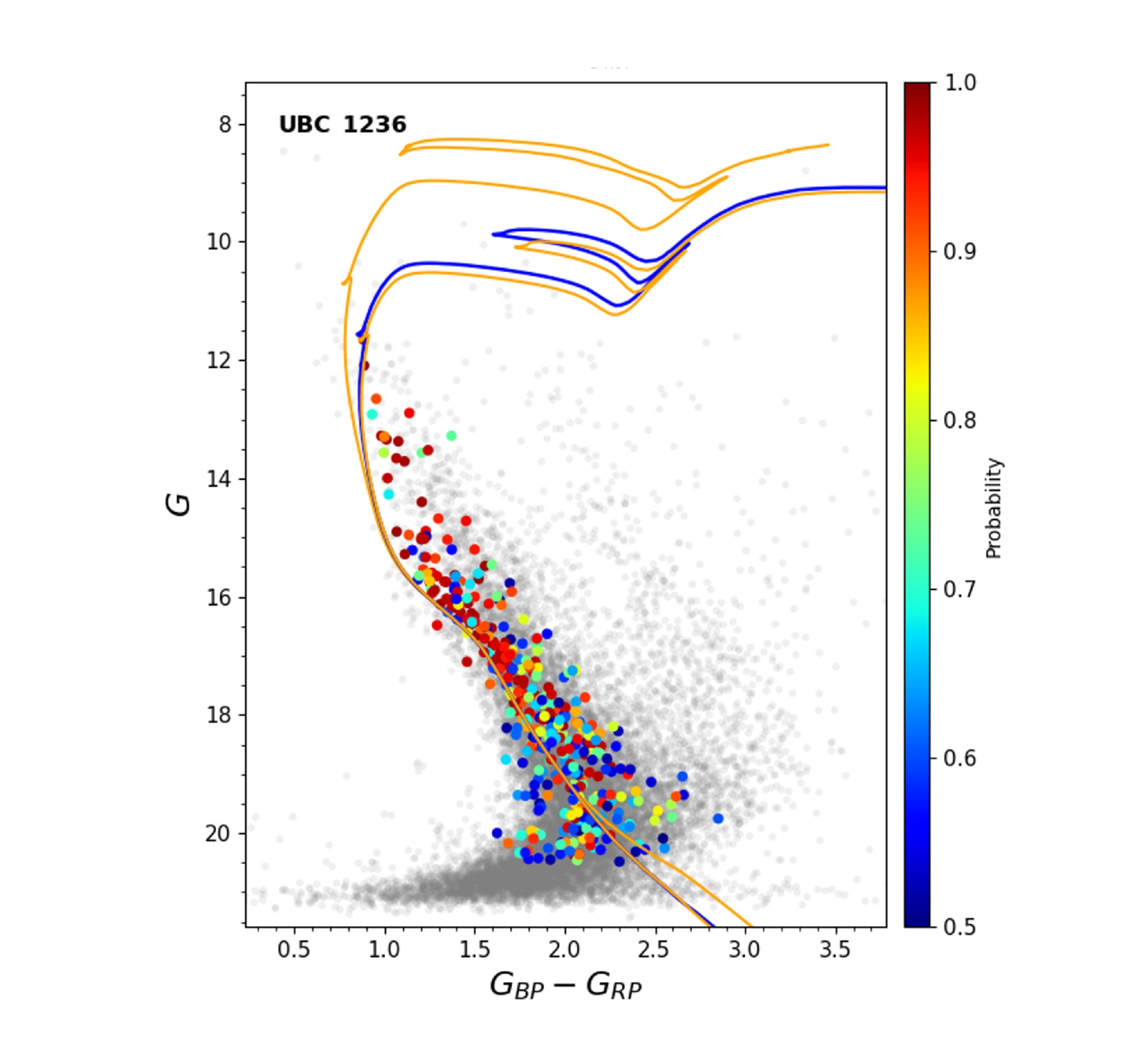}
\includegraphics[width=0.35\linewidth]{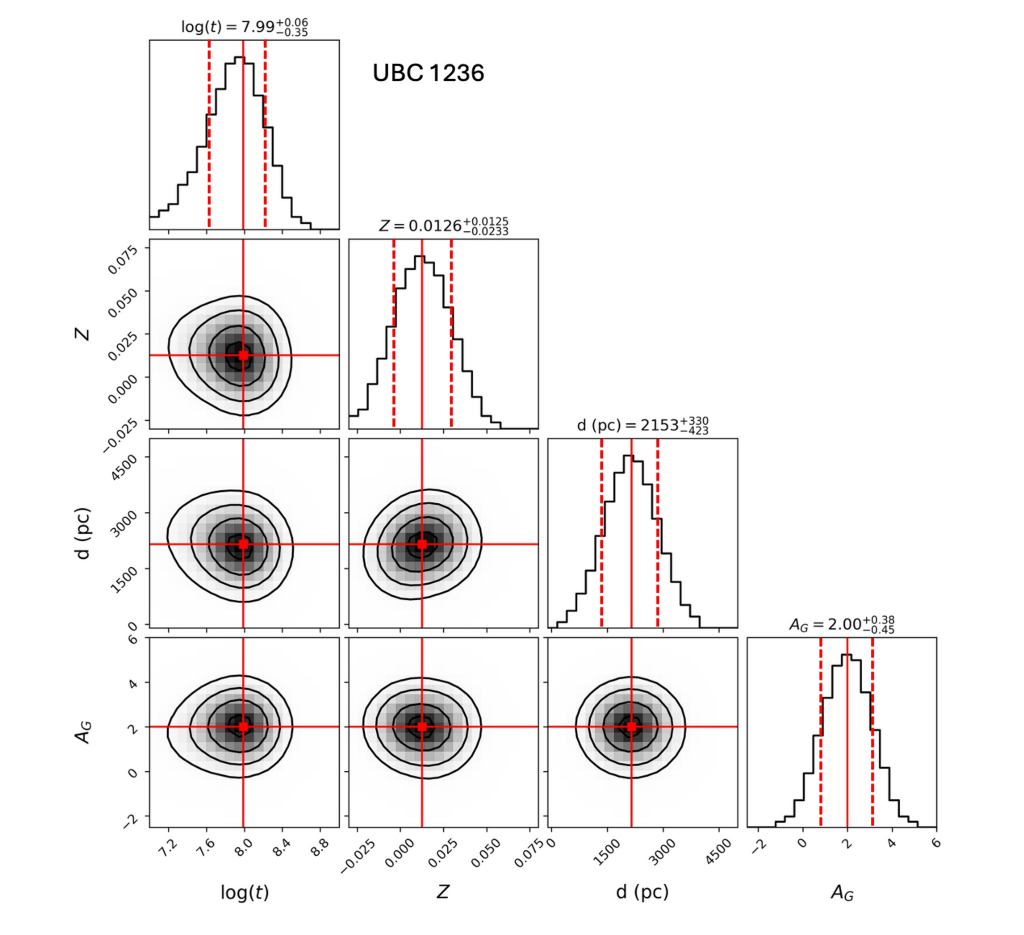}
\caption{Representative CMDs and MCMC corners of the studied OCs in this study.}
\end{figure}

\begin{figure}
    \centering
    
\includegraphics[width=0.35\linewidth]{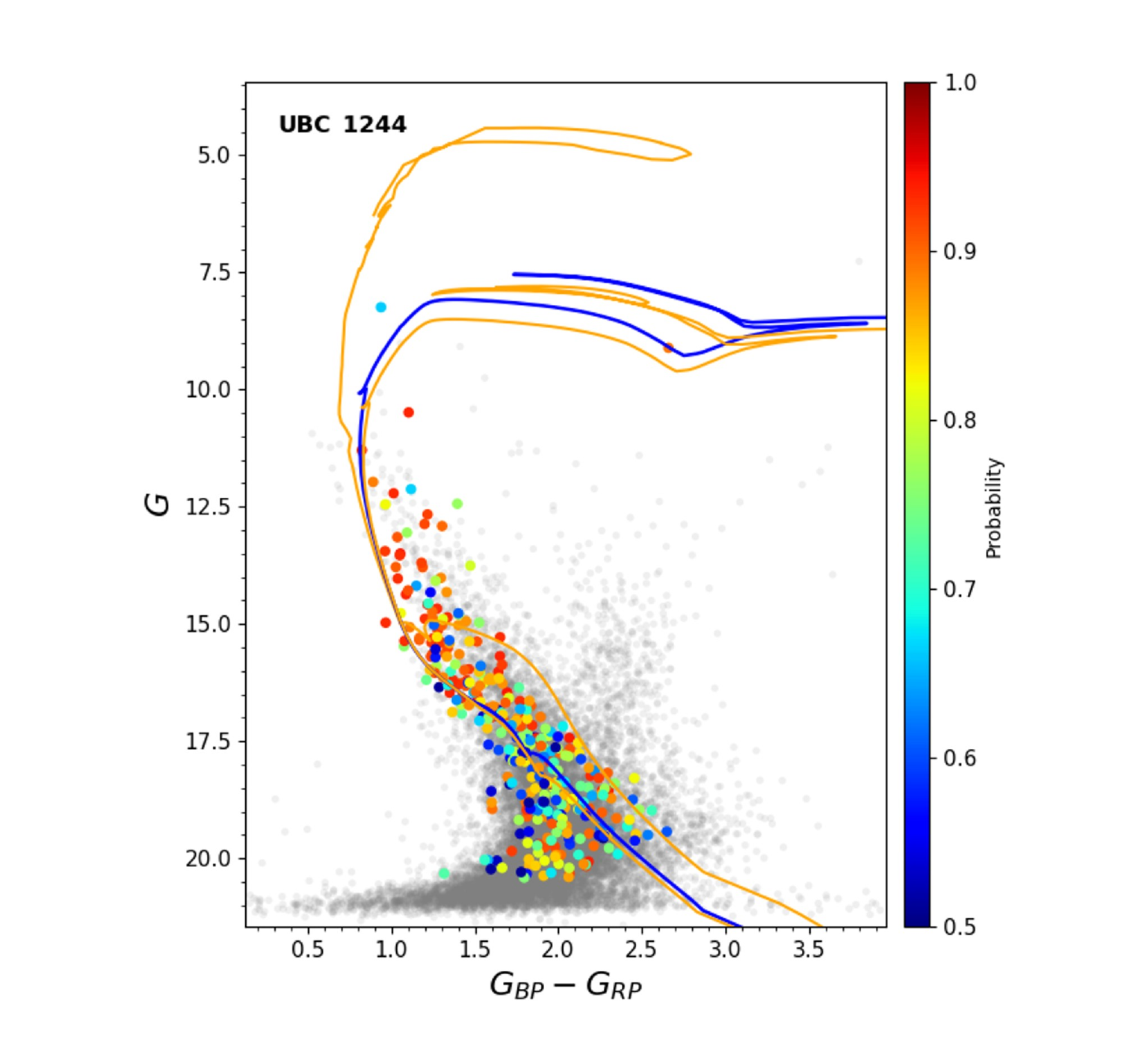}
\includegraphics[width=0.35\linewidth]{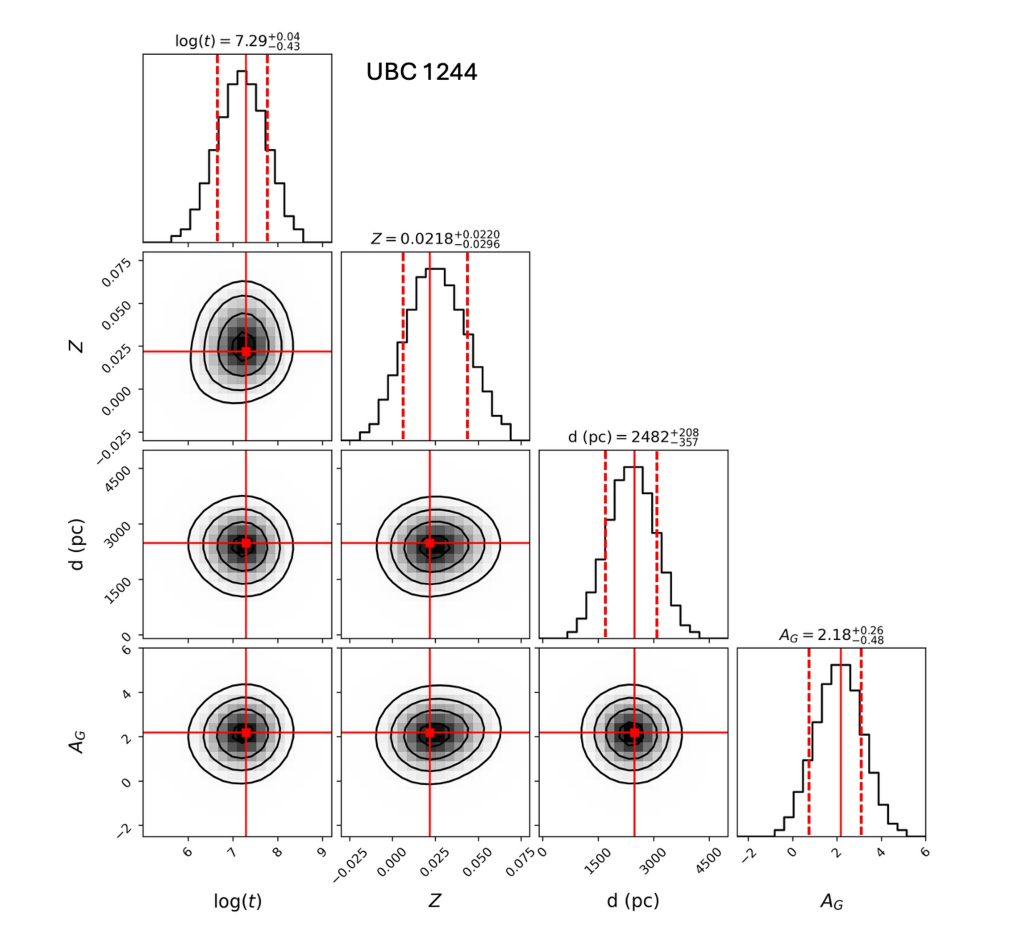}
\includegraphics[width=0.40\linewidth]{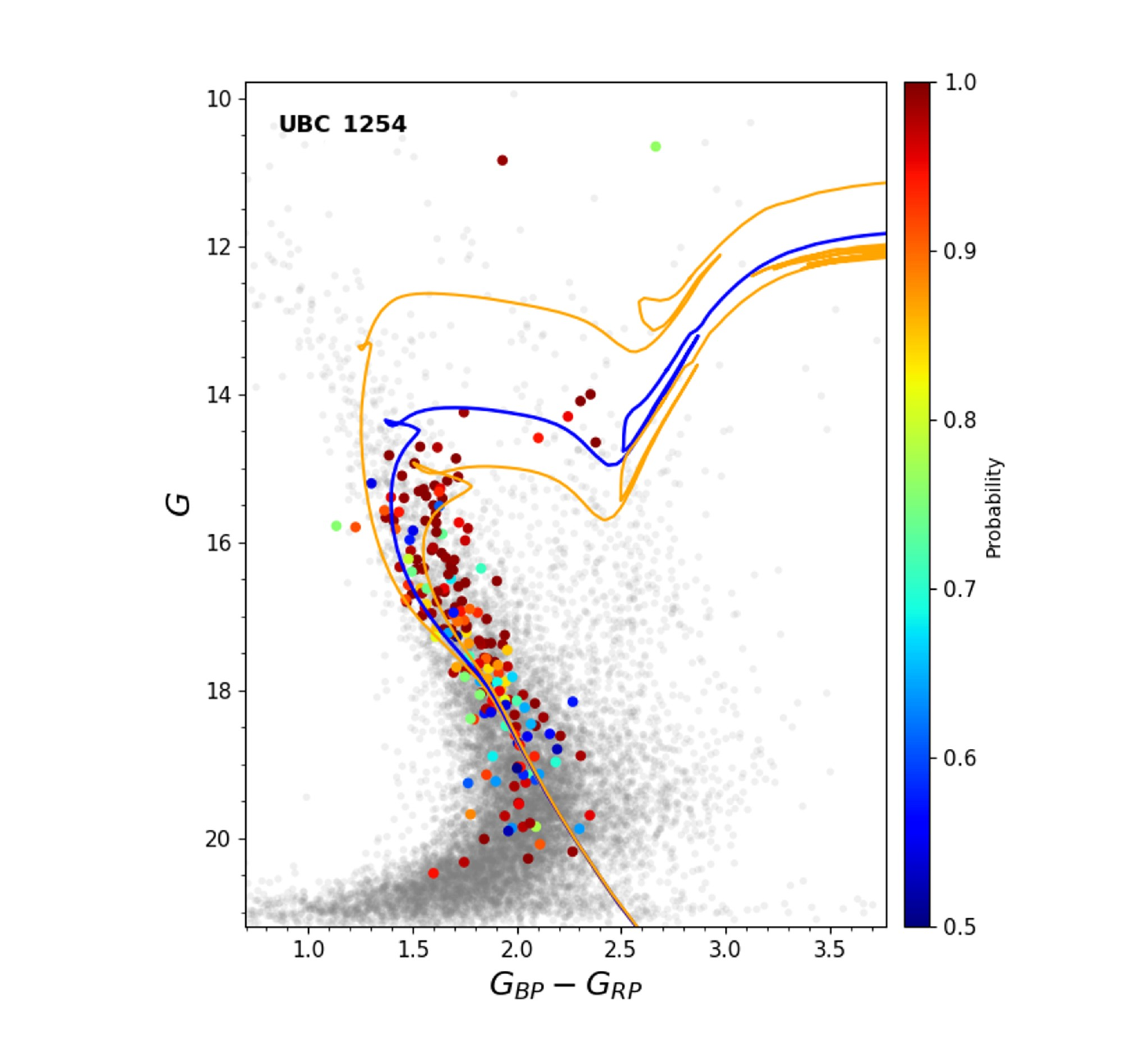}
\includegraphics[width=0.35\linewidth]{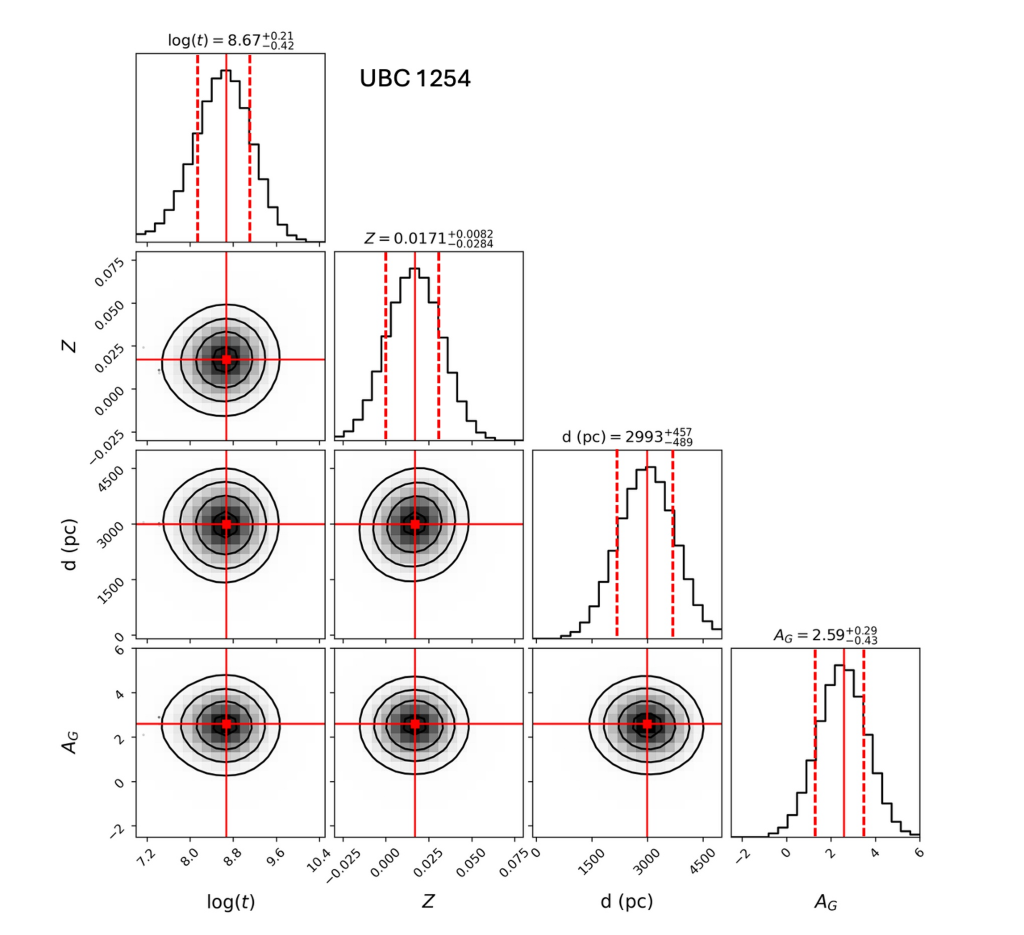}
\includegraphics[width=0.35\linewidth]{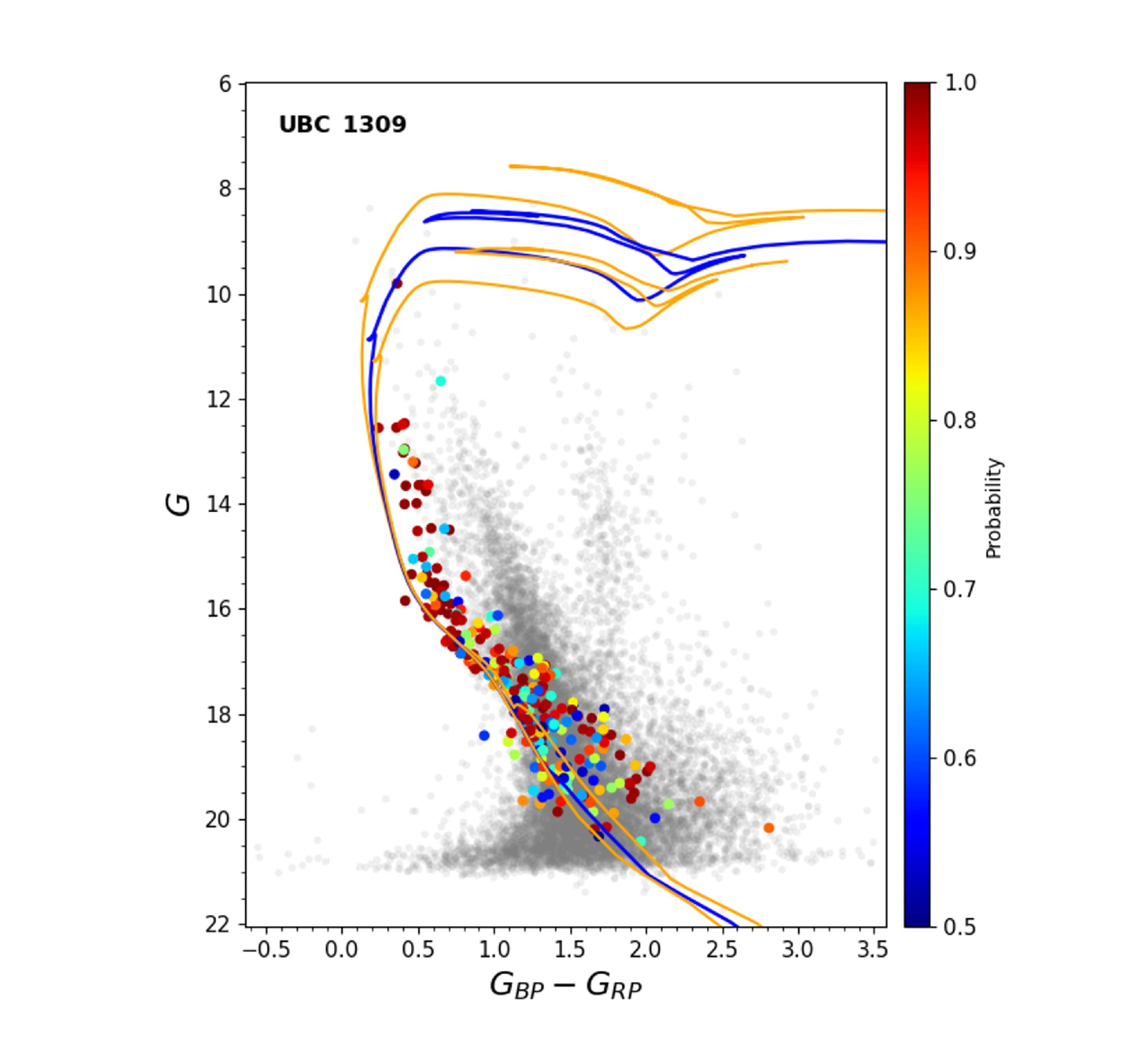}
\includegraphics[width=0.35\linewidth]{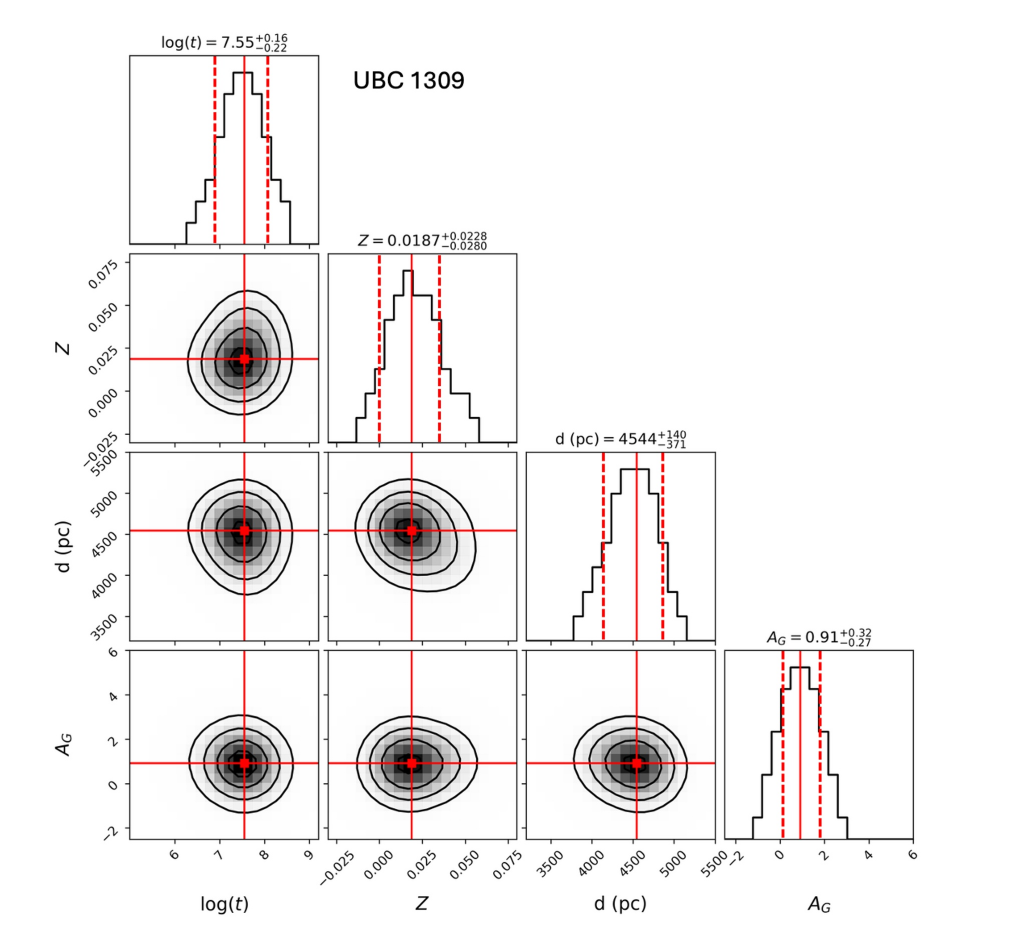}
\includegraphics[width=0.35\linewidth]{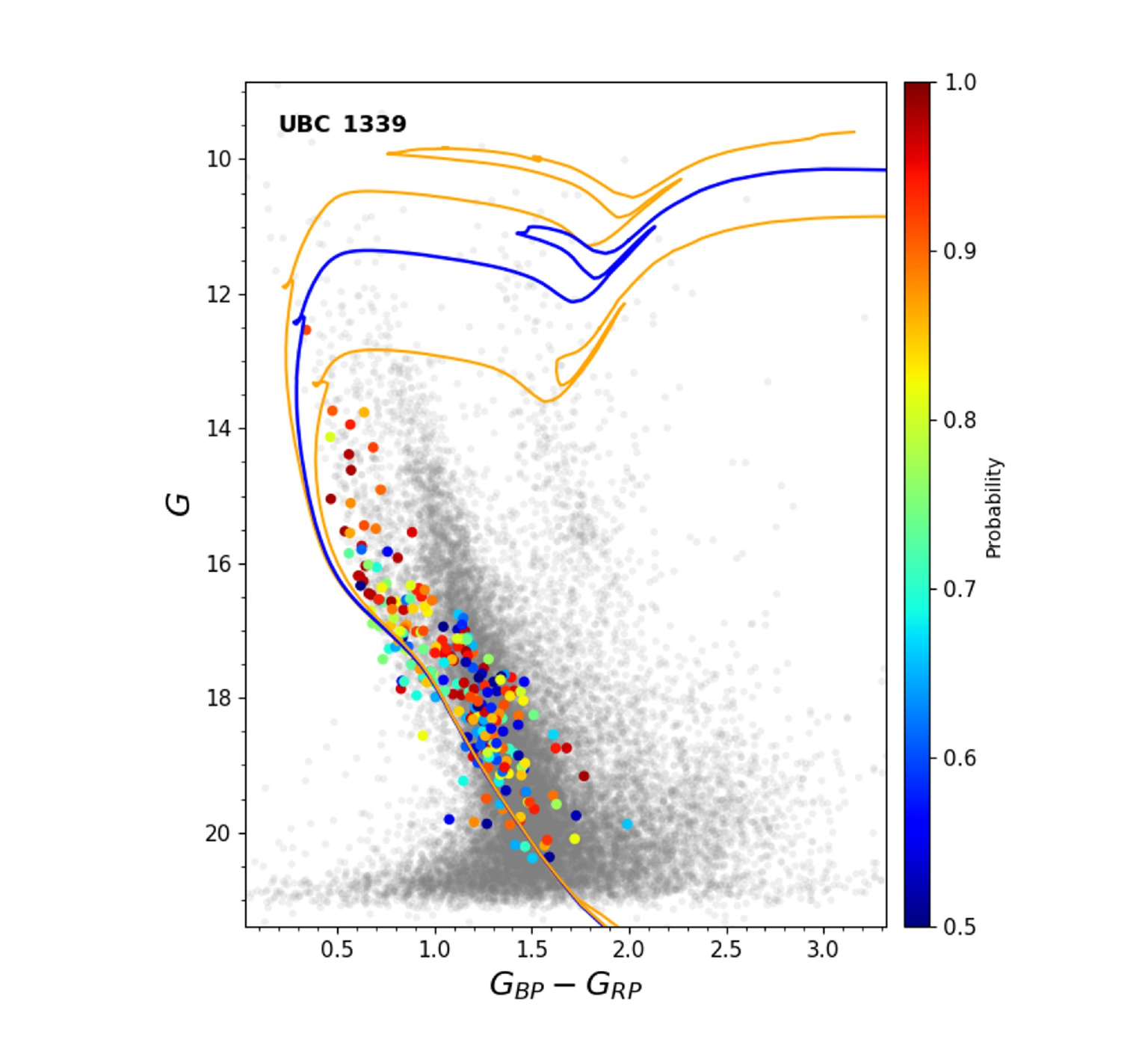}
\includegraphics[width=0.35\linewidth]{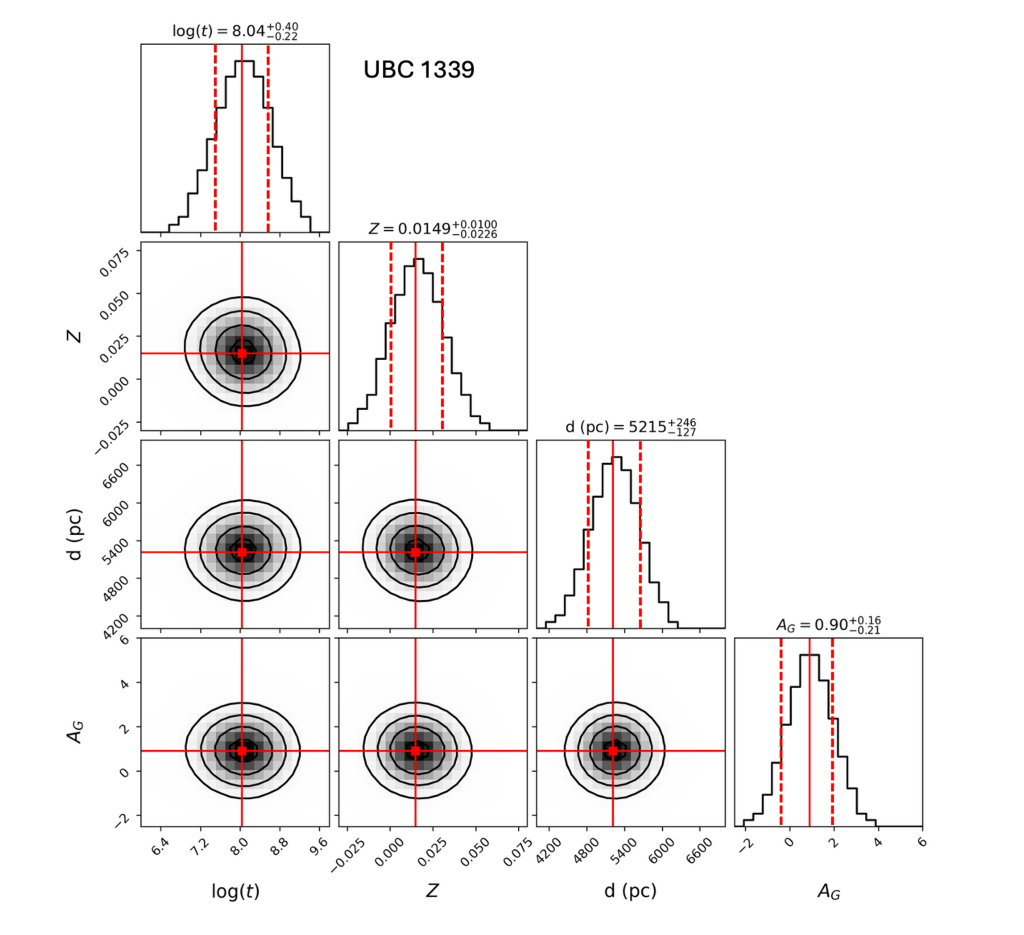}

\caption{Representative CMDs and MCMC corners of the studied OCs in this study.}
\end{figure}

\section{Orbit err} \label{fig:orbit-append-err}

\begin{figure*}[htbp]
\centering
\begin{minipage}{0.495\textwidth}
\centering
\includegraphics[width=\linewidth]{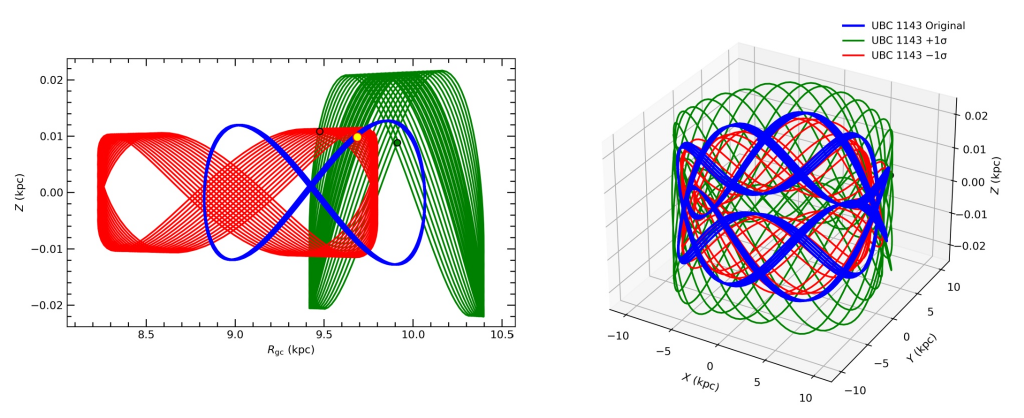}
\end{minipage}
\hfill
\begin{minipage}{0.495\textwidth}
\centering
\includegraphics[width=\linewidth]{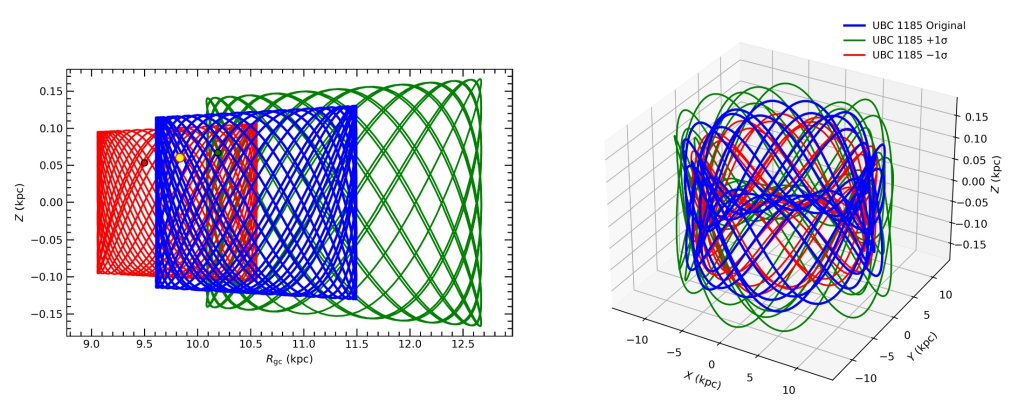}
\end{minipage}
\vspace{2mm}
\begin{minipage}{0.495\textwidth}
\centering
\includegraphics[width=\linewidth]{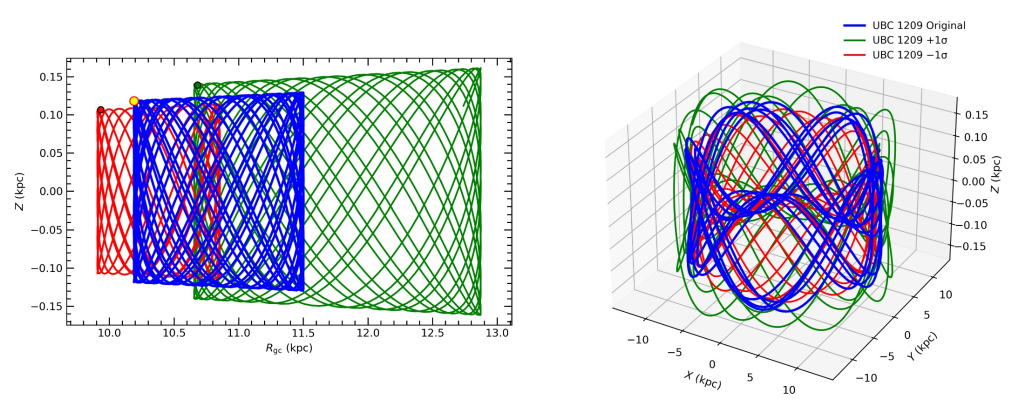}
\end{minipage}
\hfill
\begin{minipage}{0.495\textwidth}
\centering
\includegraphics[width=\linewidth]{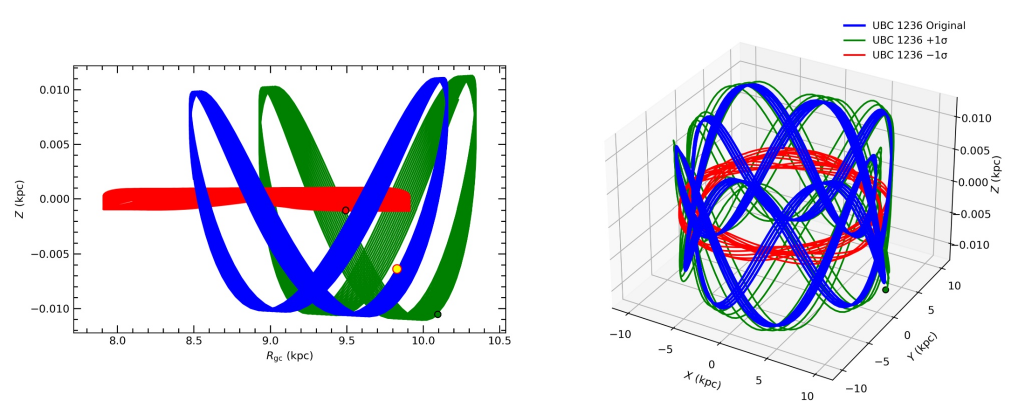}
\end{minipage}
\vspace{2mm}
\begin{minipage}{0.495\textwidth}
\centering
\includegraphics[width=\linewidth]{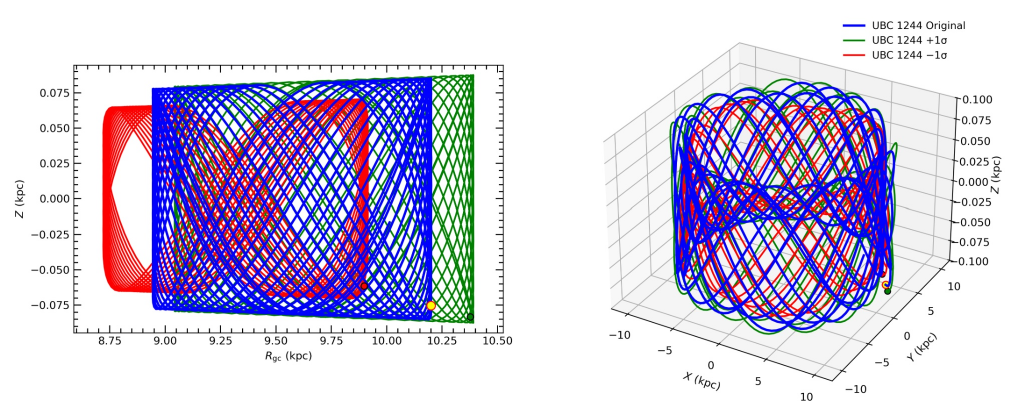}
\end{minipage}
\hfill
\begin{minipage}{0.495\textwidth}
\centering
\includegraphics[width=\linewidth]{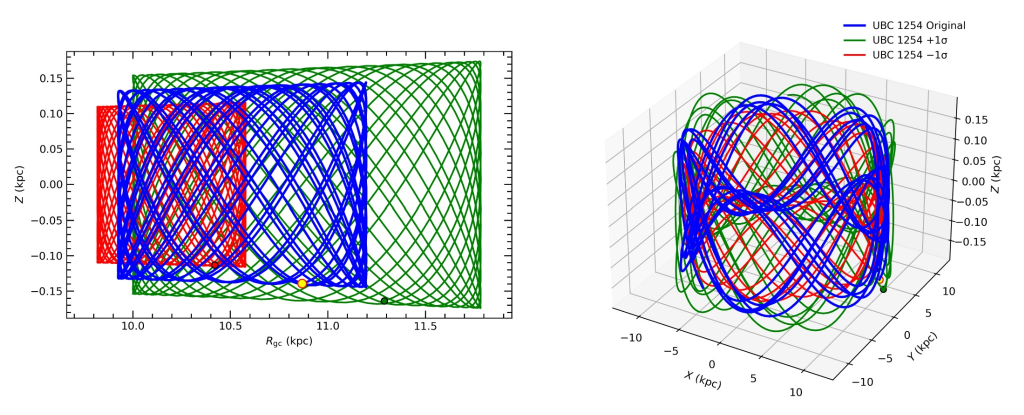}
\end{minipage}
\vspace{2mm}
\begin{minipage}{0.495\textwidth}
\centering
\includegraphics[width=\linewidth]{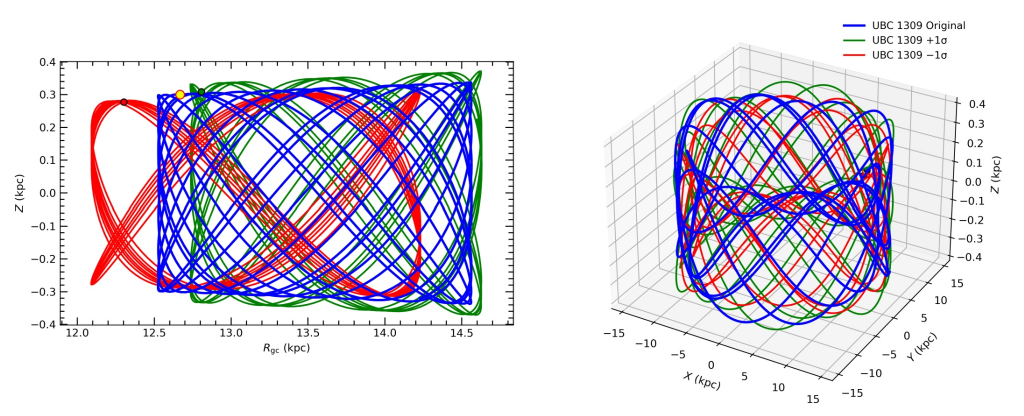}
\end{minipage}
\hfill
\begin{minipage}{0.495\textwidth}
\centering
\includegraphics[width=\linewidth]{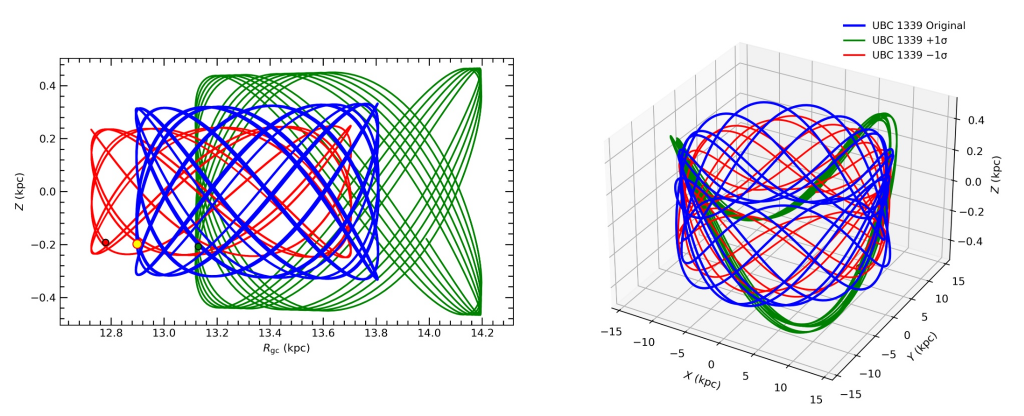}
\end{minipage}
\caption{Uncertainty-based Galactic orbital solutions for the eight UBC OCs. For each cluster, the blue curves represent the nominal/best-fit orbital solution, while the green and red curves show the orbits obtained from the upper and lower \(1\sigma\) input-parameter sets, respectively. These integrations illustrate the effect of the propagated uncertainties in the astrometric parameters, distances, and systemic radial velocities on the inferred Galactic orbital paths.}

\end{figure*}

\section{Orbit} \label{fig:orbit-append}

\begin{figure}
\centering
\includegraphics[width=0.9\linewidth]{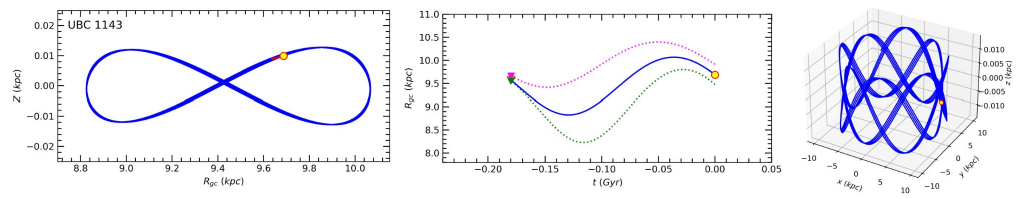}\\
\includegraphics[width=0.9\linewidth]{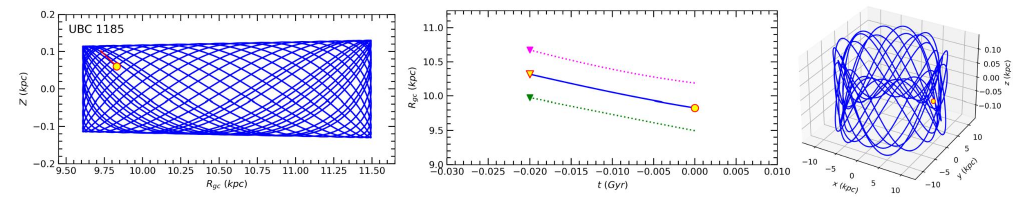}\\
\includegraphics[width=0.9\linewidth]{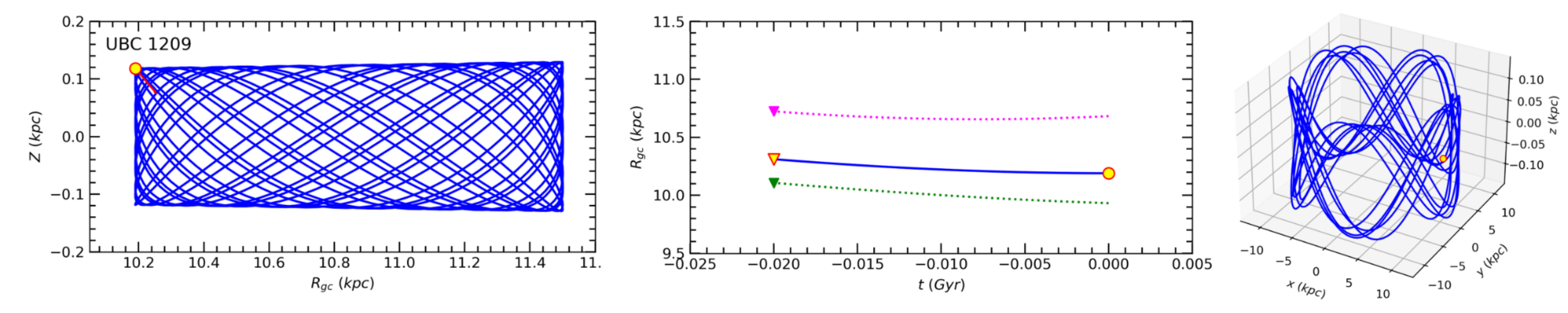}\\
\includegraphics[width=0.9\linewidth]{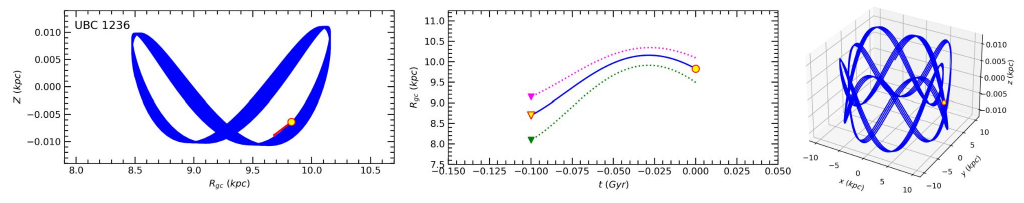}\\
\includegraphics[width=0.9\linewidth]{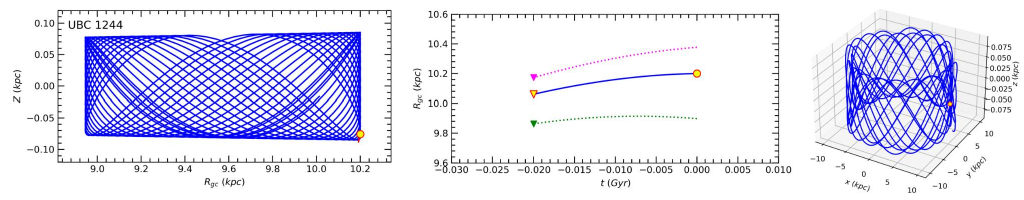}\\
\includegraphics[width=0.9\linewidth]{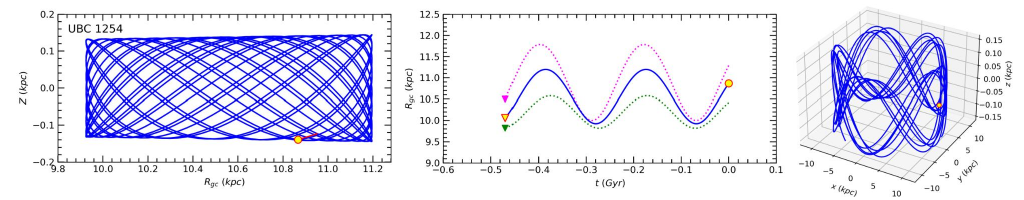}\\
\includegraphics[width=0.9\linewidth]{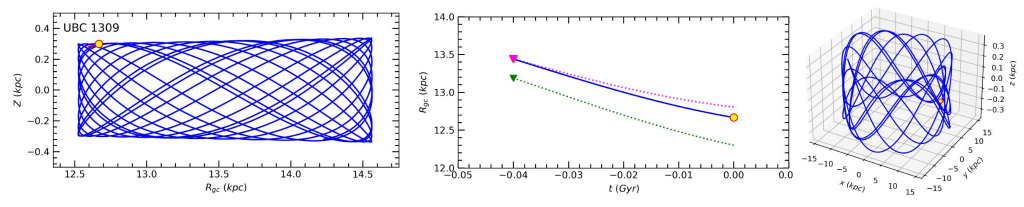}\\
\includegraphics[width=0.9\linewidth]{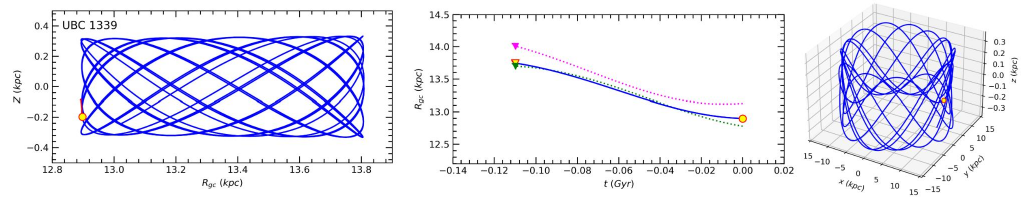}
\caption{Galactic orbits of the studied eight UBC OCs in this study.}
\end{figure}

\bibliographystyle{raa}
\bibliography{open_clusters}

\end{document}